\documentclass[11pt, oneside]{article}   	

\usepackage{jheppub}

\usepackage{amsmath}
\usepackage{amssymb}
\usepackage{amsfonts}
\usepackage{amsthm}
\usepackage{amsbsy}
\usepackage{array}
\usepackage[usenames,dvipsnames]{xcolor}
\usepackage{tikz}
\usetikzlibrary{fadings}
\usepackage{tikz-3dplot}
\usepackage{pgfplots}
\pgfplotsset{compat=1.17}

\usepackage{mathtools}

\usepackage{subcaption}

\usepackage{dsdshorthand}

\usepackage[vcentermath]{youngtab}
\newcommand{\myng}[1]{\,{\tiny\yng #1}\,}

\renewcommand\vol{\mathop{\mathrm{vol}}}

\newcommand{\thalf}{\tfrac{1}{2}}

\newcommand{\CP}{{\C \mathrm{P}}}

\newcommand{\tpsi}{{\tilde{\psi}}}
\makeatletter
\def\@fpheader{\ }
\makeatother

\makeatletter
\newcommand*{\shifttext}[2]{%
	\settowidth{\@tempdima}{#2}%
	\makebox[\@tempdima]{\hspace*{#1}#2}%
}
\makeatother

\renewcommand{\H}{{\mathbb{H}}} 

\DeclareMathOperator{\tr}{tr}

\renewcommand{\cO}{{\mathcal{O}}}

\pgfdeclareradialshading{tikz@lib@fade@circle@5}{\pgfpointorigin}{%
	color(0pt)=(pgftransparent!0); color(5bp)=(pgftransparent!0);%
	color(25bp)=(pgftransparent!100); color(50bp)=(pgftransparent!100)%
}
\pgfdeclarefading{circle with fuzzy edge 80 percent}{%
	\pgfuseshading{tikz@lib@fade@circle@5}%
}

\title{AdS $N$-body problem at large spin}
\author{Petr Kravchuk, Jeremy A. Mann}
\affiliation{
Department of Mathematics, King’s College London, Strand, London, WC2R 2LS, UK
}
\date{}
\abstract{
	Motivated by the problem of multi-twist operators in general CFTs, we study the leading-twist states of the $N$-body problem in $\AdS$ at large spin $J$. We find that for the majority of states the effective quantum-mechanical problem becomes semiclassical with $\hbar=1/J$. The classical system at $J=\oo$ has $N-2$ degrees of freedom, and the classical phase space is identified with the positive Grassmannian $\mathrm{Gr}_{+}(2,N)$. The quantum problem is recovered via a Berezin-Toeplitz quantization of a classical Hamiltonian, which we describe explicitly. For $N=3$ the classical system has one degree of freedom and a detailed structure of the spectrum can be obtained from Bohr-Sommerfeld conditions. For all $N$, we show that the lowest excited states are approximated by a harmonic oscillator and find explicit expressions for their energies.
}

\begin{document}

\maketitle

\newpage

\section{Introduction}

The spectrum of scaling dimensions $\De$ of local operators in an interacting conformal field theory is, generally speaking, extremely complex. This can be seen from basic thermodynamic considerations. Nevertheless, there are various limits in the spectrum that can be understood analytically. The focus of this work is the large spin limit $J\to \oo$ and bounded twist $\tau=\De-J$.

In this limit, simple universal structures have been observed first in perturbation theory~\cite{Callan:1973pu,Kehrein:1992fn,Kehrein:1995ia,Derkachov:1995zr,Derkachov:1996ph} and later in the general non-perturbative context~\cite{Fitzpatrick:2012yx,Komargodski:2012ek,Caron-Huot:2017vep}. Specifically, it is now largely a theorem~\cite{Pal:2022vqc, vanRees:2024xkb} that, given a pair of local primary operators $\cO_1,\cO_2$ with twists $\tau_1$ and $\tau_2$,\footnote{For simplicity, we only consider traceless-symmetric operators in this paper.} for sufficiently large spin $J$ there exists a family of local primary operators $[\cO_1\cO_2]_{J}$ of spin $J$, such that as $J\to \oo$, their twist approaches
\be
	\tau_{J}\to \tau_1+\tau_2.
\ee
There also exist subleading families whose twist asymptotes to $\tau_1+\tau_2+2n$ (see \cite{vanRees:2024xkb} for a recent mathematical proof of their existence); for simplicity, we will focus on the leading family $[\cO_1\cO_2]_J$. The operators $[\cO_1\cO_2]_J$ are referred to as the double-twist operators. A lot of work in recent years (see~\cite{Bissi:2022mrs} for a review) went into studying their properties  and, in particular, in computing the large-$J$ expansions of the anomalous dimension $\g$, which is defined as
\be
	\tau=\tau_1+\tau_2+\gamma.
\ee

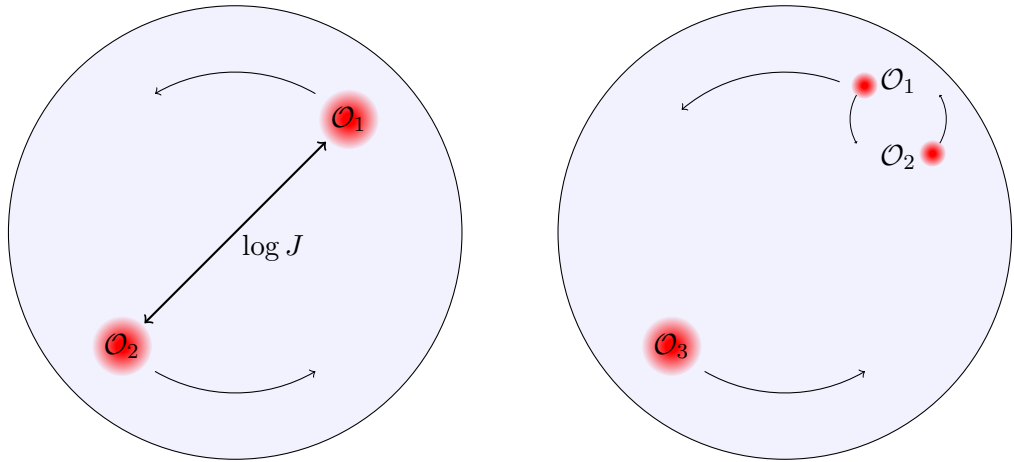
\begin{figure}[t]
	\centering
\begin{tikzpicture}
	\draw[fill=blue!5] (0,0) circle (3);
	
	\fill[fill=red,path fading=circle with fuzzy edge 80 percent] (1.5,1.5) circle (0.4);
	\fill[fill=red,path fading=circle with fuzzy edge 80 percent] (-1.5,-1.5) circle (0.4);

	\draw[->] (1.06066,1.83712) arc[start angle=60, end angle=120, radius=2.12132];
	\draw[->] (-1.06066,-1.83712) arc[start angle=225+15, end angle=225+15+60, radius=2.12132];
	
	\draw[<->,thick] (-1.2,-1.2) -- (1.2,1.2);
	\node[] at (0.5,-0.2) {$\log J$};
	\node[] at (-1.5,-1.5) {$\cO_2$};
	\node[] at (1.5,1.5) {$\cO_1$};	
\end{tikzpicture}
\hspace{1cm}
\begin{tikzpicture}
	\draw[fill=blue!5] (0,0) circle (3);
	
	\begin{scope}[transform canvas={shift={(1.5,1.5)}, scale=0.3,rotate=90}]
	\fill[fill=red,path fading=circle with fuzzy edge 80 percent] (1.5,1.5) circle (0.6);
	\fill[fill=red,path fading=circle with fuzzy edge 80 percent] (-1.5,-1.5) circle (0.6);
	\draw[->] (1.06066,1.83712) arc[start angle=60, end angle=120, radius=2.12132];
	\draw[->] (-1.06066,-1.83712) arc[start angle=225+15, end angle=225+15+60, radius=2.12132];
	\end{scope}
	\fill[fill=red,path fading=circle with fuzzy edge 80 percent] (-1.5,-1.5) circle (0.4);
	
		\draw[->] (-1.06066,-1.83712) arc[start angle=225+15, end angle=225+15+60, radius=2.12132];
	\begin{scope}[transform canvas={rotate=10}]
	\draw[->] (1.06066,1.83712) arc[start angle=60, end angle=120, radius=2.12132];
	\end{scope}	
	
	\node[] at (-1.5,-1.5) {$\cO_3$};
	\node[] at (1.5,2) {$\cO_1$};	
	\node[] at (1.5,1) {$\cO_2$};	
\end{tikzpicture}
\caption{A slice of $\AdS$ space at constant global time. Left: a two-body state in $\AdS$ at large $J$. Right: a hierarchical three-body state in $\AdS$. }
\label{fig:AdSslice}
\end{figure}

The existence of double-twist states can easily be understood holographically. The primary states created by $\cO_1$ and $\cO_2$ are each dual to localized excitations at rest in the center of the $\AdS_{d+1}$. The primary state created by $[\cO_1\cO_2]_J$ can be then viewed as the state which contains the excitation created by $\cO_1$ and the excitation created by $\cO_2$, diametrically opposed, and both orbiting around the center of $\AdS_{d+1}$, see figure~\ref{fig:AdSslice}. The geodesic distance between these excitations is proportional to $\log J$, and so in the limit $J\to \oo$ the interaction between them can be neglected, explaining the twist additivity. We will review below why it is the twist and not the scaling dimension that is additive. The fact that we are dealing with a two-body problem simplifies the calculation of the twist correction $\g$.

A natural question is whether this picture can be extended to $N$-body states or, equivalently, multi-twist operators $[\cO_1\cdots \cO_N]_J$.  While OPE coefficients of such operators have been a subject of active study in holographic CFT~\cite{Fitzpatrick:2015qma,Fitzpatrick:2019zqz,Fitzpatrick:2020yjb,Ceplak:2021wzz}, much less is known about their anomalous dimensions, especially in theories with finite central charge. An obvious approach is to build ``hierarchical'' states by iterating the double-twist construction, for example
\be\label{eq:hierarchical_states}
	[[\cO_1\cO_2]_{\ell_1}\cO_3]_J,\quad [[[\cO_1\cO_2]_{\ell_1}\cO_3]_{\ell_2}\cO_4]_J, \quad [[\cO_1\cO_2]_{\ell_1}[\cO_3\cO_4]_{\ell_2}]_J,\quad \cdots.
\ee
As long as we choose $\ell_i$ such that all the inner double-twists exist, the above expressions define double-twist families of states labeled by spin $J$. In fact, it is reasonable to expect that all Regge trajectories at large spin have a double-twist description of the above form~\cite{Henriksson:2023cnh}.

In order to be able to compute, say, the anomalous dimension of $[[\cO_1\cO_2]_{\ell_1}\cO_3]_J$ in terms of the operators $\cO_i$, we have to assume that $1\ll \ell_1\ll J$ in order for both double-twist constructions to be under analytical control. Indeed, the state $[\cO_1\cO_2]_{\ell_1}$ has size $\sim \log\ell_1$, and the separation $\log J$ between $[\cO_1\cO_2]_{\ell_1}$ and $\cO_3$ needs to be much larger than the size of either state so that the system can be viewed as a two-body problem, see figure~\ref{fig:AdSslice}. 

The condition  $1\ll \ell_1\ll J$ also guarantees that the pairwise distances between the $\cO_i$ are all large. This is important for two reasons. Firstly, it makes the interactions weak, and we can hope to understand them in a perturbative fashion. Secondly, it ensures that only the long-distance physics, at scales much larger than the AdS scale, is important. One can therefore hope that the AdS description is not essential and this class of states has calculable properties in any  (possibly non-holographic) CFT.\footnote{That the hierarchical states are calculable in non-holographic CFTs is of course known and follows from the CFT constructions of double-twist states~\cite{Fitzpatrick:2012yx, Komargodski:2012ek}. Our point here is that large AdS distances are necessary for this to be true.} An extension of this regime, where $\ell_1\sim \sqrt J$ has been studied in~\cite{Harris:2024nmr} using multi-point bootstrap.

In this paper, we study a different and the most numerous class of multi-twist states in AdS. For these states it is still true that the pairwise distances between the $\cO_i$ are all large. In particular, the interactions between the $\cO_i$ are still suppressed, and we still expect that our results should be applicable to general CFTs. However, in this class of states there is no hierarchy between the pairwise distances and therefore these states cannot be studied by iterating the solution to the two-body problem. 

\begin{figure}
	\centering
	\begin{tikzpicture}
		\draw[fill=blue!5] (0,0) circle (3);
		
		\fill[fill=red,path fading=circle with fuzzy edge 80 percent] (1.5,1.5) circle (0.4);
		\draw[->] (1.06066,1.83712) arc[start angle=60, end angle=120, radius=2.12132];
		\node[] at (1.5,1.5) {$\cO_1$};	
		
		\begin{scope}[transform canvas={rotate=120}]
			\fill[fill=red,path fading=circle with fuzzy edge 80 percent] (1.5,1.5) circle (0.4);
			\draw[->] (1.06066,1.83712) arc[start angle=60, end angle=120, radius=2.12132];
		\end{scope}
		\node[] at (-2.04904,0.549038) {$\cO_2$};	
		
		\begin{scope}[transform canvas={rotate=240}]
			\fill[fill=red,path fading=circle with fuzzy edge 80 percent] (1.5,1.5) circle (0.4);
			\draw[->] (1.06066,1.83712) arc[start angle=60, end angle=120, radius=2.12132];
		\end{scope}
		\node[] at (0.549038,-2.04904) {$\cO_3$};	
		
	\end{tikzpicture}
	\caption{A typical three-body state of the class considered in this paper.}
	\label{fig:laughlin}
\end{figure}
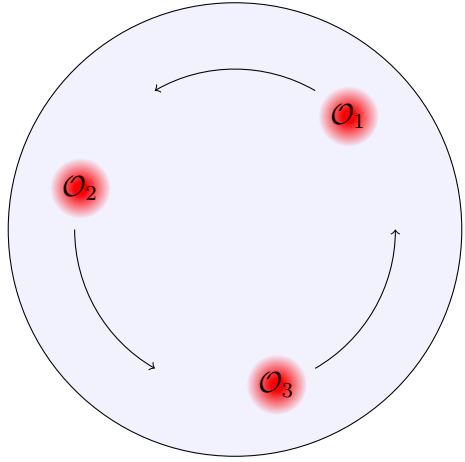

A typical representative of this class of states is illustrated in figure~\ref{fig:laughlin}, where the excitations sit at the vertices of an equilateral triangle that rotates around its center, giving rise to the total spin $J$. As we will see, when all $\cO_i$ are identical, this typically describes the state with the smallest value of $|\g|$, where the anomalous dimension $\g$ is now defined as
\be\label{eq:gammadefn}
	\tau=\tau_1+\cdots+\tau_N+\gamma.
\ee
Our goal will be to understand in detail the state in figure~\ref{fig:laughlin} as well as its excitations. We will show that such states can be described by a semiclassical quantum-mechanical problem with $\hbar=J^{-1}$. The fraction of states that can be described in this way tends to $1$ as $J\to \oo$.

\begin{figure}[t]
	\centering
	\includegraphics[scale=0.75]{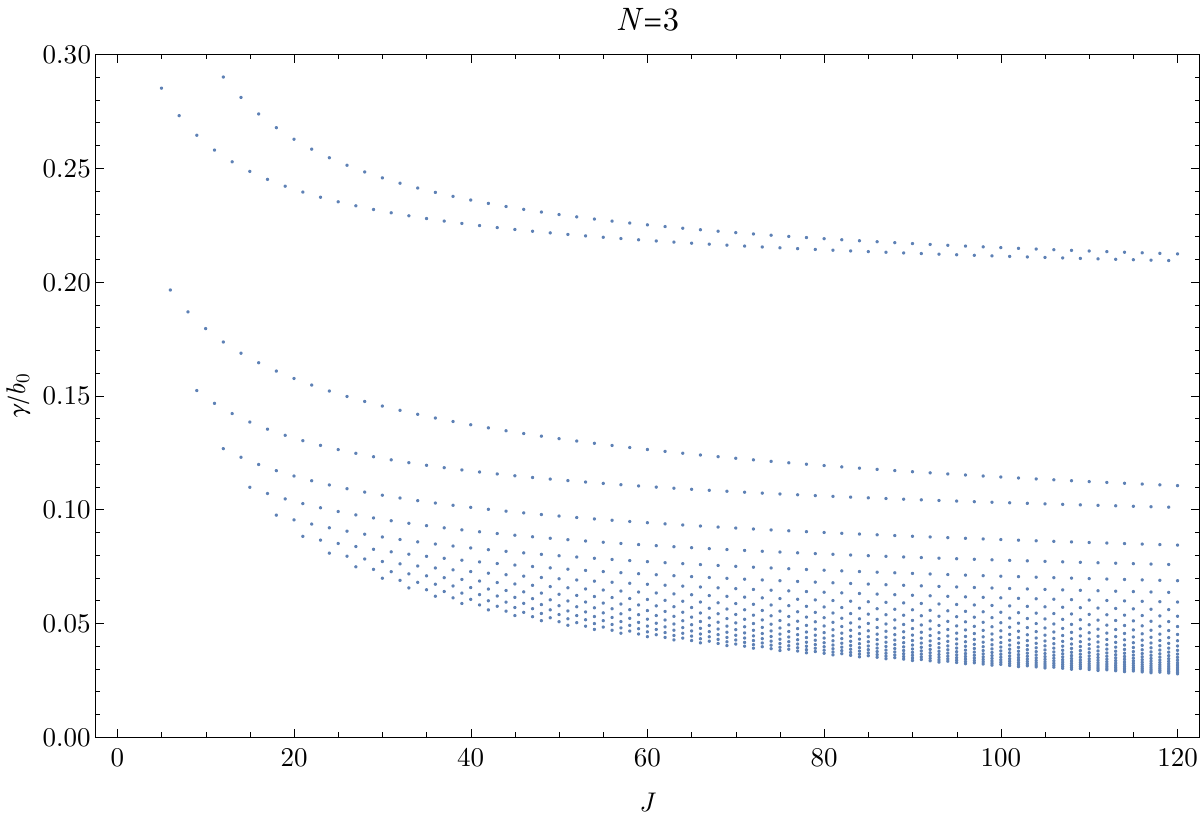}
	\caption{
		A typical spectrum of anomalous dimensions for $N=3$ leading twist states in the model~\eqref{eq:sfmodel} as a function of spin $J$. In this figure, $\De_\f = 1.234,\, \De_\s = 0.6734$, and the constant $b_0$ is defined in~\eqref{eq:b0defn}.
	}
	\label{fig:sampleN3}
\end{figure}

\begin{figure}[t]
	\centering
	\includegraphics[scale=0.75]{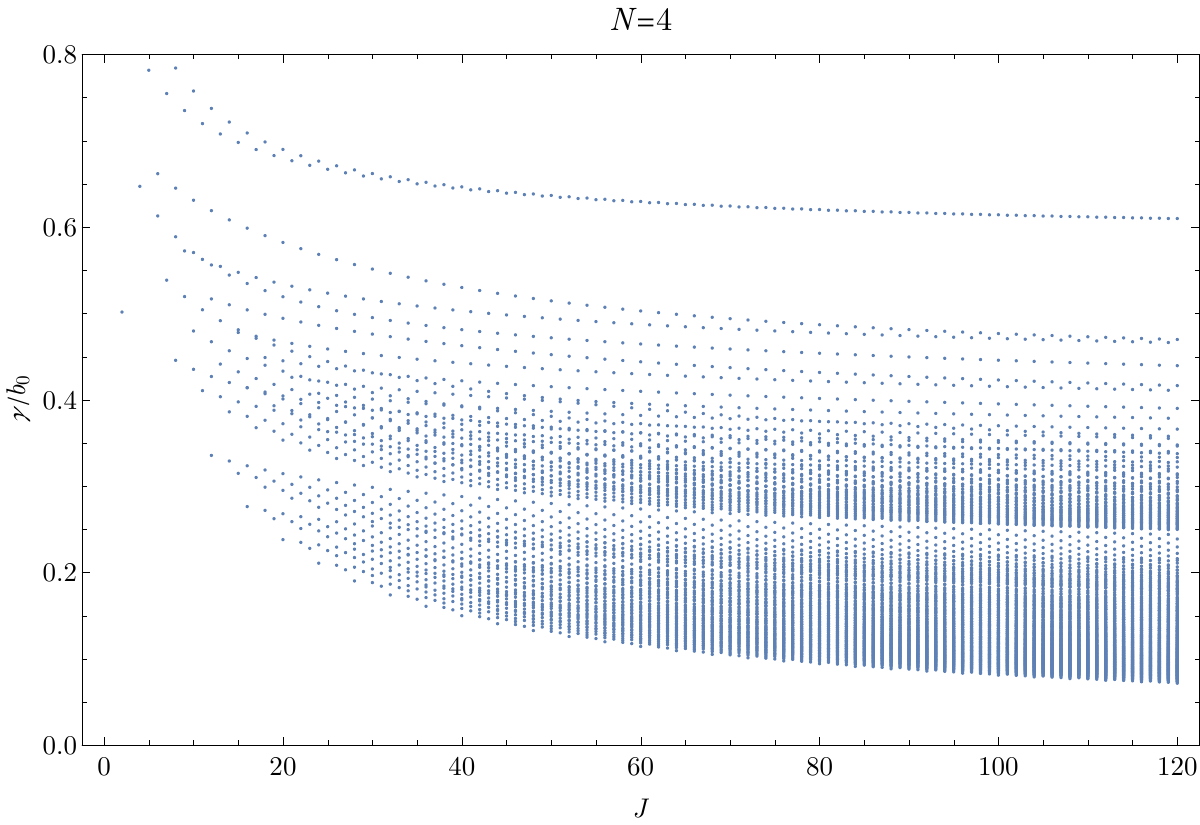}
	\caption{
		Same as figure~\ref{fig:sampleN3} but for $N=4$.
	}
	\label{fig:sampleN4}
\end{figure}

In figures~\ref{fig:sampleN3} and~\ref{fig:sampleN4} we show a typical spectrum of anomalous dimensions for $N=3$ and $N=4$ (in a model that we describe below), where all $\cO_i$ are identical scalars $\f$. The high-lying states form obvious double-twist families which approach constant values of $\g$ at large $J$. For instance, the states with the largest value of $\g/b_0$ in figure~\ref{fig:sampleN3} form the family $[[\f\f]_0\f]_J$.\footnote{The states $[[\f\f]_0\f]_J$ form two Regge trajectories: an even-spin and an odd-spin trajectory. Both are shown in figure~\ref{fig:sampleN3} and appear as separate families due to $(-1)^J$ terms in the anomalous dimension.} Lower-lying states form families $[[\f\f]_{\ell_1}\f]_J$ with even $\ell_1>0$. The limit $J\to\oo$ with $\ell_1$ fixed or growing slowly can be understood using the double-twist construction.

The state illustrated in figure~\ref{fig:laughlin} is the lowest-lying state in figure~\ref{fig:sampleN3}. In section~\ref{sec:three-body}, we will derive for $N=3$ states a Bohr-Sommerfeld rule of the form
\be
	c_0(J^{\De_\s}\g)+c_1(J^{\De_\s}\g)+\cdots = 2\pi (k+\thalf),
\ee
where each $c_n(E)$ is $J^{1-n}$ times a $J$-independent function, and $\De_\s>0$ describes the decay of interactions at large distances. Out of the $J/6+O(1)$ states at spin $J$, this condition accurately describes the spectrum of $\sim J/6$ states above and including the lowest-lying state in figure~\ref{fig:sampleN3}. We will compute the functions $c_0$ and $c_1$ explicitly, and show that this gives a very good agreement with the exact spectrum (see figure~\ref{fig:N3testRulesemiclassics} in section~\ref{sec:three-body}).

 We consider the general case $N\geq 3$ in section~\ref{sec:Nbody}. We find that unlike in the case of $N=3$, for $N>3$ the effective quantum-mechanical description has more than one degree of freedom, and only the ``leading-order Bohr-Sommerfeld rule'', i.e.\ the Weyl law can be written down. We explicitly describe the classical phase space and the classical Hamiltonian in the case of pair interactions, and we verify that this correctly predicts the semiclassical density of states (see figures~\ref{fig:montecarlo_vs_exact_4part}  and~\ref{fig:nEJ_fits} in section~\ref{sec:Nbody}). We also obtain explicit results for the lowest-lying excitations with $k\ll J$ which can be described by an effective harmonic oscillator.
 
Our methods are quite general and apply to a large class of interactions. In particular, we expect the main conclusions of this work to carry over to general, non-holographic CFTs. However, since we do not know the effective multi-twist interactions in non-holographic CFTs, in this paper we rely on an AdS toy model as an example. Specifically, we will consider a QFT of two scalars $\Phi$ and $\Sigma$ in rigid $\AdS_{d+1}$ with the action given by
\be\label{eq:sfmodel}
S=\int d^{d+1}x\sqrt{-g}\p{
	-\ptl_\mu \Phi \ptl^\mu\bar\Phi-m^2_\Phi\Phi\bar\Phi-\thalf (\ptl\Sigma)^2-\thalf m_\Sigma^2 \Sigma^2+\l \Phi\bar \Phi \Sigma
},
\ee
and where the scalar $\Phi$ is complex and carries a $U(1)$ charge $1$.\footnote{We introduce a conserved charge exclusively to avoid discussing various extraneous processes involving annihilation of pairs of $\Phi$ particles.} We denote the dual CFT operators by $\f$ and $\s$.

We will then study the leading-twist states with $U(1)$ charge $N$ at a given spin $J$, which are the states with $N$ $\Phi$ particles with Yukawa interactions mediated by $\Sigma$. Since we expect the interactions to be suppressed at large $J$, we will only focus on the leading $O(g^2)$ contribution to the anomalous dimension $\g$. One way to think about this model is that for $N=2$ it reproduces the full result for $\g$ that the Lorentzian inversion formula~\cite{Caron-Huot:2017vep,Simmons-Duffin:2017nub,Kravchuk:2018htv} gives for the $t$-channel exchange of a scalar operator $\sigma$.\footnote{Most of the time we will be interested only in the leading term in the large-$J$ expansion. In this case, there is no difference between scalar exchanges and spinning exchanges, and our results for this model can be viewed as describing the $t$-channel exchange of any local operator $\cO$, after replacing $\De_\sigma$ by $\tau_\cO$ and suitably modifying the three-point couplings.} We describe this model in more detail in section~\ref{sec:toy}.

When most of the results of this paper were in place, the work~\cite{Fardelli:2024heb} appeared which studied a similar class of AdS models. For their models, they found the spectrum of leading-twist multi-twist states numerically and compared it to the data of various CFTs. The analytical analysis of $N>2$ states in~\cite{Fardelli:2024heb} was mostly limited to the leading term in the anomalous dimension $\g$ of the state in figure~\ref{fig:laughlin}. Our focus, as mentioned above, is instead on developing a solution theory for a much more general class of states. While we do not construct models as elaborate as those of~\cite{Fardelli:2024heb}, our methods should allow an analytic calculation of many of their numerical results at large spin.

The rest of this paper is organized as follows. In section~\ref{sec:intuition} we describe an intuitive classical picture of the dynamics of leading-twist $N$-body states in AdS. In section~\ref{sec:toy} we study the model~\eqref{eq:sfmodel} in detail and verify explicitly that the AdS two-body binding energies agree with the Lorentzian inversion formula. In section~\ref{sec:three-body} we study the $N=3$ case and develop a semiclassical picture by relating it to Berezin-Toeplitz quantization. In section~\ref{sec:Nbody} we discuss the general case $N\geq 3$ and its semiclassical limit. We conclude in section~\ref{sec:discussion}. Appendices contain various details of our calculations, conventions, and a brief discussion of pseudodifferential operators.

\subsection{Some classical intuition}
\label{sec:intuition}

To gain some intuition about the dynamics of the lowest-twist states in AdS, consider the following heuristic argument. For simplicity, we first focus on an $\AdS_3$ subspace by setting some angles to $0$, and generalize to $\AdS_{d+1}$ later. In the global coordinates $(t,\r,\vf)$ the metric takes the form
\be
	ds^2_{\AdS_3}=-\cosh^2 \r \,dt^2+d\r^2+\sinh^2\r\, d\vf^2.
\ee
Here, $t\in (-\oo,+\oo)$ is the global time, $\r\in [0,\oo)$ is the radial coordinate and $\vf\in [0,2\pi)$ is the angular coordinate. The conformal boundary is at $\r=\oo$. Changing to the coordinates $(t,\r,\vf')$ where $\vf'=\vf-t$ we find
\be\label{eq:AdS3metric_2}
	ds^2_{\AdS_3}=-dt^2+d\r^2+\sinh^2\r\, d\vf'^2+2\sinh^2\r\, d\vf' dt.
\ee
The vector field $-i\ptl_t$ in the coordinates $(t,\r,\vf')$ corresponds to $-i(\ptl_t+\ptl_\vf)$ in the original coordinates $(t,\r,\vf)$. Therefore, the Hamiltonian that generates $t$ evolution in $(t,\r,\f)$ coordinates is  the twist $\tau$ (see appendix~\ref{app:conventions} and section~\ref{sec:symmetries} below for a more detailed discussion of our conventions).

Intuitively, the smallest values of $\tau$ should correspond to slowly moving particles, and so we can employ a non-relativistic approximation in the $(t,\r,\vf')$ coordinates. For a classical particle, we have the action
\be
	S&=-m\int |ds|=-m\int dt\sqrt{1-\dot\r^2-\sinh^2\r\, \dot\vf'^2-2\sinh^2\r\, \dot\vf'}\\
	&\approx \int dt \p{
		-m+\frac{m/4}{2}((2\dot\r)^2+\sinh^2\r\,\dot\vf'^2)+m\sinh^2\r\,\dot\vf'
	},
\ee
where we have expanded to the second order in velocities. The piece quadratic in velocities can be interpreted as the non-relativistic kinetic term for a particle of mass $m_\text{eff}=m/4$ in the hyperbolic disk $\D$ parameterised by the polar coordinates $(\r',\vf')$, where $\r'=2\r$ and the metric is given by
\be
	ds^2_\D=d\r'^2+\sinh^2 \r' d\vf'^2.
\ee
The term linear in derivatives can be interpreted as 
\be
	m\int dt \dot\vf'\sinh^2\r' = \int  A,
\ee
where $A=A_{\vf'} d\vf'=m\sinh^2\frac{\r'}{2} d\vf'$. This coincides with the action of a charge-one particle in the electromagnetic field with gauge potential $A$. The corresponding field strength is
\be
	F=dA=\frac{m}{2}\sinh \r' d\r'\wedge d\vf'=\frac{m}{2}{\vol}_\D,
\ee
where ${\vol}_\D=\sinh \r' d\r'\wedge d\vf'$ is the volume form on the hyperbolic disk. Thus, we find an effective magnetic field of constant magnitude $B_\text{eff}=m/2$. 

If we formally quantize this particle, we expect to find Landau levels split by the cyclotron frequency\footnote{Recall that we are working in dimensionless units since we have set the $\AdS$ radius $R$ to $1$.} $\w_c = B_\text{eff}/m_\text{eff}=2$. This is valid at least as long as $B_\text{eff}$ is large so that the particle is localized on scales smaller than hyperbolic curvature.\footnote{The exact spectrum of Landau levels in hyperbolic space is given by $n(n+B_\text{eff}-1)/m_\text{eff}$~\cite{Comtet:1986ki}.} Therefore, this naive picture predicts the low-twist spectrum of one-particle states to be given by
\be
	\tau = 2n+\text{const}.
\ee
This agrees with the expectation from a $d=2$ boundary theory, where different descendants $\ptl_-^m \ptl_+^n\cO(0)$ of a (quasi-)primary operator $\cO(0)$ have twist $\tau=\tau_\cO+2n$.
However, now we can interpret the lowest-twist states with $n=0$ as corresponding to the lowest Landau level (LLL) in the hyperbolic disk $\D$. 
\begin{figure}[t]
	\centering
	\begin{tikzpicture}[scale=1.5]
		\draw[] (-1.2,0)--(-1.2,2.6);
		\draw[] (1.2,0)--(1.2,2.6);
		\tdplotsetmaincoords{60}{120}
		\begin{scope}[tdplot_main_coords]
			\draw[red,thick, domain=0:540, samples=200, smooth, variable=\t]
			plot ({cos(\t)}, {sin(\t)}, {2*\t/360});
			\draw[green,thick, domain=0:540, samples=200, smooth, variable=\t]
			plot ({cos(\t+120)}, {sin(\t+120)}, {2*\t/360});
			\draw[blue,thick, domain=0:540, samples=200, smooth, variable=\t]
			plot ({cos(\t+240)}, {sin(\t+240)}, {2*\t/360});
			
			\draw[opacity=0.2, fill=blue!30] (0,0,0) circle [radius=1.2];
			\draw[] (0,0,3) circle [radius=1.2];
		\end{scope}
	\end{tikzpicture}
	\caption{Spiraling geodesics in AdS\label{fig:spirals}.}
\end{figure}
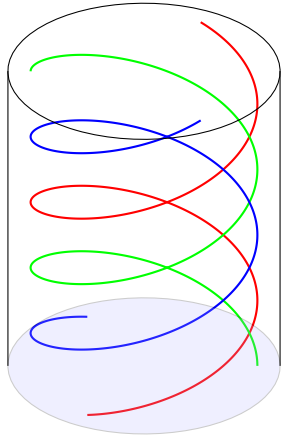

Two comments are in order. Firstly, the above discussion took place in $\AdS_3$. In $\AdS_{d+1}$ there are additional transverse degrees of freedom. It turns out that for the leading twist states these degrees of freedom are not excited and the LLL picture is still valid. Secondly, the above discussion is rather heuristic, starting with a non-relativistic limit of a classical particle. Fortunately, as we will show in section~\ref{sec:toy}, the same LLL Hilbert (sub-)space can identified in the full quantum field theory of free scalars on $\AdS_{d+1}$. In particular, none of our calculations depend on the LLL interpretation, and we only use the LLL picture to motivate our discussion. With these comments in mind, let us discuss the implications of the LLL picture for the large-spin dynamics.

Recall that the LLL is infinitely-degenerate. Physically, different LLL states can be visualized as being localized at different points of the hyperbolic disk $\D$. It is useful to think about $\D$ as being the effective phase space for the LLL particle. One then expects one state per $2\pi$ phase space volume. 

By construction, the Hamiltonian is a constant when restricted to LLL,
\be
	H=\tau=\tau_{0}.
\ee
Therefore, in the absence of interactions, the LLL particles are stationary in $(t,\r,\vf')$ coordinates. Using $\vf=\vf'+t$, we see that these states can be visualized as spiraling geodesics in $\AdS_3$, see figure~\ref{fig:spirals}.

Lowest-twist multi-particle states can be visualized as several LLL particles localized at different points on $\D$. We will see in section~\ref{sec:toy} that at least some $\AdS_{d+1}$ interactions can be reduced in LLL to instantaneous pair potentials $U_{ij}$. When such potentials are added, the effective Hamiltonian becomes non-trivial,
\be
	\tau=\tau_0+\sum_{i<j} U_{ij}.
\ee
Interpreting this classically, Hamilton's equations imply that the particles move with velocities proportional to derivatives of $U_{ij}$. This slow motion gives a small correction to eigenvalues of $\tau$, which is what we want to calculate.

\section{$\mathrm{AdS}_{d+1}$ toy model}
\label{sec:toy}

\subsection{Summary of the model and first-order perturbation theory}

The toy model in AdS$_{d+1}$ is a two-scalar field theory with cubic coupling.  We denote the bulk scalar fields by $\Phi,\Sigma$ and their boundary CFT operators by $\phi,\s$, such that the masses and the conformal dimensions are related by $m^2_\Phi = \De_\phi(\De_\phi-d)$,  $m^2_\Sigma = \De_\s(\De_\s-d)$.  To slightly simplify the discussion, we take $\Phi$ to be a complex scalar charged under a $U(1)$ symmetry. The cubic coupling is $\cL_{\text{int}} = \l \Phi\bar\Phi \Sigma$, and the full action is given by
\be
	S=\int d^{d+1}x\sqrt{-g}\p{
		-\ptl_\mu \Phi \ptl^\mu\bar\Phi-m^2_\Phi\Phi\bar\Phi-\thalf (\ptl\Sigma)^2-\thalf m_\Sigma^2 \Sigma^2+\l \Phi\bar \Phi \Sigma
	}
\ee

To study the multi-particle states of $\Phi$, it is convenient to integrate $\Sigma$ out to produce a quartic potential in $\Phi$:
\begin{equation}
e^{i S_\text{int}[\Phi]} = \int D\Sigma \,e^{i\int d^{d+1}x\sqrt{-g}\p{
	-\thalf (\ptl\Sigma)^2-\thalf m_\Sigma^2 \Sigma^2+\l \Phi\bar \Phi \Sigma
}}
\end{equation}
The result is
\be
S_\text{int}[\Phi] &= \int dt\, V(t),\\
V(t) &= \frac{\l^2}{2} \int d^{d+1} x_1 \sqrt{-g} d^{d+1} x_2 \sqrt{-g} K_\Sigma(x_1,x_2) :\mathrel{\Phi\bar \Phi(x_1)\Phi\bar\Phi(x_2)}:\de(t_1-t),\label{eq:V(t)}
\ee
where $K_\Sigma$ is the propagator for $\Sigma$.
To the leading order in $\l$, the interaction energies can be obtained from the Rayleigh-Schr\"odinger perturbation theory for $V$ in the Hilbert space of the free theory, see~\cite{Fitzpatrick:2011hh,Fardelli:2024heb}\footnote{Note that $V(t)$ is only time-dependent through the Heisenberg evolution, it has no ``explicit'' time-dependence. To stress this, we will write $V$ instead of $V(t)$ in what follows.}. In the rest of this section we will first describe the free Hilbert space and then compute the matrix elements of $V$.

Note that the above description is of our toy model, in which all approximations are easily controlled from first principles. However, as discussed in more detail in the introduction, we believe that the main results of this work apply more generally. In particular, we expect that neither the assumption of $\l$ being small, nor the restriction to simple $\Sigma$-exchange interactions are essential for describing the leading-twist states at large spin.

\subsection{Symmetries}
\label{sec:symmetries}

To simplify the forthcoming calculations, it helps to carefully consider the symmetries of the problem. The $\AdS_{d+1}$ isometries coincide with the conformal symmetries of the boundary and are generated in Euclidean signature by the standard generators $D,P_\mu, K_\nu, M_{\mu\nu}$, with indices running over $\{1,\cdots, d\}$. These operators satisfy the hermiticity conditions $D^\dagger=D,P_\mu^\dagger=K_\mu, M_{\mu\nu}^\dagger=-M_{\mu\nu}$. When acting on the Hilbert space $\cH$ of states on the unit sphere in the boundary $\R^d$, these generators have the familiar spectrum given by the operator-state correspondence. It is possible to choose coordinates (see appendix~\ref{app:conventions}) so that this unit sphere becomes the $t=0$ spatial slice of the Lorentzian cylinder $\R\x S^{d-1}$, where $t$ is the global time. Therefore, $\cH$ coincides with the Hilbert space of the AdS theory. Appropriate complex linear combinations of the above generators then span the isometries of the Lorentzian $\AdS_{d,1}$.

The $\AdS_{d,1}$ metric in global coordinates is
\be\label{eq:ds2_globalcoords}
	ds^2_{\AdS_{d,1}}=-\cosh^2\r dt^2+d\r^2+\sinh^2\r \, ds_{S^{d-1}}^2,
\ee
where $ds^2_{S^{d-1}}$ is the radius-1 round metric on the sphere $S^{d-1}$ parameterized by a unit vector $n\in \R^{d}$. Using the conventions in appendix~\ref{app:conventions}, it is easy to check that the action of $D$ on bulk operators is
\be
	[D,\cO(x)]=-i\ptl_t \cO(x),
\ee
and thus $D$ becomes the Hamiltonian for $t$ translations.

We choose a preferred rotation generator $M_{12}$ and define the twist generator as
\be
	\tau = D+i M_{12}.
\ee
We also introduce the following coordinates on $S^{d-1}$:
\be\label{eq:angles_globalcoords}
	n^1=\cos \theta \cos\vf,\quad 
	n^2=\cos \theta \sin\vf,\quad 
	n^i=\sin \theta \,\hat n^i\quad (i=3,\cdots d),
\ee
where $\theta\in [0,\pi/2]$, $\vf\in [0,2\pi)$ and $\hat n\in S^{d-3}$. In these coordinates, the twist generator acts on local fields as
\be
	[\tau,\cO(x)]=-i(\ptl_t+\ptl_\vf)\cO(x).
\ee
Therefore, if we define $\vf'=\vf-t$, then in the coordinates $(t,\r,\vf',\theta,\hat n)$ the Hamiltonian for $t$ translations coincides with $\tau$.

Under the adjoint action of $\tau$, the conformal algebra splits into eigenspaces with eigenvalues $0$ and $\pm 1$. To describe these eigenspaces, it is convenient to introduce the components $v^\pm$ of a vector $v^\mu$, defined as	$v^\pm = -i v^1\mp v^2$. The $\pm$ component has charge $\pm 1$ under $iM_{12}$, which implies
\be\label{eq:taucommutators}
	&[\tau, P^-]=[\tau, K^+]=[\tau,\bar\tau] =[\tau, M^{ij}]=0, \nn\\
	&[\tau, M^{\pm i}]=\pm M^{\pm i}, \quad [\tau, P^+] = 2P^+, \quad [\tau, K^-]=-2K^-,\nn\\
	&[\tau, P^i]=P^i,\quad [\tau, K^i]=-K^i.
\ee
where $i,j\geq 3$ and $\bar \tau = D-iM_{12}$. The generators commuting with $\tau$ generate a $\mathfrak{so}(2,1)\x\mathfrak{so}(d-2)$ subalgebra (not counting $\tau$ itself). 

\subsection{Minimal twist Hilbert space}
\label{sec:minimal-twist}

Our first goal is to describe the Hilbert space of the free theory of the field $\Phi$. For simplicity, in this paper we focus on the lowest-twist subspace of a given $U(1)$ charge. The free-theory Hilbert space splits into multi-particle sectors, which can be understood as properly symmetrized tensor products of the single-particle subspace. The single-particle subspace (say, containing a single $\Phi$ particle) forms a (generically) irreducible Verma module of the conformal algebra, which we denote by $\cH_{1}^\text{full}$. The primary of this Verma module is a scalar operator of dimension $\De_\phi$.

A useful characterization of a one-particle state $|\Psi\>$ is given by its wave-function $\Psi(x)=\<0|\bar\Phi(x)|\Psi\>$. The lowest-twist states have to be annihilated by all negative-twist generators, see~\eqref{eq:taucommutators},
\be\label{eq:lowesttwistcondition}
	K^-|\Psi\>=K^i|\Psi\>=M^{-i}|\Psi\>=0,
\ee
where $i\geq 3$. This implies that the wave-function $\Psi(x)$ satisfies a set of differential equations
\be
	\cK^-\Psi(x)&=\cK^i\Psi(x)=0,\label{eq:wfequations}\\
	\cM^{-i}\Psi(x)&=0,
\ee
where $\cK$ and $\cM$ denote the vector fields associated to the respective generators. It turns out that the constraint $\cM^{-i}\Psi(x)=0$ is redundant and follows from~\eqref{eq:wfequations}.

The solution is conveniently written in terms of
\be\label{eq:alpha_def}
	\a=e^{i\varphi'}\cos\theta\tanh\r,
\ee
which is $t$-independent and also satisfies $\cK^-\a=\cK^i\a=0$. In fact, any $t$-independent function which satisfies~\eqref{eq:wfequations} is a function of $\a$. Indeed, the vector fields $\cK^-$, $\cK^{i}$, and $\ptl_t$ are linearly-independent and there are $d=\dim \AdS_{d+1}-1$ of them. Note that by construction $\a$ takes values in the unit disk, $|\a|<1$.

We know that the lowest twist states in the one-particle Verma module satisfy $\tau|\Psi\>=\De_\phi|\Psi\>$, which implies 
\be\label{eq:wftwist}
	\ptl_t\Psi(x) = -i\De_\phi \Psi(x)
\ee
It is now not hard to find the general solution of~\eqref{eq:wfequations} and \eqref{eq:wftwist}, which we express as
\be
	\Psi(x) = C_{\De_\phi,d}^{1/2} \frac{e^{-i\De_\phi t}}{(\cosh\r)^{\De_\phi}}\psi(\a), \quad C_{\De,d}:= \frac{2 \pi^{d/2}\Gamma(\De)}{\Gamma(\De-d/2+1)},
	\label{eq:Psi1particle}
\ee
where $\psi(\a)$ is an arbitrary holomorphic function of $\a$, and $C_{\De_\phi,d}^{1/2}$ is multiplicative constant that we factor out for future convenience. We can therefore identify the states in the lowest-twist subspace $\cH_1$ of $\cH_1^\text{full}$ with the corresponding wavefunctions $\psi(\a)$.

The minimal twist subspace $\cH_1$ naturally forms a representation of the twist-0 subalgebra $\mathfrak{so}(2,1)\x \mathfrak{so}(d-2)$. The quantum number of $\mathfrak{so}(2,1)$ is the conformal spin $\bar h =\bar\tau/2$, while the $\mathfrak{so}(d-2)$ representation labels are transverse spins. Since the generators $M^{ij}$ spanning $\mathfrak{so}(d-2)$ act trivially on $\psi(\a)$, the latter form representations with zero transverse spin. The action of $\mathfrak{so}(2,1)$ is conveniently expressed in terms of the generators 
\be
	L_0=\thalf \bar\tau, \quad L_+=\tfrac{i}{2}K^{+}, \quad L_-=\tfrac{i}{2}P^{-},
\ee
which have the standard commutation relations $[L_m,L_n]=(m-n)L_{m+n}$. Acting on $\psi$, we have
\be\label{eq:Laction}
	(L_0\psi)(\a)&=(\a\ptl_\a+\De_\phi/2)\psi(\a),\nn\\
	(L_+\psi)(\a)&=\ptl_\a\psi(\a),\nn\\
	(L_-\psi)(\a)&=(\a^2\ptl_\a+\De_\f \a)\psi(\a).
\ee
As expected, this defines a lowest-weight representation of $\mathfrak{so}(2,1)$ with $\bar h=\De_\f/2$, where the lowest weight state satisfies 
\be
L_+\psi=0,\quad L_0\psi = \bar h \psi,
\ee
and is given by $\psi(\a)\equiv 1$. The states with definite $L_0$ weights $\bar h+n$ are simply the monomials $\psi(\a)=\a^n$, where $n\in \Z_{\geq 0}$.

The Hermiticity conditions $L_n^\dagger = L_{-n}$ fix the inner product in $\cH_1$ up to normalization. In terms of $\psi(\a)$ it is given by
\be
	\<\psi_1|\psi_2\> := \frac{\De_\f-1}{\pi} \int_{\mathbb{D}} d^2\a (1-\a\bar\a)^{2\De_\f-2} \,\bar{\psi_1(\a)}\psi_2(\a),
	\label{eq:scalar_prod}
\ee
where $\mathbb{D}$ is the unit disk $|\a|<1$, and the multiplicative constant $(\De_\phi-1)/\pi$ is chosen such that the lowest-weight state $\psi(\a)=1$ has unit norm. At the same, the multiplicative constant $C_{\De_\phi,d}^{1/2}$ entering the relation between $\Psi(x)$ and $\psi(\a)$ in~\eqref{eq:Psi1particle} ensures that the scalar product~\eqref{eq:scalar_prod} descends exactly from the canonical scalar product of the AdS field: $\<\Psi_1|\Psi_2\> = \<\psi_1|\psi_2\>$.

Note that although the wavefunction $\psi(\a)$ is parameterized by $\a$ in the unit disk $\D$, and there is an action of $\mathfrak{so}(2,1)$ on $\D$, there isn't a natural embedding of $\D$ into $\AdS_{d,1}$ that respects this action. For example, $L_0=\thalf \bar\tau$ necessarily generates translations along the non-compact time direction. Relatedly, even though the anti-Hermitian combinations of $L_n$ act by hyperbolic isometries on $\D$, on the Hilbert space the Lie algebra $\mathfrak{so}(2,1)$ exponentiates to the universal cover $\tl{\SO}(2,1)\subseteq \tl \SO(2,d)$, rather than the hyperbolic isometry group $\SO(2,1)$ of~$\D$.

We now consider multi-particle states, which are obtained through the usual Fock space construction as symmetrized tensor products of the single-particle states, $\cH_N=(\cH_1^{\otimes N})^{S_N}$. For the leading-twist states we find
\be\label{eq:PsiNparticle}
	\Psi(x_1,\cdots, x_N)=\psi(\a_1,\cdots,\a_n)\prod_{i=1}^N C_{\De_\f,d}^{1/2} \frac{e^{-i\De_\phi t_i}}{(\cosh \r_i)^{\De_\phi}},
\ee
where the wavefunction $\Psi$ is defined as
\be\label{eq:PsiDefn}
	\Psi(x_1,\cdots, x_N)=\<0|\bar\Phi(x_1)\cdots \bar \Phi(x_N)|\Psi\>.
\ee
Bose symmetry implies that $\psi(\a_1,\cdots,\a_n)$ is symmetric in its arguments.

Defining the spin $J$ as $J=-iM_{12}=\tfrac{\bar\tau-\tau}{2}$, we find that it acts on $\psi(\a)$ by
\be
	(J\psi)(\a_1,\cdots, \a_N)=\sum_{i=1}^N \a_i \ptl_{\a_i}\psi(\a_1,\cdots,\a_N).
\ee
In other words, states with definite spin $J$ are homogeneous polynomials in $\a_i$ of total degree~$J$. It can be verified that such $\psi$ are the highest-weight vectors in the traceless-symmetric spin-$J$ representation of $\SO(d)$. We will use $\cH_{N,J}\subset \cH_N$ to denote the Hilbert space of lowest-twist states at spin $J$.

In terms of $\psi$, the inner product on the $N$-particle states becomes simply
\be\label{eq:N-particle-inner}
	\<\psi_1|\psi_2\>=\frac{(\De_\f-1)^N}{\pi^N N!}\int_\D d^{2N}\a\prod_{k=1}^N(1-\a_k\bar \a_k)^{\De_\f - 2}\,\bar{\psi_1(\a_1,\cdots, \a_n)}\psi_2(\a_1,\cdots,\a_n).
\ee

\paragraph{Primary states} Due to~\eqref{eq:lowesttwistcondition}, the lowest-twist states are primary if and only if they are annihilated by $K^+\propto L_+$. This happens precisely when
\be
	\sum_{i=1}^N\ptl_{\a_i} \psi(\a_1,\cdots,\a_N)=0,
\ee
see~\eqref{eq:Laction}. We therefore find that the lowest-twist traceless-symmetric spin-$J$ primaries in the $N$-particle Hilbert space are in one-to-one correspondence with wavefunctions $\psi(\a_1,\cdots \a_N)$ that are symmetric, homogeneous of degree $J$, and translation-invariant polynomials in the $\a_i$. We will denote the vector space of such wavefunctions by $\cH_{N,J}^\text{primary}\subset \cH_{N,J}$.

For example, the unique lowest-twist primary in the 1-particle Hilbert space is given by $\psi(\a)=1$ and has spin-$0$. The two-particle Hilbert space has a unique primary state at every even spin $J$, given by
\be
	\psi(\a_1,\a_2)=(\a_1-\a_2)^J.
\ee
To enumerate the $N$-particle primary states, we define the translation-invariant combinations
\be
	\b_i = \a_i-\frac{1}{N}\sum_{k=1}^N \a_k.
\ee
A basis of states at spin $J$ is then given by the wavefunctions
\be\label{eq:sym_poly_basis}
	\psi(\a_1\cdots\a_N)=f_m(\a):= e_2(\b_1\cdots \b_N)^{m_2}\cdots e_N(\b_1\cdots \b_N)^{m_N},
\ee
where the non-negative integers $m_k$ satisfy $\sum_{k=2}^N k m_k=J$ and $e_n$ are the elementary symmetric polynomials.  Note that $e_1(\b_1,\cdots,\b_n)=\sum_i\b_i=0$ and thus does not appear.

The number of lowest-twist states at spin $J$ is therefore equal to the number of partitions of $J$ into integers from $2$ to $N$. We have the generating function
\be
	\sum_{J=0}^\oo \p{\dim \cH_{N,J}^\text{primary}} x^J = \prod_{k=2}^N(1-x^k)^{-1}.
\ee
Explicit expressions for a given $N$ can be obtained by computing
\be
	\dim \cH_{N,J}^\text{primary}=\frac{1}{2\pi i}\oint_{|x|=\half} dx x^{-J-1}\prod_{k=2}^N(1-x^k)^{-1}
\ee
via the residue theorem as the integration contour is deformed to infinity. It is not hard to check that the leading term at large $J$ comes from the residue at $x=1$ and gives
\be\label{eq:dimHNJ}
	\dim \cH_{N,J}^\text{primary}=\frac{J^{N-2}}{N!(N-2)!}+O(J^{N-3}).
\ee
In particular, for $N=3$ we find
\be\label{eq:N3dimension}
	\dim \cH_{3,J}^\text{primary}=J/6+O(1),
\ee
where the $O(1)$ term only depends on $J\!\!\mod 6$.

\subsection{Effective pair potential}
\label{sec:eff_pair_pot}

We now consider the problem of finding the leading correction from the potential~\eqref{eq:V(t)} to the energies of the leading-twist states with $N$ $\Phi$-particles. If we wanted to compute the correction to the energy of a state $|\Psi\>$ that is non-degenerate in the free theory, it would be as simple as computing the expectation value $\<\Psi|V|\Psi\>$. However, as discussed above, the energies of $N$-particle states are highly degenerate at large spin $J$. Therefore, we need to use the degenerate perturbation theory, which instructs us to diagonalize the restriction of $V$ to the lowest-twist degenerate $N$-particle subspace $\cH_{N,J}$, or more generally to $\cH_{N}$.

The simplest way to characterize this restriction is via the matrix elements
\be
	\<\Psi_1|V|\Psi_2\>,\quad \Psi_1,\Psi_2\in\cH_{N}.
\ee
Due to our choice of $V$, given by~\eqref{eq:V(t)}, the only non-trivial calculation to do is in the case $N=2$. We will find that in terms of wavefunctions $\psi$ the matrix elements are given by (c.f.~\eqref{eq:N-particle-inner})
\be\label{eq:two-particle-matrix-element}
	\<\Psi_1|V|\Psi_2\>=\frac{(\De_\f-1)^2}{\pi^2 2!}\int_\D d^{4}\a\prod_{k=1}^2(1-\a_k\bar \a_k)^{\De_\f - 2}\bar{\psi_1(\a_1,\a_2)}\psi_2(\a_1,\a_2) U_2(s_{12}),
\ee
where
\be
	\quad s_{12} := \sinh^2 \tfrac{\bd_{12}}{2} = \frac{(\a_1-\a_2)(\bar\a_1-\bar\a_2)}{(1-\a_1\bar\a_1)(1-\a_2\bar\a_2)} \in [0,\infty),
\ee
 $\bd_{12}$ is the hyperbolic distance between $\a_1$ and $\a_2$ in $\D$, and the function $U_2(s)$ is given in~\eqref{eq:scalar_potential_result} below. Note that when $\bd_{12}$ is large, it is double the (extremal) geodesic distance between the codimension-two surfaces of constant $\a$ in $\AdS_{d,1}$.
 
 In the case of general $N$, the matrix elements are given simply by the sum over the pairwise interactions,
\be\label{eq:matrixelementgeneral}
\<\Psi_1|V|\Psi_2\>=\frac{(\De_\f-1)^N}{\pi^N N!}\int_\D d^{2N}\a\prod_{k=1}^N(1-\a_k\bar \a_k)^{\De_\f - 2}\bar{\psi_1(\a_1,\cdots, \a_n)}\psi_2(\a_1,\cdots,\a_n)\sum_{i<j}U_2(s_{ij}).
\ee

To derive the form of $U_2(s)$, we begin with the explicit expression for the matrix element that follows from the definitions~\eqref{eq:V(t)} of $V$ and~\eqref{eq:PsiDefn} of the wave-functions $\Psi(x)$,
\be
	\<\Psi_1|V|\Psi_2\>=\frac{\l^2}{2} \int d^{d+1} x_1 \sqrt{-g} d^{d+1} x_2 \sqrt{-g} K_\Sigma(x_1,x_2) \de(t_1)\Psi_2(x_1,x_2)\bar{\Psi_1(x_1,x_2)}.
\ee
As the matrix elements are $t$-independent, we set $t=0$. Plugging in the lowest-twist wavefunctions~\eqref{eq:PsiNparticle}, we get
\be
	\<\Psi_1|V|\Psi_2\>=\frac{\l^2}{2} \int d^{d+1} x_1 \sqrt{-g} d^{d+1} x_2 \sqrt{-g} \frac{C_{\De_\f,d}^2K_\Sigma(x_1,x_2)\de(t_1)}{(\cosh\r_1)^{2\De_\phi}(\cosh\r_2)^{2\De_\phi}} \psi_2(\a_1,\a_2)\bar{\psi_1(\a_1,\a_2)}.
\ee
We therefore find
\be
	\<\Psi_1|V|\Psi_2\>= \frac{(\De_\f-1)^2}{\pi^2 2!} \int \prod_{k=1}^2d^2\a_k(1-\a_k\bar \a_k)^{\De_\f - 2}\psi_2(\a_1,\a_2)\bar{\psi_1(\a_1,\a_2)} F(\a_1,\a_2),
\ee
where the function $F$ is defined by
\be\label{eq:Fdefinition}
	F(\a_{0,1},\a_{0,2})\equiv \frac{\pi^2\l^2 C_{\De_\f,d}^2}{(\De_\phi-1)^2} \int\prod_{k=1}^2 \frac{d^{d+1} x_k \sqrt{-g}\, \de^2(\a_k-\a_{0,k})}{(1-\a_k\bar \a_k)^{\De_\f - 2}(\cosh\r_k)^{2\De_\phi}}K_\Sigma(x_1,x_2) \de(t_1).
\ee

The key point is that the function $F$ is $\mathfrak{so}(2,1)$-invariant, where $\mathfrak{so}(2,1)$ acts on $\a$'s by hyperbolic isometries of $\D$. Indeed, the factor
\be
	\frac{\de^2(\a_k-\a_{0,k})}{(\cosh\r_k)^{2\De_\phi}(1-\a_k\bar \a_k)^{\De_\f - 2}}
\ee
can be verified to be $\mathfrak{so}(2,1)$-invariant by an explicit calculation, and $K_\Sigma(x_1,x_2)$ is invariant under the full $\tl{\SO}(d,2)$. The only suspect factor is $\de(t_1)$, which is not invariant. For example, if we compute the variation of~\eqref{eq:Fdefinition} under $L_0$, it is going to be proportional to
\be\label{eq:L0variation}
-i \frac{\pi^2\l^2 C_{\De_\f,d}^2}{2(\De_\phi-1)^2} \int\prod_{k=1}^2 \frac{d^{d+1} x_k \sqrt{-g}\, \de^2(\a_k-\a_{0,k})}{(1-\a_k\bar \a_k)^{\De_\f - 2}(\cosh\r_k)^{2\De_\phi}}K_\Sigma(x_1,x_2) \de'(t_1).
\ee
Note that $\de(t_1)$ in~\eqref{eq:Fdefinition} has been replaced by $\cL_0\de(t_1)=-i\de'(t_1)/2$, where $\cL_0$ is the Killing vector corresponding to $L_0$. We can compute this integral in $(t,\r,\f,\psi,\hat n)$ coordinates. The integral over the times $t_k$ takes the form
\be
	\int dt_1 dt_2 f(t_2-t_1)\de'(t_1)=\int dt_2 f'(t_2)=0.
\ee
where the function $f$ of $t_2-t_1$ comes from $K_\Sigma(x_1,x_2)$, while all the other factors in~\eqref{eq:L0variation} are $t_k$-independent. Therefore, the $L_0$ variation~\eqref{eq:L0variation} vanishes. The same is true for $L_+$ and $L_-$ variations due to $\cL_+\de(t_1)=-i\bar\a_1\de'(t_1)/2$ and $\cL_-\de(t_1)=-i\a_1\de'(t_1)/2$. We conclude that $F(\a_1,\a_2)=U_2(s_{12})$ for some function $U_2$.\footnote{We have essentially shown that the matrix elements of $V$ on $\cH_{2,J}$ are invariant under the $\mathfrak{so}(2,1)$ transformations of the free theory. The computation was slightly non-trivial since, much like $\tau$, the $\mathfrak{so}(2,1)$ generators receive corrections. Let's say $Q$ is one of the free theory generators. Then we have $[Q,\tau]=[Q+\de Q,\tau + V]=0$, where $\de Q$ is the correction to $Q$. At the leading order, this implies only that $[Q,V]=[\tau, \de Q]$. Thus, the operator $V$ is not invariant under $Q$, but $\<\Psi_1|[Q,V]|\Psi_2\>=\<\Psi_1|[\tau, \de Q]|\Psi_2\>=0$ if $\Psi_1$ and $\Psi_2$ are $\tau$-eigenstates with the same eigenvalue.}

It is thus enough to compute $F(\a_1,0)$. For this, we rewrite $F$ as
\be\label{eq:Fsymmetry_fixed}
	F(\a_{0,1},0)=\frac{\pi^2\l^2 C_{\De_\f,d}^2}{(\De_\phi-1)^2} \int \frac{d^{d+1} x_1 \sqrt{-g}\, \de^2(\a_1-\a_{0,1})}{(1-\a_1\bar \a_1)^{\De_\f - 2}(\cosh\r_1)^{2\De_\phi}} \de(t_1)\, I(x_1),
\ee
where
\be\label{eq:def_I}
	I(x_1)=\int d^{d+1} x_2 \sqrt{-g} K_\Sigma(x_1,x_2)\frac{\de^2(\a_2)}{(\cosh\r_2)^{2\De_\phi}} .
\ee
Since $K_\Sigma$ is the Green's function of the Klein-Gordon equation, $I(x_1)$ is the solution with the source ${\de^2(\a)/}{(\cosh\r)^{2\De_\phi}}$. Therefore, instead of computing the above integral, we can simply solve the Klein-Gordon equation directly. This can be done using separation of variables and is detailed in appendix~\ref{app:klein_gordon_source}. Substituting the resulting solution~\eqref{eq:I_to_cn} into~\eqref{eq:Fsymmetry_fixed} yields (see appendix~\ref{app:U2exp_into_ks})
\be\label{eq:scalar_potential_result}
	U_2(s)= \sum_{n=0}^\infty b_n \,  k_{\De_\s+2n}\left(\frac{1}{s+1}\right),
\ee
where
\be
	k_{2h}(z)&=z^h {}_2F_1(h,h;2h;z),\\
	b_0 &=-\frac{\l^2}{16\pi^{d/2}} \frac{\G(\De_\s/2)^2}{\G(\De_\s-d/2+1)} \frac{\G(\De_\phi+\De_\s/2-d/2)^2}{\G(\De_\phi+\De_\s/2-1)^2} \frac{\Gamma(\De_\f-1)^2}{\Gamma(\De_\f-d/2+1)^2},\label{eq:b0defn}\\
	b_n &= \frac{b_0}{n!} \frac{(d/2-1)_n}{(\De_\s-d/2+1)_n} \frac{(\De_\s/2-\De_\phi+1)_n^2}{(\De_\s/2+\De_\phi-1)_n^2} \frac{(\De_\s-1)_n}{(\De_\s)_n} \frac{\De_\s-1+2n}{\De_\s-1} (\De_\s/2)_n.
\ee
Note that the expansion~\eqref{eq:scalar_potential_result} is effectively an expansion at large $s$ since each $k_{2h}(1/(s+1))$ is suppressed by $s^{-h}$. For example, the leading behaviour at large $s$ is
\be\label{eq:scalar_potential_leading}
	U_2(s)=b_0s^{-\De_\s/2}(1+O(s^{-1})).
\ee
When $J$ is large, the typical separation $s$ between the two particles will be large as well. We will therefore often only need the leading term~\eqref{eq:scalar_potential_leading}.

\subsection{Decay of two-particle and $N$-particle potentials}
\label{sec:higher_potential}

The result~\eqref{eq:scalar_potential_leading} shows exponential decay of $U_2(s)$ in the hyperbolic distance $\bd\sim \log s$. While ultimately this decay should be due to the large-distance asymptotic of the bulk-to-bulk propagator $K_\Sigma$ in~\eqref{eq:Fdefinition}, the details of this are not immediately obvious due to the Lorentzian region over which $x_1,x_2$ are integrated, which includes null and time-like separations.

A simple way to resolve such difficulties would be a Wick rotation to Euclidean $\AdS_{d+1}$. However, this seems impossible due to the presence of the delta-functions $\de^2(\a_k-\a_{0,k})$ in~\eqref{eq:Fdefinition} which are non-analytic in the global $\AdS_{d,1}$ time. Fortunately, the definition~\eqref{eq:alpha_def} of $\a$ shows that these delta-functions are time-independent and do not obstruct the Wick rotation if we work in the coordinates $(t,\r,\vf',\cdots)$. The Wick rotation in these coordinates is possible since the generator of time translations is the twist $\tau$, which is non-negative definite.

It is interesting to note that since the metric in $(t,\r,\vf',\cdots)$ coordinates contains an off-diagonal term proportional to $dtd\vf'$ (see~\eqref{eq:AdS3metric_2}), it becomes complex after the Wick rotation. It is easy to check that the resulting complex metric is allowable in the sense of~\cite{Kontsevich:2021dmb}.\footnote{This is true more generally: if, in some coordinate system, translations in time are isometries of a real Lorentzian metric (and are, of course, time-like), then the Wick rotation in this time gives an allowable complex metric~\cite{Witten:2021nzp}.}

At a more practical level, this Wick rotation can be viewed as a contour deformation in~\eqref{eq:Fdefinition} after which we are integrating over $t_E=it$; non-negativity of the twist $\tau$ ensures that $K_\Sigma$ does not have singularities that would prevent this deformation. The purpose of doing this is that we hope that the asymptotics of the integral are easier to determine with the new integration contour.

To see that this is indeed the case, we first note that $K_\Sigma(x_1,x_2)$ decays at large $\z$ as $\z^{-\De_\s}$, where
\be
	\z =&-1+ \cos t_{12}\cosh\r_1\cosh\r_2\nn\\
	&+\p{\cos\theta_1\cos\theta_2\cos(t_{12}+\vf'_{12})+\sin\theta_1\sin\theta_2\hat n_1\.\hat n_2}\sinh\r_1\sinh\r_2,
\ee
with $t_{12}=t_1-t_2$, $\vf'_{12}=\vf'_1-\vf'_2$, and $\z$ is related to the $\AdS_{d,1}$ geodesic distance $\r_{12}$ by $\z = 2\sinh^2\r_{12}/2$. Recall also that having a large hyperbolic distance between $\a_1$ and $\a_2$ requires at least one of $\a_1$ and $\a_2$ approach the boundary of the unit disk.

Suppose that at least $\a_1=e^{i\vf_1}\cos\theta_1\tanh\r_1$ is pushed to the boundary of the unit disk. This forces $\r_1$ to be large and $\theta_1$ to be small, yielding
\be
	\z\approx \frac{e^{\r_1}}{2}\p{
		\cos t_{12}\cosh \r_2-\cos\theta_2\cos(t_{12}+\vf'_{12})\sinh\r_2
	}.
\ee
Prior to the Wick rotation, this expression can oscillate in sign depending on $t_{12}$, and therefore is not guaranteed to be large.

On the other hand, after the Wick rotation this becomes
\be\label{eq:zeta_large_rho1}
\z\approx \frac{e^{\r_1}\cosh t_{E,12}\cosh\r_2}{2}\p{1-|\a_2|(\cos\vf'_{12}+i\sin\vf'_{12}\tanh t_{E,12})},
\ee
which can only become small if $\vf'_2\to \vf'_1$, and $\a_2$ is pushed to the boundary of the unit disk as well. Since $F(\a_1,\a_2)$ is invariant under $\mathfrak{so}(2,1)$ transformations of $\a$'s, we can compute  it with $\a_1=-\a_2$, and then this condition is never satisfied in the integral~\eqref{eq:Fdefinition}. Therefore, large hyperbolic distance between $\a_1$ and $\a_2$ guarantees that $\z$ is large for all points on the integration contour. This then guarantees the decay of $F$ and thus $U$ at large hyperbolic distances.

To see the rate of the decay, note for $\a_2=-\a_1$ we have 
\be
\z \propto e^{\r_1}e^{\r_2}\propto (1-|\a_1|^2)^{-1/2}(1-|\a_2|^2)^{-1/2}\propto s^{1/2}.
\ee
The decay $K_\Sigma\sim \z^{-\De_\s}$ then leads to $F(\a_1,\a_2)=U_2(s)\sim s^{-\De_\s/2}$.

\paragraph{Higher-particle interactions} This logic allows us to estimate the decay of effective potentials corresponding to higher-particle interactions. Let us first discuss briefly the role of higher-particle interactions in our model.

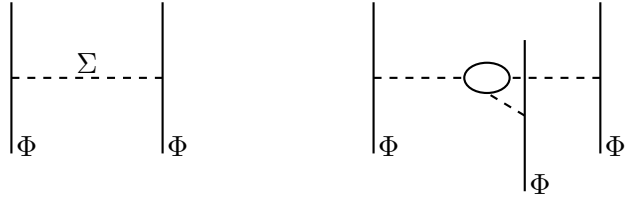
\begin{figure}
\centering
\begin{tikzpicture}[baseline=0]
	\draw[thick] (-1,0) -- (-1,2);
	\draw[thick] (1,0) -- (1,2);
	\draw[thick,dashed] (-1,1) -- (1,1);
	\node[] at (-1+0.2,0.1) {$\Phi$};
	\node[] at (+1+0.2,0.1) {$\Phi$};
	\node[] at (0,1.2) {$\Sigma$};
\end{tikzpicture}
\hspace{2cm}
\begin{tikzpicture}[baseline=0]
	\draw[thick] (-1.5,0) -- (-1.5,2);
	\draw[thick] (1.5,0) -- (1.5,2);
	\draw[thick] (0.5,-0.5) -- (0.5,2-0.5);
	\draw[thick] (0,1) ellipse (0.3 and 0.2);
	\draw[thick,dashed] (-1.5,1) -- (-0.3,1);
	\draw[thick,dashed] (1.5,1) -- (0.3,1);
	\draw[thick,dashed] (0.5,1-0.5) -- (0,1-0.2);
	\node[] at (-1.5+0.2,0.1) {$\Phi$};
	\node[] at (+1.5+0.2,0.1) {$\Phi$};
	\node[] at (+0.5+0.2,0.1-0.5) {$\Phi$};
\end{tikzpicture}
\caption{Some diagrams contributing to the effective potential between $\Phi$ particles. Left: the leading contribution to the two-particle potential. Right: a contribution to the three-particle potential.\label{fig:diagrams}}
\end{figure}

The potential $U_2(s)$ that we computed is basically due to the leading $\Sigma$-exchange diagram shown in left panel of figure~\ref{fig:diagrams}. More complicated diagrams such as that in the right of figure~\ref{fig:diagrams} would contribute to potentials which involve $n\geq 3$ $\Phi$-particles at once. Their projections onto the lowest twist subspace can be characterized by potentials
\be
	U_n(\a_1,\cdots, \a_n).
\ee
We would like to understand how quickly $U_n$ decays in typical multi-particle states at large spin $J$.

A subtlety is that in our model such potentials appear only at higher orders in the perturbation theory, in which case it is no longer justified to project the interactions to the lowest-twist subspace, and mixing effects with higher twists must be taken into account. However, we expect that taking the projection is still valid at large spin, when the interactions are weak due to large $\AdS$ separations rather than weak coupling. 

Returning to the decay of the potentials $U_n$, we expect that at least at the leading orders they can be computed by generalizations of~\eqref{eq:Fdefinition},
\be
	U(\a_{0,1}\cdots \a_{0,n})\propto \int\prod_{k=1}^n \frac{d^{d+1} x_k \sqrt{-g}\, \de^2(\a_k-\a_{0,k})K_\Sigma(x_k,y_k)}{(1-\a_k\bar \a_k)^{\De_\f - 2}(\cosh\r_k)^{2\De_\phi}} \G(y_1\cdots y_n)\de(t_1),
\ee
where $\G(y_1,\cdots,y_n)$ is some vertex function supported (after the Wick rotation) when all $y_k$ are close together on the order of $\AdS$ scale. 

We will see in the following sections that the typical configurations on which $N$-particle wavefunctions have support satisfy $1-|\a_i|^2\sim J^{-1}$, with the phases of $\a_i$ relatively evenly distributed over the unit circle. Following the logic above, this gives
\be
	\z_k \propto e^{\r_k}\propto (1-|\a_k|^2)^{-1/2}\propto J^{1/2}
\ee
for the arguments of the $K_\Sigma$, leading to the decay
\be
	U_n\sim J^{-n\De_\s/2}.
\ee

In particular, all higher-particle potentials $U_{n\geq 3}$ are suppressed at large $J$ by at least $J^{-\De_\s/2}$ relative to the contribution of $U_2(s)$. One could imagine constructing a model where the exchanges contributing to $U_2$ would be restricted compared to those contributing to $U_3$, leading to $U_3$ dominating over $U_2$. We haven't been able to devise a simple and natural (i.e.\ where all interactions allowed by symmetries exist) model of this kind, but we did not explore this question systematically.

\subsection{Toeplitz operators}

In section~\ref{sec:eff_pair_pot} we derived the expression~\eqref{eq:matrixelementgeneral} for matrix elements $\<\Psi_1|V|\Psi_2\>$, together with the expression~\eqref{eq:scalar_potential_result} for the effective pair potential $U_2(s)$. These matrix elements define some operator $V_\text{eff}:\cH_N\to\cH_N$ on the lowest-twist Hilbert space $\cH_{N}$. Formally, it can be described as
\be
	V_\text{eff}&=P_{\cH_{N}}V\vert_{\cH_{N}},\nn\\
	\<\Psi_1|V_\text{eff}|\Psi_2\>&=\<\Psi_1|V|\Psi_2\>,\quad \forall \Psi_1,\Psi_2\in \cH_{N},\label{eq:matrixelementid}
\ee
where $P_{\cH_{N}}:\cH\to \cH_{N}$ is the orthogonal projector from the full Hilbert space to the lowest-twist $N$-particle states.
As discussed above, our goal in the leading order perturbation theory is to diagonalize $V_\text{eff}$. 

Equation~\eqref{eq:matrixelementid} and the matrix elements~\eqref{eq:matrixelementgeneral} fully define the operator $V_\text{eff}$, albeit somewhat implicitly. Here, we would like to point out that the resulting $V_\text{eff}$ belongs to the class of so-called Toeplitz operators~\cite{de1981spectral,guillemin1995star}.

Indeed, according to section~\ref{sec:minimal-twist}, the minimal-twist subspace $\cH_N$ was identified with the Hilbert space of holomorphic functions on the polydisk $\D^N$ with respect to the inner product~\eqref{eq:N-particle-inner}. The Hilbert space $\cH_N$ can be viewed as a closed subspace of the larger space $L^2(\D^N)$ of all square-integrable functions with the same inner product. The total energy $\sum_{i<j}U_2(s_{ij})$ appearing in~\eqref{eq:matrixelementgeneral} can be understood as a (unbounded) multiplication operator acting from $\cH_{N}$ to $L^2(\D^N)$. It is then easy to check that $V_\text{eff}$ can be described as
\be\label{eq:Veff_pairpotentials}
	V_\text{eff}=P_\text{hol} \sum_{i<j} U_2(s_{ij}),
\ee
where $P_\text{hol}:L^2(\D^N)\to \cH_N$ is the orthogonal projector onto the subspace of holomorphic functions. That is, if we want to compute the action of $V_\text{eff}$ on a wavefunction $\psi(\a_1,\cdots,\a_N)$, we first compute
\be
	\sum_{i<j} U_2(s_{ij})\psi(\a_1,\cdots,\a_N).
\ee
This is not a holomorphic function of $\a_1,\cdots,\a_N$ anymore, and thus does not live in $\cH_N$. We can fix this by applying the orthogonal projection $P_\text{hol}$ to $\cH_N$ inside of $L^2(\D^N)$, which yields the desired holomorphic wavefunction $(V_\text{eff}\psi)(\a_1,\cdots,\a_N)$. 

Operators that act in this way are known as Toeplitz operators, and the function $\sum_{i<j}U_2(s_{ij})$ is called the symbol\footnote{More specifically, it is the covariant symbol as defined by Berezin~\cite{berezin1975quantization}. Since we do not introduce any other kind of symbol for Toeplitz operators in this work, we can and will omit this qualifier without ambiguity.} of the Toeplitz operator $V_\text{eff}$. In a well-defined sense, $V_\text{eff}$ can be viewed as a quantization of $\sum_{i<j} U_2(s_{ij})$. There exists an extensive literature on Toeplitz operators and in particular on their semiclassical behavior, see~\cite{ma2008generalized,schlichenmaier2010berezin,le2018brief} and references therein. We will find this useful later when we analyze the large-$J$ spectrum of $V_\text{eff}$. 

\subsection{Two-body binding energies and Lorentzian inversion formula}
\label{sec:binding_energies}
Using~\eqref{eq:two-particle-matrix-element} and~\eqref{eq:scalar_potential_result}, it is straightforward to compute the two-particle binding energy. Indeed, given the unique primary two-particle state at spin $J$:
\be
	\psi_J(\a_1,\a_2)=(\a_1-\a_2)^J,
\ee
the binding energy is simply
\be	\label{eq:def_2bodybinding}
	\gamma_{2,J}=\frac{\<\psi_J|V_\text{eff}|\psi_J\>}{\<\psi_J|\psi_J\>}.
\ee

To compute the inner product and the matrix element, we can consider a more general integral of a generic function $f(s)$:
\be\label{eq:favg_def}
	\<f\>_{\De_1,\De_2}\equiv \int d^4\a (1-|\a_1|^2)^{\De_1-2}(1-|\a_2|^2)^{\De_2-2} f(s_{12}).
\ee
Using (see appendix~\ref{app:proof_geodesic_to_LCB})
\be \label{eq:geodesic_to_LCblocks}
&\int d^4\a (1-|\a_1|^2)^{\De_1-2}(1-|\a_2|^2)^{\De_2-2}\de(s-s_{12})\nn\\
&=\frac{\pi^2}{\De_1+\De_2-1}{}_2F_1(\De_1,\De_2;\De_1+\De_2;-s),
\ee
we find
\be\label{eq:favg_formula}
	\<f\>_{\De_1,\De_2}=\frac{\pi^2}{\De_1+\De_2-1}\int_0^\oo ds f(s)\,{}_2F_1(\De_1,\De_2;\De_1+\De_2;-s).
\ee

It is now easy to check from~\eqref{eq:N-particle-inner} and~\eqref{eq:matrixelementgeneral} that the two-body binding energies take the form
\be\label{eq:two-body-binding}
	\gamma_{2,J}=\frac{\<s^{J}U_2(s)\>_{\De_\f+J,\De_\f+J}}{\<s^{J}\>_{\De_\f+J,\De_\f+J}}.
\ee
For example, using the expansion~\eqref{eq:scalar_potential_leading} of $U_2$ at large separation, we get the leading large-spin asymptotics
\be\label{eq:leading_gamma_lll}
\gamma_{2,J}\sim  b_0\frac{\<s^{J-\De_\s/2}\>_{\De_\f+J,\De_\f+J}}{\<s^{J}\>_{\De_\f+J,\De_\f+J}}\sim b_0\frac{\G(\De_\f+\De_\s/2-1)^2}{\G(\De_\f-1)^2}J^{-\De_\s}.
\ee
We can also obtain finite-$J$ results by fully resumming~\eqref{eq:scalar_potential_result} and plugging it into~\eqref{eq:two-body-binding}.

The same binding energies can be obtained by computing the leading correction to the boundary four-point function $\<\phi\phi\bar\phi\bar\phi\>$ and examining the resulting conformal block expansion. The most convenient way of doing this is using the Lorentzian inversion formula (LIF). It is well-known that since the LIF is not sensitive to double-twist exchanges, the anomalous dimensions can be obtained by applying LIF to the $t$-channel conformal block for the exchange of $\sigma$. The resulting correction is~\cite[section~2.3]{Albayrak:2019gnz}
\be\label{eq:gamma_lif}
\gamma_{LIF} = -2 \frac{C_{\phi\phi\s}^2\G(\De_\s)}{\G(\De_\s/2)^2} \frac{\sin^2 \pi(\De_\phi-\De_\s/2)}{\sin^2\pi\De_\phi} \frac{\int_0^1 \frac{d z}{z^2}\left(\frac{z}{1-z}\right)^{\De_\phi} k_{2\De_\phi+2J}(z) k_{\De_\s}^{(d)}(1-z)}{\int_0^1 \frac{d z}{z^2}\left(\frac{z}{1-z}\right)^{\De_\f} k_{2\De_\phi+2J}(z)},
\ee
where we defined
\be
k_{2h}^{(d)} (z) := z^h {}_2F_1(
	h,h,2h-d/2+1,z), \quad k_{2h}^{(2)}\equiv k_{2h},
\ee
and the OPE coefficient $C_{\f\f\s}$ at leading order in $\l$ is given by (see appendix~\ref{app:Cphiphisigma})
\be\label{eq:Cphiphisigma}
C_{\phi\phi\s}=  \l \frac{\pi^{d/2} }{2} \G\left(\frac{2\De_\phi+\De_\s-d}{2}\right) \frac{C_{\De_\phi,d}}{\G(\De_\phi)^2} \frac{C_{\De_\s,d}^{1/2}}{\G(\De_\s)} \G\left(\frac{\De_\s}{2}\right)^2\G\left(\De_\phi-\frac{\De_\s}{2}\right)+O(\l^2),
\ee
where $C_{\De,d}$ was defined in~\eqref{eq:Psi1particle}. It is well known that the large-spin expansion stems from the $z\rightarrow 1$ limit of the double-discontinuity in the inversion formula. In the case of~\eqref{eq:gamma_lif}, inserting the leading-order asymptotics $k_{\De_\s}^{(d)} = (1-z)^{\De_\s/2}(1+O(1-z))$ yields
\begin{equation}
\g_{LIF}\sim  -2 \frac{C_{\phi\phi\s}^2\G(\De_\s)}{\G(\De_\s/2)^2} \frac{\Gamma(\De_\phi)^2}{\G(\De_\phi-\De_\s/2)^2} J^{-\De_\s}.
\label{eq:leading_gamma_lif}
\end{equation}
At finite spin, the integrals in~\eqref{eq:gamma_lif} can be expressed as a linear combination of two ${}_4 F_3$ hypergeometric functions using \cite[eqs.~(2.33),(2.44)]{Albayrak:2019gnz}.

Comparing $b_0$ in~\eqref{eq:scalar_potential_result} and $C_{\phi\phi\s}$ in~\eqref{eq:Cphiphisigma}, we find that the leading large-spin asymptotics of $\g_{2,J}$ in~\eqref{eq:leading_gamma_lll} and $\g_{LIF}$ in~\eqref{eq:leading_gamma_lif} are identical. More generally, we prove in appendix~\ref{app:LIF_to_LLL} the exact equality $\g_{2,J}=\g_{LIF}$ as analytic functions of spin.

\subsection{Numerical diagonalization at finite spin}
\label{sec:diagonalization}

The Toeplitz operator~\eqref{eq:matrixelementid} acts as a linear operator on the finite-dimensional subspace $\cH_{N,J}^{\mathrm{primary}}\subset \cH_N$. If its matrix elements in a basis can be computed explicitly, then we can extract the exact eigenvalues from numerical diagonalization.

At leading order in perturbation theory, the symbol of the Toeplitz operator $V_{\mathrm{eff}}$ is the symmetrization of a pair potential $U_2(s_{12})$. In this section, for any Toeplitz operator of this form, we devise a general method to compute matrix elements of $V_{\mathrm{eff}}$ in the basis~\eqref{eq:sym_poly_basis} of $\cH_{N,J}^{\mathrm{primary}}$. The same kind of numerical diagonalization problem was recently studied in~\cite[section~4]{Fardelli:2024heb}. These authors computed the matrix elements in a different basis, using the decomposition of $N$-twist states into iterated double-twist states in $\cH_1^{\otimes N}$.

To compute the matrix elements of $V_{\mathrm{eff}}$, our strategy is to first determine the eigenvalues and a basis of eigenvectors for the pair potentials $P_{\mathrm{hol}} U_2(s_{ij})$. We then compute the basis-change matrix from these pair potential eigenvectors to the functions in~\eqref{eq:sym_poly_basis}. The basis-change matrices and the pair potential eigenvalues fully determine the matrix elements of $V_{\text{eff}}$.

Without loss of generality, we consider the Toeplitz operator $P_{\mathrm{hol}}U_2(s_{12})$ of the $(12)$ pair. Since the pair potential breaks the $S_N$ symmetry down to $S_2\times S_{N-2}$, the latter is actually an operator on $\cH_2 \otimes \cH_{N-2}$, represented by holomorphic square-integrable functions $\psi(\a_1,\a_2,\a_3,\dots,\a_N)$ that are symmetric under $\a_1\leftrightarrow\a_2$ and permutations of $(\a_3,\dots,\a_N)$. The eigenspaces of $P_{\mathrm{hol}}U_2(s_{12})$ inside $\cH_2\otimes \cH_{N-2}$ can be characterized using the representation theory of $\mathfrak{so}(2,1)$. In this context, recall that $\cH_1\equiv \cH_1(\De_\phi/2)$ is an irreducible lowest-weight representation with of lowest weight $\bar h=\De_\phi/2$, as follows from the action~\eqref{eq:Laction} of the generators. Then $\cH_2\equiv \cH_2(\De_\phi/2)$ is the symmetrized tensor product of two such lowest-weight representations, which is known to decompose into an infinite direct sum of lowest-weight representations with lowest-weight $\De_\phi+\ell$, where $\ell$ is an even spin label:
\begin{equation}
\cH_2(\De_\phi/2)\cong \bigoplus_{\ell\in 2\N} \cH_1(\De_\phi+\ell).
\label{eq:tensor_product_dec}
\end{equation}
Each vector space in the direct sum is spanned by the action of the raising operator $(L_1+L_2)_-$ on the lowest-weight vector $\psi_{2,\ell}(\a_1,\a_2)=(\a_1-\a_2)^\ell$. Since $U_2(s_{12})$ is invariant under $\mathfrak{so}(2,1)$, its corresponding Toeplitz operator commutes with all $\mathfrak{so}(2,1)$ generators, such that it must act as a constant on the irreducible subspaces. By acting on the lowest-weight vector, we deduce that $\cH_1(\De_\phi+\ell)$ are the eigenspaces of $P_{\mathrm{hol}} U_2(s_{12})$ with eigenvalues $\g_{2,\ell}$ corresponding to the two-body binding energies computed in section~\ref{sec:binding_energies}. 

Now, just like we saw previously, the subspace $(\cH_2 \otimes \cH_{N-2})_J^{\mathrm{primary}}$ of spin-$J$ primaries is obtained by imposing that its wavefunctions $\psi$ are translation-invariant and homogeneous of degree $J$. The tensor product decomposition~\eqref{eq:tensor_product_dec} then implies that $P_{\mathrm{hol}} U_2(s_{12})$ has eigenspaces $(\cH_1(\De_\phi+\ell) \otimes \cH_{N-2})_J^{\mathrm{primary}}$ with eigenvalues $\g_{2,\ell}$. To find an explicit basis of functions $\psi_\ell^{(12)}(\a_1,\dots,\a_N)$ for these eigenspaces, we use the quadratic Casimir operator for the action of $\mathfrak{so}(2,1)$ on $\cH_2\subset \cH_2\otimes \cH_{N-2}$. As a quadratic form in the generators, it is given by
\begin{equation}
L_{ij} := \mathcal{C}^2(L_i+L_j),\quad \mathcal{C}^2(L):= L_0(L_0-1)-L_-L_+.
\label{eq:casimir_def}
\end{equation}
Moreover, it shares the same eigenspaces $\cH_1(\De_\phi+\ell)\subset \cH_2$ with eigenvalues $(\De_\phi+\ell)(\De_\phi+\ell-1)$. It is convenient to express the action of $L_{12}$ in terms of the following translation-invariant variables:
\begin{equation}
	r:=\a_1-\a_2,\quad s:=\a_1+\a_2-2 \frac{\a_3+\dots+\a_N}{N-2},\quad \gamma_k := \a_k-\frac{\a_3+\dots+\a_N}{N-2},
\end{equation}
where $k=1,\dots,N$ and $\g_3+\dots+\g_{N}=0$. We can always find a translation gauge where $\a_{1,2}=(r\pm s)/2$ and $\a_{3,\dots,N}=\g_{3,\dots,N}$. Denoting $\psi(\a):= \chi(r,s,\g)$, it is then easy to check that the quadratic Casimir acts as
\begin{equation}
(L_{12}\psi)\left(\frac{r+s}{2},\frac{r-s}{2},\g\right) = r^{2-\De_\phi} (\ptl_r^2-\ptl_s^2) r^{\De_\phi} \chi(r,s,\g).
\label{eq:L12_sr}
\end{equation}
Since the above differential operator is completely independent of $\g_3,\dots,\g_N$, it admits a basis of eigenfunctions on $(\cH_2\otimes \cH_{N-2})_J^{\text{primary}}$ with the factorized form
\begin{equation}\label{eq:pairpot_eigenfunctions}
	\psi_{\ell,m}^{(12)}(\a_1,\dots,\a_N) := P_{\ell,J-|m|}(r,s)\, e_m(\gamma_3,\dots,\gamma_N), 
\end{equation}
where the first factor is an eigenfunction of the Casimir operator~\eqref{eq:L12_sr}, while the second factor is a product of elementary symmetric polynomials
\begin{equation}
e_m(\g):=\prod_{k=2}^{N-2} e_{k}(\g)^{m_k},\quad |m|:=\sum_{k=2}^{N-2} k\,m_k \leq J.
\end{equation}
Note that $P_{\ell,J-|m|}$ must be homogeneous of degree $J-|m|$ in $(r,s)$ for $\psi_{\ell,m}^{(12)}$ to be homogeneous of degree $J$ in $\a$. This constraint reduces the eigenvalue equation~\eqref{eq:L12_sr} with eigenvalue $(\De_\phi+\ell)(\De_\phi+\ell-1)$ to an ODE in $r/s$ with polynomial solution
\begin{equation}
	P_{\ell,j}(r,s) = r^\ell s^{j-\ell}\, {}_2F_1\left(\frac{1+\ell-j}{2},\frac{\ell-j}{2},\De_\phi+\ell+\frac{1}{2};\frac{r^2}{s^2} \right).
\end{equation}

Now that we have determined the eigenvalues $\g_{2,\ell}$ and a basis of eigenvectors $\psi_{\ell,m}^{(12)}$ for the two-particle operators $U_2(s_{12})$, we can find the action of $U_2(s_{12})$ in the basis $f_m(\a)$ given by~\eqref{eq:sym_poly_basis}. For this, we determine the transformation matrices between the basis $f_m$ and $\psi_{\ell,m}^{(12)}$ and conjugate by them the diagonal matrix that represents $U_2(s_{12})$ in the $\psi_{\ell,m}^{(12)}$ basis.

A subtlety is that the transition matrices are not square since the basis $\psi_{\ell,m}^{(12)}$ is not $S_N$-invariant, but rather only $S_2\x S_{N-2}$-invariant. The matrix that maps from the $S_2\x S_{N-2}$-invariant wavefunctions to $S_N$-invariant wavefunctions is therefore obtained by first applying $S_N$ symmetrization. 

\section{Three-body problem at large spin}
\label{sec:three-body}

In the previous section, we saw in our toy model that the subspace $\cH_{N}$ of leading-twist $N$-particle states can be described by the wavefunctions $\psi(\a_1,\cdots,\a_N)$, holomorphic in $\a_i$, which range over the hyperbolic disk $\D$ equipped with the inner product~\eqref{eq:N-particle-inner}. We also saw that the twist Hamiltonian restricted to this subspace is given by a Toeplitz operator with the symbol
\be\label{eq:tau_with_U2_only}
	\tau = N\De_\f+\sum_{i<j}U_2(s_{ij}),
\ee
where the effective potential is given in~\eqref{eq:scalar_potential_result}. Equivalently $\tau= N\De_\f+V$, with the matrix elements of $V$ given by~\eqref{eq:matrixelementgeneral}. For $N=2$, we explicitly verified that this twist Hamitonian exactly reproduces the Lorentzian inversion formula result for the $t$-exchange of the scalar $\s$.

In the rest of this paper, we will analyze the spectral problem of the above type for $N\geq 3$ particles, focusing on $N=3$ in this section. We will only be interested in the large-spin limit, i.e.\ the spectrum of $\tau$ on $\cH_{N,J}$ with $J\gg 1$. The methods we develop will not rely on the specific form of the twist Hamiltonian $\tau$. While we will keep using~\eqref{eq:tau_with_U2_only} in examples for concreteness, the same methods apply to the more general twist Hamiltonian with the symbol
\be\label{eq:tau_generic_model}
	\tau = N\De_\f+U_N(\a_1,\cdots,\a_N),
\ee
as long as the (non-holomorphic) $N$-body potential $U_N$ has suitable behavior at large distances. Although we have no proof of this, we expect that quite generally the multi-twist operators in CFT can be described by a model of type~\eqref{eq:tau_generic_model} at large spin, with suitably chosen $U_N$. In fact, as we discuss in section~\ref{sec:higher_potential}, generically one should expect the two-body interactions of the type~\eqref{eq:tau_with_U2_only} to dominate at large spin.

The anomalous dimension operator $\g=\tau-N\De_\f$ is the Toeplitz operator with symbol $U_N$, schematically
\be
	\g=U_N(\a_1,\cdots,\a_N).
\ee
We will usually discuss the spectrum of $\g$ rather than $\tau$, which is more convenient due to the absence of the constant $N\De_\f$ shift. 

The key to understanding the spectrum of $\g$ at large spin $J$ is to realize that the problem becomes semi-classical with $\hbar=J^{-1}$. We do not know of a simple and rigorous way of demonstrating this. In particular, the classical system obtained in the $J\to \oo$ limit is not easy to construct directly. In the context of planar conformal gauge theories, the semiclassical nature of the large-spin limit was derived explicitly using methods from integrability~\cite{Korchemsky:1995be,Korchemsky:1997yy,Braun:1999te,Belitsky:2003ys,Dorey:2008zy,Belitsky:2008mg}. 

Heuristically, we can look at the classical problem of $N$ particles in constant magnetic field in the hyperbolic disk $\D$ that was discussed in the introduction. Spin $J$ is just a charge associated to one of the generators of $\mathfrak{so}(2,1)$ isometries of the hyperbolic disc $\D$. To make $J$ more explicit, we can perform the symplectic reduction~\cite{marsden1974reduction} of the classical phase space with respect to these isometries.\footnote{Note that after the restriction to the LLL, the effective classical phase space becomes $\D^N$ in which the Landau centers move. } One can then check that the reduced Poisson bracket $\{f,g\}$ becomes proportional to $J^{-1}$ as $J\to \oo$. Since the quantization condition is $[f,g]\approx i\hbar \{f,g\}$, this is equivalent to having a small $\hbar\sim J^{-1}$. This semiclassical behavior is not apparent prior to the symplectic reduction, partly because the spin $J$ is implicit, and partly because the trivial ``center-of-mass motion'' that is modded out by the reduction is not semiclassical.
We do not reproduce this calculation in detail since it is in any case non-rigorous and would require an otherwise unnecessary discussion of symplectic reduction.

Nevertheless, the idea of performing some sort of reduction with respect to the isometries of $\D$ will be useful in our fully quantum problem. We will carry it out in the next subsections. We will see that the Hamiltonian of the reduced problem is still a Toeplitz operator, albeit now in a more general setting of Berezin-Toeplitz quantization with $\hbar = J^{-1}$.

\subsection{Line bundles on $\CP^{N-2}$ and Berezin-Toeplitz quantization}
\label{sec:linebundles_and_BT}

The analogue of symplectic reduction in our quantum problem is simply the restriction of $\tau$ to the subspace of primary wavefunctions with spin $J$, i.e.\ to $\cH^\text{primary}_{N,J}$. In terms of the wavefunctions $\psi(\a)$, this means that $\psi$ is a homogeneous degree-$J$ symmetric polynomial that is also translationally-invariant, see section~\ref{sec:minimal-twist}.

For the moment, let us ignore the restriction that $\psi$ is symmetric. Translation invariance means that $\psi$ is, in effect, a function on $\C^{N-1}$. If $\psi$ were homogeneous of degree $0$, i.e.\ invariant under the action of the multiplicative group $\C^\x$ on $\C^{N-1}$, this would imply that $\psi$ can be seen as a function on $\CP^{N-2}=\C^{N-1}/\C^\x$. However, $\psi$ is homogeneous with generally non-zero degree $J$. It is well-known that rather than being functions on $\CP^{N-2}$, such $\psi$ describe holomorphic sections of the holomorphic line bundle $\cO(J)$ on $\CP^{N-2}$.\footnote{In this notation, the tautological line bundle of $\CP^{N-2}$ is $\cO(-1)$ and the canonical line bundle (the bundle of holomorphic $(N-2)$-forms) is $\cO(-(N-1))$. The line bundle $\cO(m)$ is dual to $\cO(-m)$, and $\cO(0)$ is the trivial bundle.} We denote $\cL=\cO(1)$, so that $\cO(J)=\cL^{\otimes J}$. 

Restoring the permutation invariance, we can write
\be\label{eq:CPhilbert}
	\cH^\text{primary}_{N,J}\simeq  \G(\CP^{N-2},\cL^{\otimes J})^{S_N}.
\ee
In words, we have identified $\cH^\text{primary}_{N,J}$ with the space of $S_N$-invariant holomophic sections of the line bundle $\cL^{\otimes J}$ over $\CP^{N-2}$. Somewhat related is the fact that the symplectic reduction of the classical phase space discussed earlier also yields a phase space homeomorphic to $\CP^{N-2}$.

Inspecting our expressions~\eqref{eq:N-particle-inner} and~\eqref{eq:matrixelementgeneral} for the inner product on $\cH^\text{primary}_{N,J}$ and for the matrix elements of $V_\text{eff}$, we see that they are not written as integrals over $\CP^{N-2}$, but rather as integrals over the higher-dimensional $\D^N$. In other words, four real integrations in~\eqref{eq:N-particle-inner} and~\eqref{eq:matrixelementgeneral} can be performed using the homogeneity and translation-invariance of the wavefunctions $\psi$, leaving only the integration over $\CP^{N-2}$. 

After this procedure, we expect the inner product to take the form
\be\label{eq:CPinner}
	\<\psi_1|\psi_2\> = \int_{\CP^{N-2}} d\mu_J h_J(\psi_1,\psi_2),
\ee
where $d\mu_J$ is some (in general, $J$-dependent) measure on $\CP^{N-1}$ and $h_J=h_1^{\otimes J}$, where $h_1$ is a Hermitian inner product on $\cL$. Similarly, for the matrix elements of $V_\text{eff}$ we expect
\be\label{eq:CPVeff}
	\<\psi_1|\g|\psi_2\> = \int_{\CP^{N-2}} d\mu_J h_J(\psi_1,\psi_2)\cU_{N,J},
\ee
where $\cU_{N,J}$ is a function on $\CP^{N-2}$.

We will find that the measure $d\mu_J$ and the effective potential $\cU_{N,J}$ have good series expansions in powers of $J^{-1}$. This turns our problem, in the formulation of equations~\eqref{eq:CPhilbert},~\eqref{eq:CPinner} and~\eqref{eq:CPVeff}, into a Berezin-Toeplitz quantization setup~\cite{LeFlochElliptic,CharlesRegular} with $\hbar=J^{-1}$. This interpretation allows us to immediately identify the classical problem that arises in the large-$J$ limit. In particular, the classical Hamiltonian is simply the leading term in $\cU_{N,J}$, while the classical symplectic form is given by the Chern curvature of $h_1$. We discuss this in more detail in section~\ref{sec:BT_general}, after computing $d\mu_J$ and $\cU_{N,J}$ in section~\ref{sec:dmu_h1_cU}.

A caveat to the above discussion is that this procedure works literally only for $N=3$. For $N\geq 4$ and at large $J$, the classical phase space localizes onto an infinitesimal neighborhood of  a positive-codimension locus in $\CP^{N-2}$ and the discussion becomes more subtle. We will discuss the case of general $N$ in section~\ref{sec:Nbody} using an alternative description of the Hilbert space. However, the general strategy will remain the same. For the rest of this section we mostly specialize to $N=3$, although we often keep $N$ as a parameter to clarify where it enters. Some of the discussion applies to $N\geq 4$ and will be reused in section~\ref{sec:Nbody}.

\subsection{Computation of $d\mu_J$, $h_1$ and $\cU_{N,J}$}
\label{sec:dmu_h1_cU}

In order to cast~\eqref{eq:N-particle-inner} into the form~\eqref{eq:CPinner}, we need to perform the integration over the orbits of $\C^\x\ltimes\C$ that acts on $\a$ by complex translations and rescalings. Since $\C^\x \ltimes \C$ is not unimodular, there is a difference between left- and right-invariant measures, making the procedure somewhat subtle. Let us first consider a slightly more general and abstract version of the problem.

Specifically, we focus on the integral 
\be
	\int_\cM d\mu(x)A(x)B(x),
\ee
where $d\mu$ is some measure on $\cM$, and there is an action of  a group $G$ on $\cM$. We assume
\be
	B(gx) = \chi(g)^{2J}B(x),\quad d\mu(gx) = \chi(g)^{2N}d\mu(x),\quad \forall g\in G
\ee
for some multiplicative character $\chi$ of $G$, while $A(x)$ does not have any nice transformation properties. We would like to rewrite the integral over $\cM$ as an integral over $G$-orbits. To parameterize the $G$-orbit integral explicitly, we introduce a gauge-fixing function $F(x)$ such that $F(x)=0$ precisely once on each $G$-orbit. We then write\footnote{In general, $F$ is valued in $\R^n$ for some $n$ and the delta-function in this equation should be the $n$-dimensional delta-function.}
\be
	\cD_F(x)^{-1} = \int_G dg\, \de(F(gx)),
\ee
where $dg$ is the right-invariant measure. By construction, $\cD_F(gx)=\cD_F(x)$. A simple manipulation now yields
\be
	\int_\cM d\mu(x)A(x)B(x)&=\int_Gdg\int_\cM d\mu(x)A(x)B(x)\cD_F(x)\de(F(gx))\nn\\
	&=\int_Gdg\int_\cM d\mu(g^{-1}x)A(g^{-1}x)B(g^{-1}x)\cD_F(g^{-1}x)\de(F(x))\nn\\
	&=\int_Gdg\int_\cM d\mu(x)\chi(g)^{-2J-2N}A(g^{-1}x)B(x)\cD_F(x)\de(F(x))\nn\\
	&=\int_\cM d\mu(x)\tl A(x)B(x)\cD_F(x)\de(F(x)),\label{eq:orbitintegral}
\ee
where
\be\label{eq:alongorbit}
	\tl A(x) = \int_Gdg\,\chi(g)^{-2J-2N}A(g^{-1}x).
\ee
The integral~\eqref{eq:orbitintegral} is a gauge-fixed integral over orbits, so we have succeeded. The subtlety alluded to above is that~\eqref{eq:alongorbit} contains $A(g^{-1}x)$ integrated with the right-invariant measure $dg$. If we try to simplify $\tl A(hx)$ for $h\in G$, this will involve comparing $d(hg)$ with $dg$ which are related by the modular function of $G$. In particular, the product $d\mu(x)\tl A(x)B(x)$ is not formally invariant under $G$ but transforms according to the modular function of $G$. Although it is easy to derive, we will not need the explicit transformation rule.

In the case at hand, we have $\cM=\C^{2N}$, $G=\C^\x\ltimes \C$. We parameterize $g=(\l,\b)$ with $\l\in \C^\x$ and $\b\in \C$. The composition law is $(\l_1,\b_1)(\l_2,\b_2)=(\l_1\l_2,\l_1\b_2+\b_1)$ and the action on $\C^N$ is $(g \a)_i=\l\a_i +\b$ for $\a\in\C^\x$. We will often abuse notation by writing this formally as $g\a=\l\a+\b$. The right-invariant measure is
\be\label{eq:rightinvmeasure}
dg=d^2\l d^2\b |\l|^{-2}.
\ee
We identify 
\be
	d\mu(\a)=d^{2N}\a,\quad B(\a)=\bar\psi(\a)\psi(\a),\quad \chi((\l,\b)) = |\l|.
\ee
It is convenient to define $(x)^a_+=x^a\theta(x)$, so that we can extend the integration in~\eqref{eq:N-particle-inner} over $\a_i$ from $\D$ to $\C$,
\be
\<\psi_1|\psi_2\>=\frac{(\De_\f-1)^N}{\pi^N N!}\int d^{2N}\a\prod_{k=1}^N(1-|\a_k|^2)_+^{\De_\phi - 2}\bar{\psi_1(\a)}\psi_2(\a).
\ee
We can now identify $A$ as
\be
	A(\a) = \frac{(\De_\f-1)^N}{\pi^N N!}\prod_{k=1}^N(1-|\a_k|^2)_+^{\De_\phi - 2}.
\ee
Following~\eqref{eq:alongorbit}, we then define
\be\label{eq:Mexpression}
	M_J(\a)=\tl A(\a)=\frac{(\De_\phi-1)^N}{\pi^N N!}\int d^2\l d^2\b |\l|^{-2-2N-2J}\prod_{k=1}^N(1-|\l^{-1}(\a_k-\b)|^2)_+^{\De_\f - 2}.
\ee
It is easy to check that $M_J(\a)$ is translation-invariant and satisfies
\be
	M_J(\l\a)=|\l|^{2-2N-2J}M_J(\a).
\ee
The shift by $2$ from the naive value $-2N-2J$ in the exponent of $|\l|$ comes from the modular function of $\C^\x\ltimes \C$, as discussed above. Using~\eqref{eq:orbitintegral}, we get
\be
	\<\psi_1|\psi_2\>=\int d^{2N}\a\cD_F(\a)\de(F(\a)) M_J(\a)\bar{\psi_1(\a)}\psi_2(\a),
\ee
which is now an integral over $\CP^{N-2}$ (we leave the choice of the gauge-fixing function $F$ implicit for now). However, the integrand has not yet been fully factorized into the form~\eqref{eq:CPinner}.

To proceed, let $r(\a)$ be the radius of the smallest disk that contains all the points $\a_i$. By construction, $r(\l\a+\b)=|\l|r(\a)$. Therefore, if $\psi_1,\psi_2$ are sections of $\cL=\cO(1)$, i.e.\ are translation-invariant homogeneous functions of $\a$ of degree $1$, then 
\be\label{eq:h1defn}
	h_1(\psi_1,\psi_2) \equiv r^{-2}(\a)\bar{\psi_1(\a)}\psi_2(\a)
\ee
is a $\C^\x\ltimes\C$-invariant function of $\a$, and thus a function on $\CP^{N-2}$. Therefore, thus defined, $h_1$ can be regarded as a Hermitian inner product on $\cL$. In a similar manner, $h_J=h_1^{\otimes J}$ can be identified with $r(\a)^{-2J}$. We will see in a moment that this choice of $h_1$ and $h_J$ is forced in our setting.

We can now factorize
\be
	d^{2N}\a \cD_F(\a)\de(F(\a)) M_J(\a)\bar{\psi_1(\a)}\psi_2(\a)=d^{2N}\a \cD_F(\a)\de(F(\a))M_J(\a)r(\a)^{2J} h_J(\psi_1,\psi_2),
\ee
so that with the identification
\be\label{eq:muJexpression}
	d\mu_J = d^{2N}\a \cD_F(\a)\de(F(\a)) M_J(\a)r(\a)^{2J},
\ee
we find
\be
	\<\psi_1|\psi_2\>=\int_{\CP^{N-2}}d\mu_J h_J(\psi_1,\psi_2),
\ee
as desired. Explicit coordinate expressions can be obtained by selecting a suitable gauge-fixing function $F$. 

Repeating this derivation for $\<\psi_1|\g|\psi_2\>$, we find that $\cU_{N,J}$ can be computed as
\be\label{eq:VeffJexpresion}
&\cU_{N,J}(\a)M_J(\a)\nn\\
&=\frac{(\De_\phi-1)^N}{\pi^N N!}\int d^2\l d^2\b |\l|^{-2-2N-2J}\prod_{k=1}^N(1-|\l^{-1}(\a_k-\b)|^2)_+^{\De_\f - 2}U_N(\l^{-1}(\a-\b)),
\ee
where $\l^{-1}(\a-\b)$ has components $\l^{-1}(\a_i-\b)$.

Expressions~\eqref{eq:Mexpression} and~\eqref{eq:VeffJexpresion} in principle determine the measure $d\mu_J$ and the symbol $\cU_{N,J}$ of the effective potential. We would also like to know their asymptotic expansions at large $J$. Looking at~\eqref{eq:Mexpression}, we see that due to the $(1-|\l^{-1}(\a_k-\b)|^2)^{\De_\f-2}_+$ factors, the smallest $|\l|$ that contributes to the integral is precisely $|\l|=r(\a)$. Due to the factor $|\l|^{-2J}$ in the integrand, at large $J$ the integral is dominated by such values of $\l$ and we find the general structure
\be
	M_J(\a)&\sim r(\a)^{-2J}J^\#(\# + \# J^{-1} + \cdots),\\
	\cU_{N,J}(\a)&\sim J^\#(\# + \# J^{-1} + \cdots).
\ee
This, in particular, explains why the choice~\eqref{eq:h1defn} for $h_1$ that leads to the combination $M_J(\a)r(\a)^{2J}$ in~\eqref{eq:muJexpression} is necessary. 

We perform the explicit calculation of the leading terms of these expansions in the case $N=3$ in appendix~\ref{app:orbit_integrals}. To describe the result, let us first note that the function $r(\a)$ is piecewise-smooth and has different expressions depending on whether the $N=3$ points $\a_i$ form an acute or an obtuse triangle. Recall that $r(\a)$ is the radius of the smallest disk that contains all three points $\a_i$. For an acute triangle, all three points $\a_i$ lie on the boundary of this disk, and for obtuse triangles only two points do. For this reason, the asymptotic analysis of~\eqref{eq:Mexpression} and~\eqref{eq:VeffJexpresion} is different in the acute and obtuse regions.

\paragraph{Acute region}

In the acute region, the function $r(\a)$ takes the form
\be\label{eq:acute_r}
	r(\a) = \left|\frac{\a_{12}\a_{23}\a_{31}}{\a_{31}\bar\a_{21}-\a_{21}\bar\a_{31}}\right|,
\ee
where $\a_{ij}=\a_i-\a_j$. To state the results for $M_J(\a)$ and $\cU_{N,J}$, it is convenient to assume that $\a_i$ lie on the unit circle, which can always be achieved using a $\C^\x \ltimes \C$ transformation. To stress this convention, we will write $\a_i=\xi_i$ with $|\xi_i|=1$. We have
\be
	M_J(\xi) = \frac{J^{6-3\De_\f}\Gamma(\De_\phi)^3}{12\pi^2}\left|\frac{\xi_1\xi_2\xi_3}{\xi_{12}\xi_{13}\xi_{23}}\right| \prod_{i=1}^3 R_i^{1-\De_\phi} \left(1+O(J^{-1})\right),
	\label{eq:Macute}
\ee
where
\be\label{eq:Rfunction}
R_1 = -\frac{\xi_1(\xi_2+\xi_3)}{(\xi_{1}-\xi_2)(\xi_{1}-\xi_{3})},
\ee
and $R_2,R_3$ are defined by permutations. Note that $R_i>0$ when $\xi_1,\xi_2,\xi_3$ form an acute triangle.\footnote{Indeed, without loss of generality we can assume that $\xi_1=1$ and $\xi_i=e^{i\f_i}$ for $i=2,3$. In this case $R_1 = \thalf\cos\tfrac{\f_2-\f_3}{2}/(\sin\tfrac{\f_2}{2}\sin\tfrac{\f_3}{3})$. Furthermore, since the triangle is acute, we can assume that $0<|\f_i|<\pi$ and that $\f_2$ and $\f_3$ have opposite signs. Finally, the angle at $\xi_1$ being acute implies $\pi < |\f_2|+|\f_3| = |\f_2-\f_3|$. This is enough to establish that $R_1>0$.} 

To recover $M_J(\a)$, we write
\be\label{eq:MacuteGeneric}
	M_J(\a) = r(\a)^{-4-2J}M_J(\xi),
\ee
where 
\be\label{eq:xi}
	\xi_i=r(\a)^{-1}(\a_i-b(\a)),
\ee
and the circumcenter $b(\a)$ is given by
\be
	b(\a)= \frac{\a_1\bar\a_1\a_{23}+\a_2\bar\a_2\a_{31}+\a_3\bar\a_3\a_{12}}{\a_{31}\bar\a_{21}-\a_{21}\bar\a_{31}}.
\ee

For $\cU_{3,J}$ we find
\be
	\cU_{3,J}(\xi) =  \frac{b_0 J^{-\De_\s}\Gamma(\De_\phi+\De_\s/2-1)^2}{\Gamma(\De_\phi-1)^2} \sum_{i=1}^3 D_{i}^{\De_\s/2}\p{1-\De_\s C_i J^{-1}/2+O(J^{-2})},
	\label{eq:acute_Veff}
\ee
where
\be\label{eq:Ddefn}
	D_1 = \frac{(\xi_2-\xi_1)(\xi_3-\xi_1)}{(\xi_2+\xi_1)(\xi_3+\xi_1)},
\ee
and $D_2,D_3$ are defined by cyclic permutations. Similarly to $R_i$, it can be checked that $D_i>0$ when $\xi_1,\xi_2,\xi_3$ form an acute triangle.  The $O(J^{-1})$ correction factors $C_i$ are given by
\be\label{eq:Cdefn}
	C_1=
	&\p{
		\frac{2(1-\De_\s/2-\De_\f)\xi_1^2\xi_2}{(\xi_1+\xi_2)(\xi_2+\xi_3)(\xi_3+\xi_1)}
		+\frac{(1-\De_\s/2-2\De_\f)\xi_1^2}{(\xi_1+\xi_2)^2}
		+\frac{(\De_\s+3\De_\f-2)\xi_1}{(\xi_1+\xi_2)}+(2\leftrightarrow3)
	}\nn\\
	&+4\De_\f-3+\frac{4(\De_\f-1)\xi_1\xi_2\xi_3}{(\xi_1+\xi_2)(\xi_2+\xi_3)(\xi_3+\xi_1)}.
\ee

It is easy to check that at least one $D_i$ blows up on the boundary between acute and obtuse configurations. Thus, since $\De_\s>0$, the leading term in the symbol $\cU_{N,J}$ blows up on the boundary of the acute region, signaling the failure of the $1/J$ expansion---see section~\ref{sec:breakdown}. Note that the acute region has two connected components which differ by the cycling ordering of $\xi_i$ on the unit disk.

\paragraph{Obtuse region}

The obtuse region is characterized by the fact that only two of the $\a_i$ is on the boundary of the smallest disk that contains all $\a_i$. The obtuse region splits into three connected components which are identified by which $\a_i$ is in the interior. We will consider the connected component where $\a_3$ is in the interior, and the formulas for the remaining two components can be obtained by permutations.

The function $r(\a)$ is given simply by
\be\label{eq:obtuse_r}
	r(\a)=|\a_1-\a_2|/2.
\ee
To specify $M_J(\a)$ and $\cU_{N,J}(\a)$, it is convenient to use $\C^\x\ltimes \C$ symmetry and assume that $\a_1=1, \a_2=-1$, such that $|\a_3|<1$. Then
\be
M_J(\a) =\frac{(\De_\phi-1)^3}{12\pi^2} 4^{-J} \<s^{J+1}\>_{\De_\phi+J+1,\De_\phi+J+1}\, (1-|\a_3|^2)^{\De_\phi-2}\left(1+O(J^{-1})\right),
\label{eq:Mobtuse}
\ee
where $\<s^J\>_{\De_1,\De_2}$ is an $\a_3$-independent constant given by~\eqref{eq:sJavg}. The leading term in the symbol $\cU_{3,J}$ is given by (again $\a_1=1, \a_2=-1$, $|\a_3|<1$)
\begin{align}
\cU_{3,J}(\a) =& \left(\frac{2}{J}\right)^{\De_\s/2} \frac{\Gamma(\De_\phi+\De_\s/2-1)}{\Gamma(\De_\phi-1)} \left[ \left(\frac{1-\a_3\bar \a_3}{|1-\a_3|^2}\right)^{\De_\s/2}+ \left(\frac{1-\a_3\bar \a_3}{|1+\a_3|^2}\right)^{\De_\s/2}  \right]\nonumber\\
&+O(J^{-(\De_\s/2+1)},J^{-\De_\s}).
\label{eq:obtuse_Veff}
\end{align}

Note that in the obtuse region, the scaling of $\cU_{3,J}$ is $J^{-\De_\s/2}$ (see~\eqref{eq:obtuse_Veff}), while in the acute region $\cU_{3,J}$ scales as $J^{-\De_\s}$ (see~\eqref{eq:acute_Veff}). In other words, for $J\gg 1$, the absolute value of the symbol $\cU_{3,J}$ is much larger in the obtuse region than in the acute region. This matches with the fact that the leading term of $\cU_{3,J}$ blows up on the boundary of the acute region.

This same $J^{-\De_\s}$ scaling was observed for the large-spin, triple-twist anomalous dimension operator of~\cite{Harris:2024nmr}, obtained from six-point lightcone bootstrap. In fact, the leading large-spin limit of the Toeplitz operator in the obtuse region is the same as theirs at vanishing transverse spin ($\kappa=0$ in their notation). Embedding this result in the AdS three-body problem is helpful in understanding when and how this regime can be observed in the large-spin spectrum---see section~\ref{sec:breakdown} for one example. Conversely, \cite{Harris:2024nmr} is a first step toward a bootstrap derivation of the three-body problem at large spin for general multi-twist operators. 

To show that the Toeplitz operator in the obtuse region matches the operator in~\cite{Harris:2024nmr}, note that the latter is defined by its matrix elements $\gamma_{\ell_1\ell_2}(J,\kappa=0)$ in the eigenbasis~\eqref{eq:pairpot_eigenfunctions} of the pair potential $U_2(s_{12})$, labeled by even spins $0\leq \ell_1,\ell_2\leq J$. The large-spin limit they consider is $\ell_1,\ell_2=O(\sqrt{J})$. In this regime, the pair potential eigenfunctions in obtuse configuration $(1,-1,\a_3\in\D)$ tend to a basis of plane waves on the upper half-plane isomorphic to $\D$, which is orthogonal with respect to the scalar product defined by the measure $M_J(1,-1,\a_3)$ in~\eqref{eq:Mobtuse}. Evaluating the matrix elements of $\cU_{3,J}(1,-1,\a_3)$ in~\eqref{eq:obtuse_Veff} in this plane wave basis, we indeed retrieve \cite[equation~(6.19)]{Harris:2024nmr}.

\subsection{Berezin-Toeplitz quantization and Bohr-Sommerfeld conditions}
\label{sec:BT_general}

In this subsection we review Berezin-Toeplitz (BT) quantization \cite{berezin1975quantization,de1981spectral} and Bohr-Sommerfeld conditions. In the next subsection we will interpret the equations~\eqref{eq:CPhilbert}-\eqref{eq:CPVeff} as a BT quantization and compute the spectrum of $\g$ at large $J$.

We will be mostly following~\cite{LeFlochElliptic, CharlesRegular}, see the end of this section for further references. We will find that our setup is more singular than usually considered in mathematical literature, so we will not attempt to be fully rigorous. Let $\cM$ be a K\"ahler manifold whose symplectic form is $\w$. Let $\cL,\cK$ be Hermitian holomorphic line bundles on $\cM$, and assume that the Chern connection of $\cL$ has curvature $-i\w$. For a positive integer $J$, we define the BT Hilbert space to be
\be
	\cH_J^\text{BT}=\G(\cM,\cL^{\otimes J}\otimes \cK),
\ee
the Hilbert space of holomorphic sections of $\cL^{\otimes J}\otimes \cK$. The inner product is defined by
\be\label{eq:BTinner}
	\<\psi_1|\psi_2\> = \int_\cM \frac{\w^{\wedge n}}{n!}h'_J(\psi_1,\psi_2),
\ee
where $h'_J$ is the Hermitian inner product on $\cL^{\otimes J}\otimes \cK$ constructed from those of $\cL$ and $\cK$ and $n=\dim \cM$. One is then interested in the spectrum of a Hamiltonian $H$ defined by
\be\label{eq:BThamiltonian}
	\<\psi_1|H|\psi_2\>=\int_\cM \frac{\w^{\wedge n}}{n!}h'_J(\psi_1,\psi_2)H_\text{symb},
\ee
where $H_\text{symb}:\cM\to \R$ is a function on $\cM$ called the symbol of $H$.

The semiclassical limit is obtained by taking $J\to \oo$ with the identification
\be
	\hbar=J^{-1},
\ee
and under the assumption that the symbol $H_\text{symb}$ admits an asymptotic expansion
\be
	H_\text{symb}=H_\text{symb}^{(0)}+\hbar H_\text{symb}^{(1)}+\cdots.
\ee
It is convenient to introduce the normalized symbol $H_\text{norm}=\p{1+\hbar \De/4}H_\text{symb}$, where $\De$ is the (negative-semidefinite) Laplace-Beltrami operator. In other words,
\be\label{eq:normsymb}
	H_\text{norm}^{(0)}=H_\text{symb}^{(0)},\quad H_\text{norm}^{(1)}=H_\text{symb}^{(1)}+\tfrac{1}{4}\De H_\text{symb}^{(0)},\quad\cdots.
\ee
Note that the leading symbol $H_\text{norm}^{(0)}=H_\text{symb}^{(0)}$ has the interpretation of the classical Hamiltonian.

Bohr-Sommerfeld conditions allow one to compute the energy levels $E_k$ of $H$. To the leading order in $\hbar=J^{-1}$ one imposes, as expected,
\be\label{eq:leadingBS}
	A_0(E_k)=(2\pi \hbar)^n k,\quad k=0,1,\cdots,
\ee
where $A_0(E)$ is the phase-space volume enclosed by the constant-energy surface $\G_E=\{x\in \cM| H_\text{symb}^{(0)}(x)=E\}$, measured using the Liouville volume form $\w^{\wedge n}/n!$.  This is the standard statement that there is one quantum state per every $(2\pi\hbar)^n$ units of phase-space volume.

In the case $n=1$, which is the dimension relevant for the ($N=3$)-body problem with $\cM=\CP^{N-2}=\CP^1$, one can also easily compute the subleading correction. For this, one first views $c_0(E)=\hbar^{-1}A_0(E)$ as the monodromy angle of the Chern connection on $\cL^{\otimes J}$ along the curve $\G_E$. The correction term $c_1(E)$ is defined in a similar way as the monodromy around $\G_E$ of a connection on a version of $\cK$~\cite{LeFlochElliptic,CharlesRegular}. Extra care needs to be taken when $\G_E$ is not connected, which is the case in our setting. We describe the precise recipe in section~\ref{sec:subleadingBS} below. The resulting quantization condition is, in the simplest case,
\be\label{eq:subleadingBS}
	c_0(E_k)+c_1(E_k)=2\pi (k+\thalf)+O(\hbar).
\ee
Note that $c_0(E)=O(\hbar^{-1})$ and $c_1(E)=O(1)$. This means that while~\eqref{eq:leadingBS} essentially only predicts the density of states, the accuracy in~\eqref{eq:subleadingBS} is enough to resolve individual energy levels.

Before applying the above machinery to our problem, we add a brief summary of past and ongoing research on Berezin-Toeplitz quantization for the interested reader. As an example of geometric quantization~\cite{kostant1970,souriau1966quantification}, this framework was first developed in 1975 by Berezin~\cite{berezin1975quantization} for general K\"ahler manifolds, with a detailed study of Hermitian symmetric spaces in \cite{berezin1975general,berezin1975symmetric}. Using Boutet de Monvel and Guillemin's theory of Toeplitz operators~\cite{de1981spectral}, later works such as~\cite{Bordemann:1993zv} and \cite{guillemin1995star} put the Berezin-Toeplitz quantization of general compact K\"ahler manifolds on a rigorous footing---see~\cite{schlichenmaier2010berezin,le2018brief} for recent reviews of this topic. Note also the alternative approach in~\cite{ma2008generalized} and references therein, based on the asymptotics of the projection kernel, which generalizes to non-compact K\"ahler manifolds. In this context, recent works such as that of Charles~\cite{charles2003quasimodes,charles2003berezin,CharlesRegular}, Le Floch~\cite{LeFlochElliptic,LeFlochHyperbolic} and Deleporte~\cite{DeleporteHO} provide explicit semiclassical expansions of the spectrum and eigenfunctions. Theirs can be seen as a generalization of Voros's work~\cite{voros1989wentzel} on the WKB expansion in the Bargmann representation.

\subsection{Leading-order semiclassics for $N=3$}
\label{sec:LO_semiclass}

To interpret~\eqref{eq:CPhilbert}-\eqref{eq:CPVeff} as a BT quantization, we can simply set $\cK=\cO(0)$ to be the trivial line bundle with the Hermitian inner product
\be\label{eq:id_h_K}
	h_\cK(\psi_1,\psi_2)=\frac{J^{\#}d\mu_J}{\w}\bar\psi_1\psi_2.
\ee
Here, $-i\w$ is the Chern curvature of $h_1$, with $h_1$ defined in~\eqref{eq:h1defn}, while the power of $J$ is chosen to make $h_\cK$ have a finite limit at $J\to \oo$. With this definition,
\be
	\w\. h_J'(\psi_1,\psi_2) =\w\. (h_J\otimes h_\cK)(\psi_1,\psi_2) = J^{\#} d\mu_J h_J(\psi_1,\psi_2),
\ee
making~\eqref{eq:BTinner} and~\eqref{eq:CPinner} agree up to overall normalization.
Although the BT setup reviewed in the previous section does not allow the inner product on $\cK$ to depend on $J$, to the desired level of accuracy in~\eqref{eq:subleadingBS} only the leading $O(\hbar)=O(J^0)$ term in $h_\cK$ matters, and thus we can ignore the $J$-dependence. Similarly, to match~\eqref{eq:CPVeff}  and~\eqref{eq:BThamiltonian} (up to normalization) we can define 
\be\label{eq:id_H}
	H_\text{symb} = J^\#\cU_{N,J},
\ee
thereby factoring out the leading power of $J$. Note also that we are correcting~\eqref{eq:CPhilbert} to
\be\label{eq:CP1hilbert}
	\cH^\text{primary}_{3,J}\simeq (\cH^\text{BT}_J)^{S_3}= \G(\CP^{N-2},\cL^{\otimes J}\otimes \cK)^{S_3}.
\ee
Since $\cK$ is trivial, adding it does not modify the space of sections, only the inner product.

One issue overlooked in the above discussion is that the powers of $J$ that are natural to factor out in~\eqref{eq:id_h_K} and~\eqref{eq:id_H} are different in the obtuse and in the acute regions. If we define
\be
	H_\text{symb} \equiv J^{\De_\s}\cU_{N,J},
\ee
then the leading symbol $H_\text{symb}^{(0)}$ is finite in the interior of the acute region, blows up on the boundary of the acute region, and is formally infinite in the obtuse region (see the discussion around~\eqref{eq:acute_Veff} and~\eqref{eq:obtuse_Veff}). In other words, the obtuse region is classically inaccessible for all energies, and we should expect the wavefunctions to decay exponentially there. This is indeed what we observe in the exact diagonalization, see the discussion below. For future reference, using~\eqref{eq:acute_Veff} we find the leading and subleading symbols in the acute region:
\be
	H_\text{symb}^{(0)} &= \frac{b_0 \Gamma(\De_\phi+\De_\s/2-1)^2}{\Gamma(\De_\phi-1)^2} \sum_{i=1}^3 D_{i}^{\De_\s/2},\label{eq:Hsymb0}\\
	H_\text{symb}^{(1)} &=-\frac{\De_\s}{2}\frac{b_0 \Gamma(\De_\phi+\De_\s/2-1)^2}{\Gamma(\De_\phi-1)^2} \sum_{i=1}^3 C_i D_{i}^{\De_\s/2},\label{eq:Hsymb1}
\ee
where $D_i$ and $C_i$ are defined in~\eqref{eq:Ddefn} and~\eqref{eq:Cdefn}.

\begin{figure}[t]
	\centering
	\begin{tikzpicture}[scale=0.9,baseline={(0,0)}]
		\draw[->,opacity=1] (-3.2,0) -- (3.2,0) node[right] {$\Re(z)$};
		\draw[->,opacity=1] (0,-3.2) -- (0,3.2) node[above] {$\Im(z)$};
		
		\fill[red, opacity=0.05] (0,0) circle (1);
		
		\fill[red,opacity=0.05] (1,3) -- (1,-3) -- (3,-3) -- (3,3) -- cycle;
		
		\fill[red,opacity=0.05] (-3,3) -- (-3,-3) -- (-1,-3) -- (-1,3) -- cycle;
		
		\fill[blue,opacity=0.15] (1,0) arc[start angle=0, end angle=180, radius=1] -- (-1,3) -- (1,3) -- cycle; 
		\fill[blue,opacity=0.15] (1,0) arc[start angle=0, end angle=-180, radius=1] -- (-1,-3) -- (1,-3) -- cycle;
		
		\draw[dotted] (0,0) circle (1);
		
		\draw[dotted] (1,-3) -- (1,3);
		\draw[dotted] (-1,-3) -- (-1,3);
		
		\node[blue,opacity=0.5] at (0, 2) {Acute};
		\node[blue,opacity=0.5] at (0, -2) {Acute};
		\node[red,opacity=0.5] at (-2, 0.2) {Obtuse};
		\node[red,opacity=0.5] at (0, 0.2) {Obtuse};
		\node[red,opacity=0.5] at (2, 0.2) {Obtuse};
				
		\draw[thick] (-1,0) -- (0.6,1.5) -- (1,0) -- cycle;
		\node[right] at (0.6, 1.5) {$z$};
	\end{tikzpicture}
	\begin{tikzpicture}[scale=0.9,baseline={(0,0)}]
		\draw[->,opacity=1] (-3.2,0) -- (3.2,0) node[right] {$\Re(w)$};
		\draw[->,opacity=1] (0,-3.2) -- (0,3.2) node[above] {$\Im(w)$};
		
		\node (v1) at (1.5*2, 0.) {};
		\node (v2) at (-1.5, 1.5*1.7320508075688772) {};
		\node (v3) at (-1.5, -1.5*1.7320508075688772) {};
		
		\draw[opacity=0.5] (0,0) circle (1.5*2);
		\draw[dotted] (v1) arc[start angle=-90, end angle=-150, radius=1.5*3.4641] (v2);
		\draw[dotted] (v2) arc[start angle=30, end angle=-30, radius=1.5*3.4641] (v3);
		\draw[dotted] (v3) arc[start angle=150, end angle=90, radius=1.5*3.4641] (v1);
		\fill [blue,opacity=0.15] (v2) arc[start angle=210, end angle=270, radius=1.5*3.4641]  arc[start angle=90, end angle=150, radius=1.5*3.4641]  arc[start angle=-30, end angle=30, radius=1.5*3.4641] -- cycle;
		\fill [red, opacity=0.05] (v2) arc[start angle=210, end angle=270, radius=1.5*3.4641] arc[start angle=0, end angle=120, radius=3]  -- cycle;
		\fill [red, opacity=0.05] (v1) arc[start angle=90, end angle=150, radius=1.5*3.4641] arc[start angle=-120, end angle=0, radius=3]  -- cycle;
		\fill [red, opacity=0.05] (v3) arc[start angle=-30, end angle=30, radius=1.5*3.4641] arc[start angle=120, end angle=240, radius=3]  -- cycle;
		
		\draw[fill=black] (0,0) circle (0.04); 
		\draw[fill=black] (-0.803848, 0.) circle (0.04);

		\draw[fill=black] (-1.65594, 0.) circle (0.04); 
		\draw[fill=black] (2.1142, 0.) circle (0.04); 
		
		\begin{scope}[draw opacity=0.8,shift={(0,0.3)}]
		\begin{scope}[shift={(2.1142,0)}, scale = 0.05]
			\draw[] (-1,0) -- (1,0) -- (0,10) -- cycle;
		\end{scope}
		\begin{scope}[shift={(0,0)}, scale = 0.25]
				\coordinate (A) at (-1, 0); 
				\coordinate (B) at (1, 0);  
				\coordinate (C) at (0,1.73205);  
				\draw[] (A) -- (B) -- (C) -- cycle;
			    \draw (A) ++(0.5,0) arc[start angle=0, end angle=60, radius=0.5];   
				\draw (B) ++(-0.5,0) arc[start angle=180, end angle=120, radius=0.5]; 
				\draw (C) ++(-0.25,-0.433013) arc[start angle=-120, end angle=-60, radius=0.5]; 
		\end{scope}
		\begin{scope}[shift={(-0.804,0)}, scale = 0.3]
			\draw[] (-1,0) -- (1,0) -- (0,1) -- cycle;
			\draw[] (-0.3,0.7) -- (0,0.4) -- (0.3,0.7);
		\end{scope}
		\begin{scope}[shift={(-1.656,0)}, scale = 0.3]
			\draw[] (-1,0) -- (1,0) -- (0,0.5) -- cycle;
		\end{scope}
	\end{scope}
	\end{tikzpicture}
	\caption{Left: complex $z$-plane, separated into obtuse (red) and acute (blue) regions. We also show a sample triangle whose shape is parameterized by $z$. Right: the upper-half plane of $z$ mapped to the unit disk of the $w$ variable, where $\Z_3$ symmetry is manifest. Examples of triangles are shown together with their location on the unit disk.\label{fig:acuteobtuse}
	}
\end{figure}
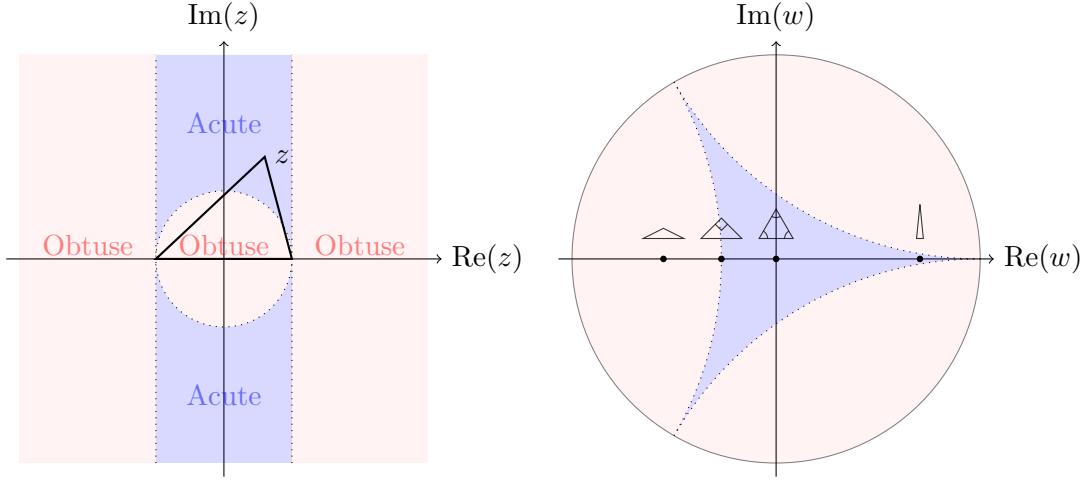

 There is in fact another reason to discard the obtuse region: to see it, let us compute the symplectic form $\w$ associated with $h_1$. For convenience, we work in the coordinates on $\CP^1$ where $\a_1=-\a_2=1$ and write $\a_3=z\in \C$. The acute region is given by
\be
	\text{Acute}=\{z\in \C|\Re z<1,\, \Re z>-1,\, |z|>1\},
\ee
while the obtuse region is
\be
	\text{Obtuse} = \{z\in \C|\Re z>1\}\cup \{z\in \C|\Re z<-1\}\cup \{z\in \C||z|<1\},
\ee
see figure~\ref{fig:acuteobtuse}. In the acute region we find, using~\eqref{eq:acute_r},
\be\label{eq:racute}
	r(z) = \left|\frac{z^2-1}{z-\bar z}\right|.
\ee
In the obtuse region, let us focus on the connected component with $|z|<1$. There, we simply have
\be\label{eq:robtuse_3}
	r(z) = 1.
\ee
From the definition~\eqref{eq:h1defn}, the curvature of the Chern connection $\nabla$ on $\cL$ is given by
\be\label{eq:curvexpression}
	\mathrm{curv}(\nabla)=\bar\ptl\ptl \log r^{-2} = \ptl_z\ptl_{\bar z} \log r^2 dz\wedge d\bar z.
\ee
Recall that the symplectic form $\w$ in BT quantization is given by $\w=i\mathrm{curv}(\nabla)$. Thus, we find
\be
\w=\begin{cases}
	\frac{-2idz\wedge d\bar z}{(z-\bar z)^2},& z\in\text{Acute},\\
	0,&z\in \text{Obtuse}.
\end{cases}\label{eq:symplecticform}
\ee
In the acute region, we recognize 
\be
\w = \frac{-2idz\wedge d\bar z}{(z-\bar z)^2}=\frac{dx\wedge dy}{y^2},
\ee
where $z=x+iy$, as the hyperbolic volume form in upper (lower) half-plane. That $\w$ vanishes in the connected component of the obtuse region with $|z|<1$ follows immediately from~\eqref{eq:robtuse_3} and~\eqref{eq:curvexpression}. Vanishing in the other connected components can be obtained either by permutation symmetry or by a direct computation.

We therefore see that not only is the obtuse region classically inaccessible, but the Liouville measure $\w$ vanishes there. On the other hand, the Liouville measure in the acute region becomes precisely the hyperbolic measure, provided we identify both the lower and the upper half-planes of $z$ with the hyperbolic plane $\H^2$. Under this identification, the acute region becomes two copies of the ideal triangle in $\H^2$, i.e.\ the hyperbolic triangle with all angles equal to zero---see figure~\ref{fig:acuteobtuse}. Note that this $\H^2$ plays the role of the configuration space of three cyclic-ordered points and is different from the $\H^2$ that appears in the Landau level analogy.

To make permutation symmetries more apparent, it is convenient to define a new coordinate $w$ by
\be\label{eq:wdef}
	w=\frac{z-i\sqrt 3}{z+i\sqrt 3}.
\ee
The upper half-plane of $z$ is mapped to the unit disk of $w$, see figure~\ref{fig:acuteobtuse}. The cyclic permutations of $\a_i$ act on $w$ by $2\pi/3$ rotations, while transpositions act by inversions, exchanging the two connected components of the acute region.

\begin{figure}[t]
	\centering
	\includegraphics[scale=0.25]{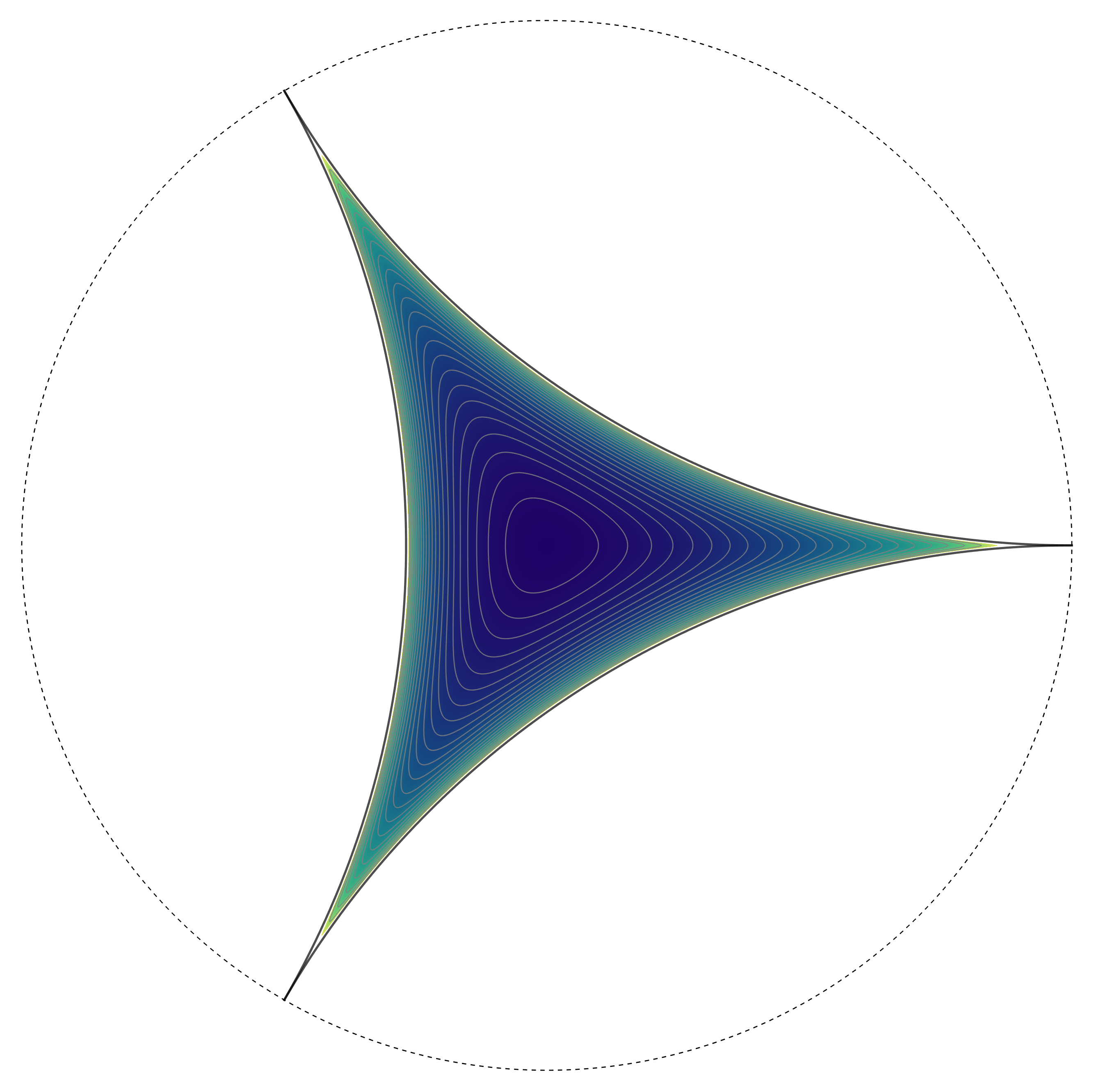}
	\caption{
		Density plot of the classical potential $H^{(0)}_\text{symb}$ in a connected component of the acute region, in the $w$ coordinate. Note that  $H^{(0)}_\text{symb}$ diverges on the boundary of the acute region and is infinite in the obtuse region.  The gray curves are the level sets of $H^{(0)}_\text{symb}$ corresponding to the exact numerical eigenvalues of $J^{\De_\s}\g$. There is no level set for the ground state since the exact ground state energy is slightly below the minimum of $H^{(0)}_\text{symb}$ due to $O(\hbar)$ corrections. In this figure, $\De_\f = 1.234,\, \De_\s = 0.6734$.
	}
	\label{fig:potential}
\end{figure}

Overall, we find a classical system whose phase space is two copies of the ideal triangle in $\H^2$, and the classical Hamiltonian is given by the leading term in~\eqref{eq:acute_Veff}. Note that the phase space has a finite volume equal to $2\pi$ (each ideal triangle has hyperbolic area $\pi$). Dividing by $3!$ to account for permutation symmetry between $\a_i$, we find the semiclassical Hilbert space dimension
\be
	\frac{2\pi/3!}{2\pi \hbar}+O(1)=J/6+O(1),
\ee
in agreement with~\eqref{eq:N3dimension}. We therefore expect all but a vanishing fraction of states to be well described by the semiclassical picture.

Semiclassically, $\g$ eigenstates $\psi$ should be localized near the level sets of the classical Hamiltonian $H_\text{symb}^{(0)}$. These level sets coincide with the classical phase space trajectories.\footnote{Note that the wavefunctions $\psi$ here are defined on the classical phase space rather than just the position space, and the discussion is more analogous to the WKB approximation in Bargmann representation~\cite{voros1989wentzel} rather than the textbook position-space WKB. } In particular, we expect the ground state $\psi$ to be localized near the minima of $H_\text{symb}^{(0)}$ at $w=0$ and $w=\oo$. The density plot of $H_\text{symb}^{(0)}$ is shown in figure~\ref{fig:potential}, together with the level sets corresponding to the exact eigenvalues of $J^{\De_\s}\g$. Here and below, we only make the plots in the unit disk of $w$. The picture outside the unit disk is exactly the same after replacing $w\to w^{-1}$, which is ensured by the $S_3$ permutation symmetry among $\a_i$.

\begin{figure}[t]
	\centering
	\includegraphics[scale=0.2]{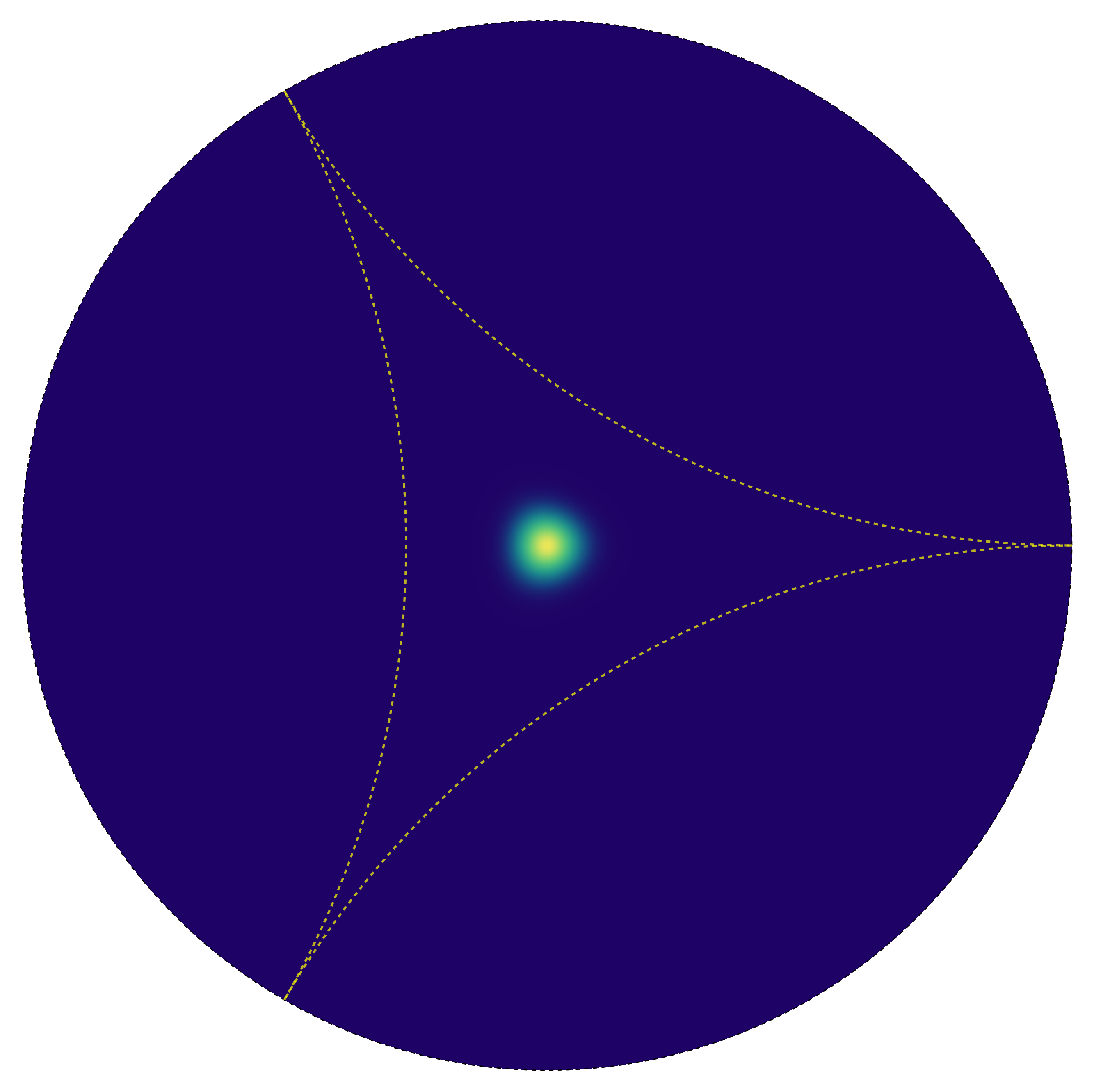}~
	\includegraphics[scale=0.2]{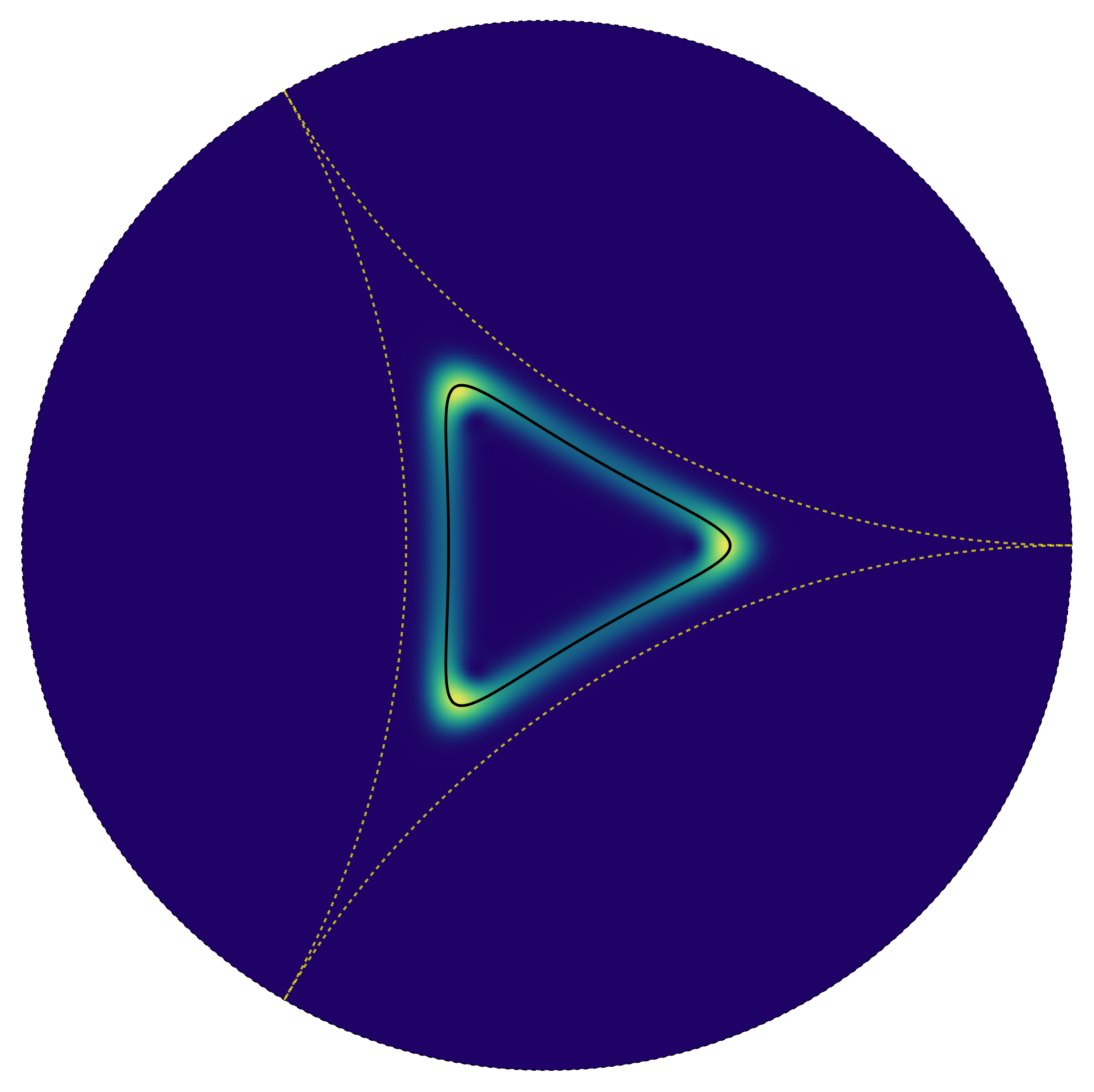}
	\caption{
		Density plot of $h_J(\psi,\psi)$ for eigenfunctions $\psi$ of $\g$ at $J=162$ in the unit disk of $w$, cf.\ figure~\ref{fig:acuteobtuse}. The wavefunctions have been obtained using exact numerical diagonalization as described in section~\ref{sec:diagonalization}. The left figure shows the ground state. The right figure shows the 7-th excited state out of the 27 excited states available at this spin (counting only fully permutation-symmetric wavefunctions). The overlaid black curve is the corresponding level set of the classical Hamiltonian $H_\text{symb}^{(0)}$. The dashed yellow curves show the boundary between the acute and the obtuse regions. In this figure, $\De_\f = 1.234,\, \De_\s = 0.6734$.
	}
	\label{fig:GS_and_excited}
\end{figure}

A natural measure of the magnitude of $\psi$ is given by the Hermitian inner product $h_J(\psi,\psi)$. In figure~\ref{fig:GS_and_excited} we show the density plots of $h_J(\psi,\psi)$ for the ground state and for an excited state, which confirm our expectations. The ground state is clearly localized around $w=0$, corresponding to an equilateral triangle configuration of $\a_i$. The excited state is nicely localized around the corresponding level set of $H_\text{symb}^{(0)}$. 

\begin{figure}[t]
	\centering
	\includegraphics[scale=0.2]{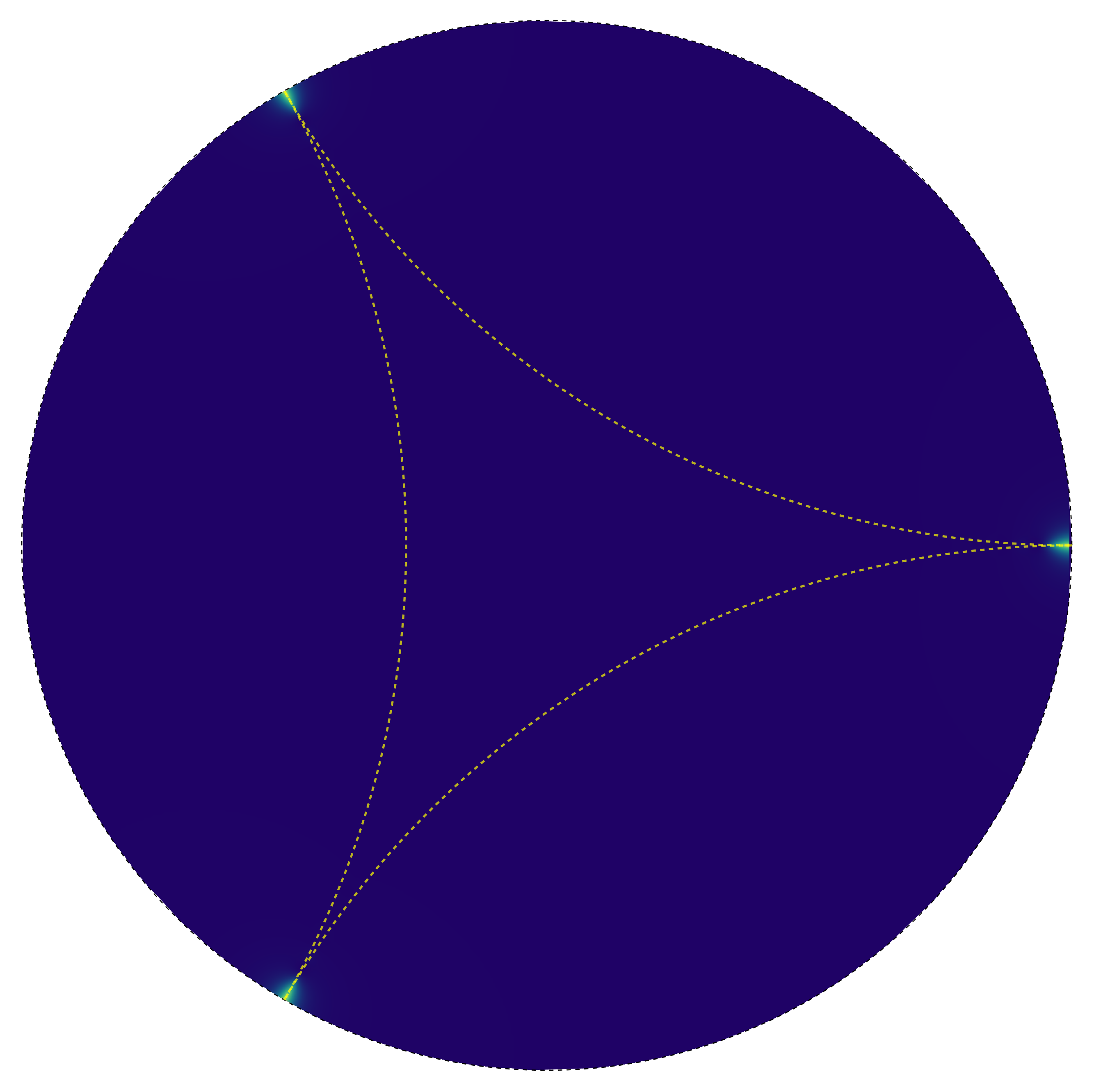}
	\caption{
		Same as figure~\ref{fig:GS_and_excited} but showing a highly excited state (25-th excited state out of 27). The wavefunction is localized around degenerate triangles where two vertices coincide, see figure~\ref{fig:acuteobtuse}.
	}
	\label{fig:UV_wavefunction}
\end{figure}

On the other hand, in figure~\ref{fig:UV_wavefunction} we show the density plot $h_J(\psi,\psi)$ for one of the most excited states at the same value of $J$. This state is localized around $w=e^{2\pi i k/3}$ for $k=0,1,2$. In terms of $\a_i$, these correspond to configurations where two of the $\a_i$ coincide---cf.\ figure~\ref{fig:acuteobtuse}. This is to be expected, since the most excited states correspond to operators of the form $[\f,[\f,\f]_\ell]_J$ with small $\ell$. This state is not in the semiclassical regime since it has non-trivial support in the obtuse region. More technically, the assumptions that we made in deriving the large-$J$ expansions of $M_J$ and $\cU_{N,J}$ are violated---we would need to make the differences $\a_{ij}$ scale non-trivially with $J$ to include such wavefunctions in our analysis.

Returning to the semiclassical states, as discussed in section~\ref{sec:BT_general}, the energy levels at leading order in $J$ can be determined using the Bohr-Sommerfeld condition~\eqref{eq:leadingBS},
\be
	A_0(E)=2\pi \hbar k=2\pi k/J, \quad k = 0,1,2, \cdots.
\ee
In our case, $A_0(E)$ becomes the hyperbolic area enclosed by the level set of $H_\text{symb}^{(0)}$. For example, in the case of the equal-energy contour shown for the excited state in figure~\ref{fig:GS_and_excited}, we find
\be
	\frac{A_0(E)}{2\pi \hbar} \approx 41.18.
\ee
This includes the area in both the connected components of the acute region. To compare this with the expected level number $k=7$, we need to recall that we are only interested in permutation-symmetric wavefunctions. At this level of accuracy, we can simply divide $A_0(E)$ by $3!$ to get
\be
	\frac{A_0(E)}{3!\x2\pi \hbar} \approx 6.86,
\ee
which is close enough to the expected $k=7$ level number. We delay the further quantitative comparison to the following subsections, where we compute the subleading correction to the Bohr-Sommerfeld condition and discuss the more precise implementation of permutation symmetries. 

\subsection{Subleading semiclassics for $N=3$}
\label{sec:subleadingBS}

To compute the subleading correction to the Bohr-Sommerfeld conditions, we first review the general case of BT quantization with 1 degree of freedom, as stated in~\cite{CharlesRegular, LeFlochElliptic} and using the notation from section~\ref{sec:BT_general}. We sketch in appendix~\ref{app:BSderivation} how this and further subleading corrections can be systematically derived using pseudodifferential operator techniques.

We temporarily assume that $\G_E$ is connected. We first carefully define $c_0(E)$. Let $U_E=\{x\in\cM|H_\text{symb}^{(0)}(x)<E\}$ be the region of the phase space where the classical energy is less than $E$, so that $\G_E=\ptl U_E$. We endow $\cM$ and thus $U_E$ with the orientation defined by $\w$. The orientation on $\G_E$ is induced from $U_E$ in the standard way compatible with Stokes's theorem. We define $c_0(E)$ so that $e^{ic_0(E)}$ is the holonomy in $\cL^{\otimes J}$ around $\G_E$. This only determines $c_0(E)$ modulo $2\pi$. However, most important to us is the case when $\G_E$ is contractible, and thus $\cL^{\otimes J}$ can be trivialized in $U_E$. In this case, $c_0(E)$ is defined unambiguously by
\be
	c_0(E) = i\oint_{\G_E}\b_{\cL^{\otimes J}}=iJ\oint_{\G_E}\b_{\cL}=iJ\int_{U_E}d\b_{\cL}=J\int_{U_E}\w,
\ee
where $\b_{\cL}$ is the connection form for the line bundle $\cL$, and similarly for other bundles. Using that the curvature $d\b_{\cL}$ is $-i\w$, we see that $\hbar c_0(E)=A_0(E)$ as promised earlier. 

Let $X$ be the Hamiltonian vector field on $\cM$ associated to $H_\text{symb}^{(0)}$.  That is to say
\be\label{eq:Xdef}
	\w(X,\.)=dH_\text{symb}^{(0)},
\ee
and the classical equation of motion is simply $\dot x = X$. It is easy to check that $X$ is tangent to the curves $\G_E$ and is oriented oppositely to the orientation of $\G_E$. Define a $(1,0)$-form $\kappa\in \L^{1,0}T^*\cM$ so that
\be
	\kappa(X) = H_\text{norm}^{(1)}.
\ee
Furthermore, let $\de$ be a half-form bundle, i.e.\ a line bundle such that $\de^{\otimes 2}$ is isomorphic to the line bundle of holomorphic 1-forms, $\L^{1,0}T^*\cM$. Define $\cL_1$ so that $\cK = \cL_1\otimes \de$, and let $\nabla_1$ be its Chern connection. Define a new connection on $\cL_1$ via
\be
	\tl \nabla_1=\nabla_1-i\kappa.
\ee
The correction $c_1(E)$ in~\eqref{eq:subleadingBS} is defined so that the holonomy of $\tl\nabla_1$ around the constant-energy contour $\G_E$ is $e^{ic_1(E)}$. Similarly to $c_0(E)$, when $\G_E$ is contractible, $c_1(E)$ can be defined unambigously as
\be
	c_1(E) = i\oint_{\G_E} \tl\b_{\cL_1}=i\oint_{\G_E} \p{\b_{\cL_1}-i\kappa},
\ee
where $\tl\b_{\cL_1}$ is the connection 1-form of $\tl\nabla_1$ and $\b_{\cL_1}$ is that of $\nabla_1$. 

We note that the part of $c_1(E)$ coming from $\kappa$ can be interpreted in the Bohr-Sommerfeld conditions as shifting energy levels by $E\to E+J^{-1}\<H^{(1)}_\text{norm}\>_E$, where $\<f\>_E$ is the time-average of $f$ over the classical trajectory of $H^{(0)}_\text{symb}$ with energy $E$. To see this, note that $\oint_{\G_E}\kappa = -\<H_\text{norm}^{(1)}\>_E\,T$ and $J^{-1}\ptl c_0(E)/\ptl E = T$, where $T$ is the period of motion of the classical system with energy $E$. In this sense, the $\<H^{(1)}_\text{symb}\>_E$ part of $\<H^{(1)}_\text{norm}\>_E$ accounts for the deformation of the classical Hamiltonian, while $\<\tfrac{1}{4}\Delta H^{(0)}_\text{symb}\>_E$ gives an intrinsically quantum correction. Similarly, the holonomy of $\nabla_1$ (i.e.\ the contribution of $\b_{\cL_1}$) can be split as the holonomy in $\cK$ minus the holonomy in $\de$. The holonomy in $\cK$ can be combined with $c_0(E)$ to be interpreted as the holonomy in $\cK\otimes \cL^{\otimes J}$. The latter is simply the bundle in which the wavefunctions live, so in some sense this can be viewed as a deformation of the ``classical'' bundle $\cL$. The holonomy in $\de$ can be seen as an intrinsically quantum correction.

When $\G_E$ is contractible and $dH_\text{symb}^{(0)}$ vanishes only at the unique minimum in $U_E$,\footnote{In a more general setting, $k+\thalf$ might need to be replaced by $k$, and $k$ may not count the level number anymore due to $2\pi$ ambiguities in $c_0(E)$ and $c_1(E)$. See~\cite{LeFlochElliptic} and also~\cite{LeFlochHyperbolic} for semiclassical analysis near singular values of $H^{(0)}_\text{symb}$.} the Bohr-Sommerfeld condition takes the form 
\be
	c_0(E)+c_1(E) = 2\pi (k+\thalf)+O(\hbar),\quad k=0,1,2,\cdots.
\ee
Furthermore, $c_0(E)+c_1(E)-\pi=2\pi k$ can be interpreted as the total phase that the wavefunction $\psi$ picks up when going around $\G_E$---see the discussion in appendix~\ref{app:BSderivation}, identifying $k$ with the number of zeroes of $\psi$ in $U_E$.

When $\G_E$ is disconnected, we effectively have several energy wells in the phase space, and the wavefunctions can be localized in any of these wells. Therefore, in this case, we have to treat each connected component of $\G_E$ independently and take the union of the resulting energy spectra.\footnote{When there are degeneracies between the spectra coming from individual connected components, there can still be exponentially-suppressed mixing due to instanton corrections and the exact eigenstates are not necessarily localized in individual wells. We discuss this in more detail in the next subsection.}

Let us now compute $c_1(E)$ for our problem. We only need to perform the calculation in the acute region. For $dz(X)$, equation~\eqref{eq:Xdef} implies
\be
	dz(X)=\frac{i(z-\bar z)^2}{2}\ptl_{\bar z} H^{(0)}_\text{symb},
\ee
where we used the symplectic form~\eqref{eq:symplecticform}. Therefore, we can take
\be\label{eq:kappa}
	\kappa = \frac{-2i H^{(1)}_\text{norm}}{(z-\bar z)^2\ptl_{\bar z} H^{(0)}_\text{symb}}dz.
\ee
Note that the form $\kappa$ behaves as $(z-z_0)^{-1}dz$ near the minimum $z_0=i\sqrt{3}$ of $H^{(0)}_\text{symb}$ since, as can be checked, $H_\text{symb}^{(0)}=A+B(z-z_0)(\bar z-\bar z_0)+\cdots$ for some constants $A,B$.\footnote{In our particular case, terms of the form $(z-z_0)^2$ and $(\bar z-\bar z_0)^2$ are forbidden by cyclic $\Z_3$ permutation symmetry. However, one can check that even when such terms are present at the minimum, the monodromy contribution is still non-zero for small $\G_E$.}  The contribution of this singularity to the monodromy of $\tl \nabla_1$ is non-zero even for very small contours $\G_E$. 

The hyperbolic Laplace-Beltrami operator is
\be
	\Delta= -(z-\bar z)^2\ptl_z\ptl_{\bar z}
\ee
and thus (see~\eqref{eq:normsymb})
\be
	H_\text{norm}^{(1)}=H_\text{symb}^{(1)}-\tfrac{1}{4}(z-\bar z)^2 \ptl_z\ptl_{\bar z} H_\text{symb}^{(0)},
\ee
where $H^{(0)}_\text{symb}$ and $H^{(1)}_\text{symb}$ are given by~\eqref{eq:Hsymb0} and~\eqref{eq:Hsymb1} (see also~\eqref{eq:xi}).

The half-form line bundle $\de$ can be taken to be $\cO(-1)$ since $\L^{1,0}T^*\cM=\cO(-2)$. In this case we have to set $\cL_1=\cO(1)$ to get $\cK=\cL_1\otimes \de=\cO(0)$. The inner product $h_\de$ on $\de$ is chosen so that the induced inner product on $\de^{\otimes 2}$ coincides with the inner product $h_{\cO(-2)}$ on $\cO(-2) =\L^{1,0}T^*\cM$, which in turn satisfies
\be
	h_{\cO(-2)}(\mu,\eta)=\frac{\bar\mu \wedge \eta}{i\w}. 
\ee
In other words, 
\be
	h_{\cO(-2)}(f dz,g dz)=\frac{|z-\bar z|^2}{2}\bar f g.
\ee
Let $dz^{1/2}$ be a section of $\de$ that satisfies $(dz^{1/2})^{\otimes 2} = dz$. Then we have
\be
	h_\de(f dz^{1/2},g dz^{1/2})=\frac{|z-\bar z|}{\sqrt{2}}\bar f g.
\ee
Let $\r$ be a section of $\cL_1=\cO(1)$, viewed as the dual bundle of $\de=\cO(-1)$, such that $\r(dz^{1/2})=1$. Then the inner product $h_{\cL_1}$ on $\cL_1$ is given by
\be\label{eq:L1inner}
	h_{\cL_1}(f\r,g\r) = \frac{\sqrt 2}{|z-\bar z|}\frac{J^{3\De_\f-6}d\mu_J}{|\w|}\bar f g.
\ee
This choice is necessary for the inner product induced on $\cK$ to be given by \eqref{eq:id_h_K}. We also plugged in the power of $J$ appropriate for the acute region, from~\eqref{eq:Macute}. Using $\r$ as the basis for sections of $\cL_1$, the connection $\tl \nabla_1$ is given by
\be
	\tl\nabla_1=\ptl + \ptl \log h_{\cL_1}(\r,\r)-i\kappa.
\ee
The term $\ptl \log h_{\cL_1}(\r,\r)$ is non-singular in the acute region and its contribution to $c_1(E)$ goes to zero for small contours $\G_E$.

It only remains to derive an explicit expression for $d\mu_J$ in $z$ coordinate. We do this by choosing the gauge-fixing function $F$ so that
\be
	\de(F(\a))=\de^2(\a_1-1)\de^2(\a_2+1)
\ee
and identifying $\a_3=z$. Since $\cD_F(\a)$ can only depend on $\a_1,\a_2$ and is $\C^\x\ltimes \C$-invariant, it has to be a constant which is readily verified to be $\cD_F(\a)=4$. We therefore find
\be
	d\mu_J=4d^2z M_J(1,-1,z)r^{2J}(1,-1,z).
\ee
In the acute region, $r(z)=r(1,-1,z)$ is given by~\eqref{eq:racute} and $M_J(\a)$ is given by~\eqref{eq:MacuteGeneric}. Overall, from~\eqref{eq:L1inner} we find
\be\label{eq:L1rhorho}
	h_{\cL_1}(\r,\r) = \sqrt{2}J^{3\De_\f-6} |z-\bar z|M_J(1,-1,z)r^{2J}(1,-1,z).
\ee

We also have to remember that $\G_E$ and $U_E$ each have two connected components---one in the upper half-plane of $z$, and one in the lower half-plane. We denote them by $\G_E^+,U_E^+$ and $\G_E^-,U_E^-$ respectively. The components $\G_E^+$ and $\G_E^-$ are exchanged by the $\Z_2$ transposition $(12)$, and thus lead to exactly the same spectrum. Focusing on $\G_E^+$, the correction $c_1(E)$ can now be written explicitly as
\be
c_1(E) = i\oint_{\G_E^+} \p{\ptl \log h_{\cL_1}(\r,\r)-i\kappa},
\ee
where the orientation on $\G_E^+$ is chosen to be counter-clockwise, $h_{\cL_1}(\r,\r)$ is given by~\eqref{eq:L1rhorho}, and $\kappa$ is defined in~\eqref{eq:kappa}. Similarly, $c_0(E)$ is given by
\be
	c_0(E) = iJ\oint_{\G_E^+}\b_{\cL}=J\int_{U_E^+}\w,
\ee
and the Bohr-Sommerfeld rule is
\be\label{eq:BSsubleading_connected}
	c_0(E)+c_1(E) = 2\pi (k+\thalf)+O(\hbar^{-1}),\quad k=0,1,2,\cdots.
\ee
The full semiclassical energy spectrum (not imposing the $S_3$ permutation invariance on the wavefunctions) is given by the solutions $E_k$ of the above equation, with each energy level being doubly-degenerate, at least up to exponentially-small corrections. We discuss the status of this degeneracy and the implementation of $S_3$ permutation invariance in the next subsection.

For now, taking the excited energy level in figure~\ref{fig:GS_and_excited} at $J=162$ as an example, we find (recall $c_0$ and $c_1$ are now defined using $\G_E^+$ only)
\be
	k=\frac{c_0(E)+c_1(E)-\pi}{2\pi}\approx 3\x7.002,
\ee
very close to the expected value $k=7$ if we interpret the factor of $3$ as taking care of the permutation symmetries. We will see in the next subsection that this is indeed the correct interpretation for $J=162$.

\subsection{Accounting for permutation symmetries}
\label{sec:permutations}

The physical wavefunctions for $N=3$ identical particles have to be permutation-invariant under $S_3$. Here we would like to study this condition from the semiclassical point of view, which amounts to determining the $S_3$ transformation properties of the semiclassical states discussed in the previous subsection.  

In the previous subsection, we saw that the spectrum of $\g$ without imposing $S_3$ invariance appears to be doubly-degenerate. This was because the classical Hamiltonian $H^{(0)}_\text{symb}$ has two minima separated by a large (technically infinite at $J=\oo$) potential barrier (see figures~\ref{fig:acuteobtuse} and~\ref{fig:potential}). In fact, we can view this situation as spontaneous symmetry breaking of $S_3$ down to the cyclic $\Z_3$. Indeed, each connected component of the acute region corresponds to a particular cyclic ordering of the three particles, with the minima of $H^{(0)}_\text{symb}$ located at $\Z_3$ symmetric equilateral configurations. 

Spontaneous symmetry breaking does not normally occur in quantum mechanics, and therefore one should generically expect the double degeneracies to be broken by instanton corrections, exponentially small in $\hbar^{-1}=J$. However, we shall see that some of the degeneracies are protected and remain exact, in a manner somewhat similar to the quantum-mechanical example in appendix D of~\cite{Gaiotto:2017yup}. Specifically, we will determine how the spectrum organizes into representations of $S_3$ and show that some naively degenerate pairs of states have to form doublets of $S_3$, which are therefore exactly degenerate. Of course, these states do not appear in the physical $S_3$-invariant spectrum.

First, we will determine the transformation properties of the energy eigenstates under the cyclic $\Z_3$. Since the action of $\Z_3$ preserves the connected component $U_E^+$, we can determine the $\Z_3$ charge of a state $\psi$ just from the behavior of $\psi$ in $U_E^+$. As before, we parameterize $\psi$ by
\be
	\psi(z) \equiv \psi(1,-1,z).
\ee
In terms of $\psi(z)$, invariance under cyclic permutations can be stated as
\be\label{eq:Z3invariance}
	\psi(z) = \psi(1,-1,z) = \psi(z,1,-1) = \p{\tfrac{z-1}{2}}^J\psi(1,-1,\tfrac{z+3}{1-z})=\p{\tfrac{z-1}{2}}^J\psi(\tfrac{z+3}{1-z}),
\ee
where we used the homogeneity and translation-invariance of $\psi(\a)$. It is convenient to rephrase this condition terms of the $w$ coordinate defined in~\eqref{eq:wdef}. Introducing
\be
	\eta(w) = \p{\frac{dz}{dw}}^{-J/2}\psi(z)
\ee
we get the condition
\be\label{eq:Z3invarianceW}
	\eta(w) = e^{2\pi i J/3}\eta(e^{2\pi i/3}w).
\ee
More generally, if $\psi$ has charge $m\mod 3$ under $\Z_3$, we have
\be\label{eq:Z3invarianceWwithCharge}
\eta(w) = e^{2\pi i (J+m)/3}\eta(e^{2\pi i/3}w).
\ee
Equation~\eqref{eq:Z3invarianceWwithCharge} implies that the zeros of $\psi$ in $U_E^+$ at $\w\neq 0$ appear in groups of $3$. Furthermore, around $w=0$ we must have an expansion 
\be
	\eta(w) = \sum_{n}\eta_n w^n,
\ee
where the sum is over $n\geq 0$ satisfying $J+m+n\equiv 0\mod 3$, which constrains the multiplicity of a possible zero at $w=0$. Altogether, this implies that the number $n_0$ of zeroes of $\psi$ in $U_E^+$ (counted with multiplicity) satisfies $J+m+n_0\equiv 0\mod 3$. Recalling that $n_0$ coincides with the integer $k$ that appears in the Bohr-Sommerfled condition~\eqref{eq:BSsubleading_connected}, we find
\be\label{eq:Z3charge}
	m\equiv -J-k\mod 3.
\ee

We see that the $\Z_3$ charge is fully determined by $J$ and the level number $k$. When $m\not\equiv 0\mod 3$, the state must be a part of the two-dimensional representation of $S_3$. Indeed, $S_3$ has three irreducible representations: the trivial and the sign representations, which are both one-dimensional, and one two-dimensional representation. In both the trivial and the sign representations the $\Z_3$ subgroup acts trivially. Since there are precisely two states for a given value of $k$, it is these two states that must form the $S_3$ doublet.

For the values of $k$ for which the $\Z_3$ charge is trivial, $m\equiv 0\mod 3$, the state $\psi$ can form either the trivial or the sign representation of $S_3$. Naively, for a given value of $k$, we can construct two wavefunctions, localized either in $U_E^+$ or in $U_E^-$. Since nothing enforces two-fold degeneracy, we generically expect that instanton corrections will break it, with exact energy eigenstates becoming the symmetric and anti-symmetric linear combinations of the localized wavefunctions. In other words, for such values of $k$ we expect one eigenstate in the trivial representation of $S_3$ and one in the sign representation.

For a more principled approach, note that the semiclassical picture predicts the representation of $S_3$ realized on the nearly-degenerate eigenstates at energy $E$ to be the induced representation $\mathrm{Ind}_{\Z_3}^{S_3}e^{2\pi i m/3}$. Indeed, we have one localized semiclassical eigenstate for every element of $S_3/\Z_3$, and it is easy to check that the action of $S_3$ agrees with the definition of the induced representation. We have 
\be
	\mathrm{Ind}_{\Z_3}^{S_3}e^{2\pi i m/3} = \begin{cases}
		\myng{(3)}\oplus \myng{(1,1,1)},&\quad m\equiv 0\mod 3\\
		\myng{(2,1)},&\quad m\not\equiv 0\mod 3
	\end{cases},
\ee
where $\myng{(3)}$ is the trivial representation, $\myng{(1,1,1)}$ is the sign representation, and $\myng{(2,1)}$ is the two-dimensional representation of $S_3$. We note that although representations other than the trivial representation $\myng{(3)}$ do not interest us here, they would be relevant if the constituent operators $\f$ were charged under a non-abelian global symmetry but the leading-twist exchange was agnostic to this charge.\footnote{This would be the case, for example, if $\Phi$ in our toy model was a fundamental of $U(N_f)$ flavor symmetry. Or, in a more general CFT context, if the leading-twist exchange between $\f$'s was due to a neutral scalar $\s$ with $\De_\s<d-2$ (exchange of a global symmetry current is always expected and is at twist $d-2$).}

In summary, the $S_3$-invariant states exist for $k$ such that
\be
	J+k\equiv 0\mod 3,
\ee
and there is exactly one such state for each suitable value of $k$. For such values of $k$, there is also a state in the sign representation of $S_3$, whose energy differs by an amount that is exponentially small in $\hbar =J^{-1}$. For all other values of $k$, there is a pair of exactly degenerate states transforming in the two-dimensional representation of $S_3$.

\begin{figure}[t!]
	\includegraphics[scale=0.37]{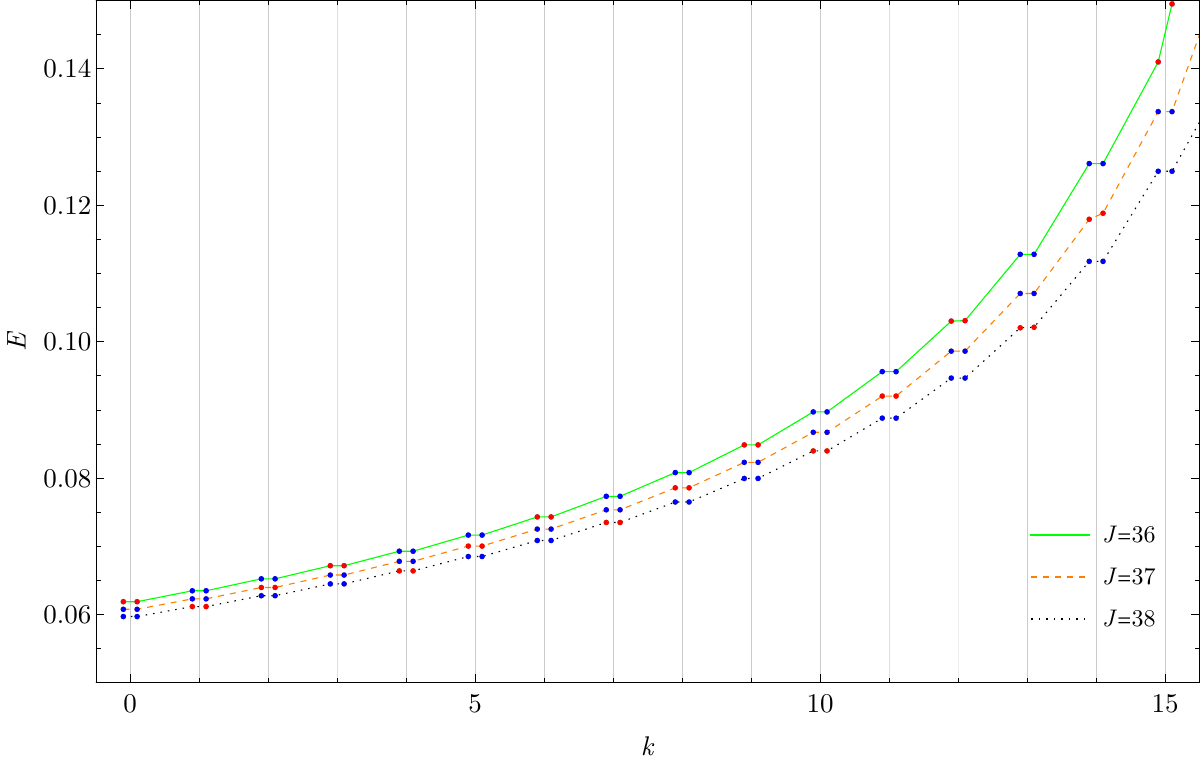}~
	\includegraphics[scale=0.37]{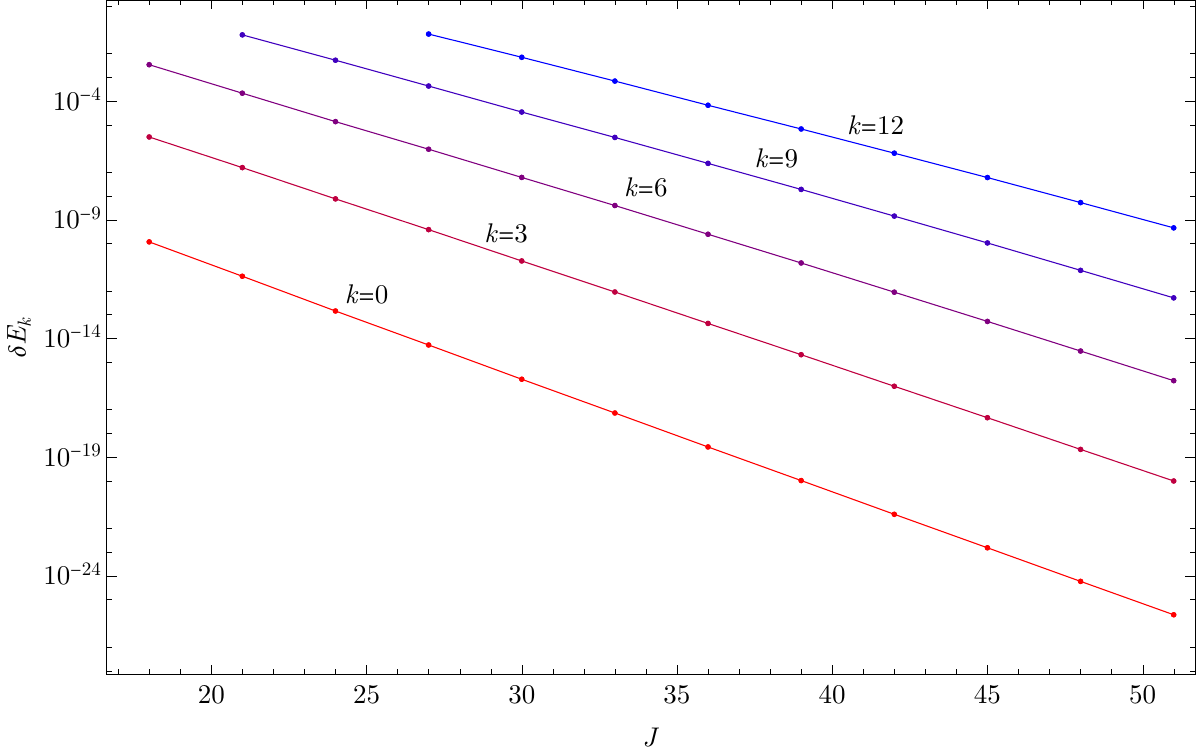}
	\caption{
	Left: the exact spectrum for $N=3$ and three values of $J$, ordered by the level number $k$. For each $k$, there is a nearly-degenerate pair of states. Pairs of states shown in blue are exactly degenerate, while the pairs shown in red are only approximately degenerate. Right: the splittings between nearly-degenerate states (in $\log$-scale) for several values of $k$, as functions of $J$. This plot only contains points with $J\equiv 0\mod 3$. In both panels, the values of $\De_\f$ and $\De_\s$ are the same as in figure~\ref{fig:GS_and_excited}.
	\label{fig:S3checks}
	}
\end{figure}

In figure~\ref{fig:S3checks} we present the exact spectra at $J=36,37,38$. These spectra clearly show approximate two-fold degeneracies, with exact degeneracies appearing precisely as predicted above. In the same figure we also show the dependence of the level splittings between approximately-degenerate states as functions of $J$, confirming the exponential decay.

It is interesting to mention that for single-trace operators built out of several fundamental fields, such as
\be
	\tr\p{D_+^{j_1}ZD_+^{j_2}ZD_+^{j_3}Z}
\ee
in $\cN=4$ Super Yang-Mills (SYM) theory, the meaning of the $S_3$ permutation group changes. Only the cyclic $\Z_3$ part arises from the permutations of identical bosons, while the $\Z_2$ transpositions are related to a global charge conjugation symmetry. In this case, both the trivial and the sign representations of $S_3$ appear in the physical spectrum. Therefore, we can expect the spectrum of such operators to be approximately doubly-degenerate, with exponentially small level splittings. In the case of planar $\cN=4$ SYM, however, these degeneracies are exact thanks to the existence of a conserved charge $Q_3$ which is odd under the $\Z_2$---see e.g.~\cite[section~4.2]{Braun:1999te}.

\subsection{Final semiclassical spectra}
\begin{figure}[t]
	\centering
	\includegraphics[scale=0.7]{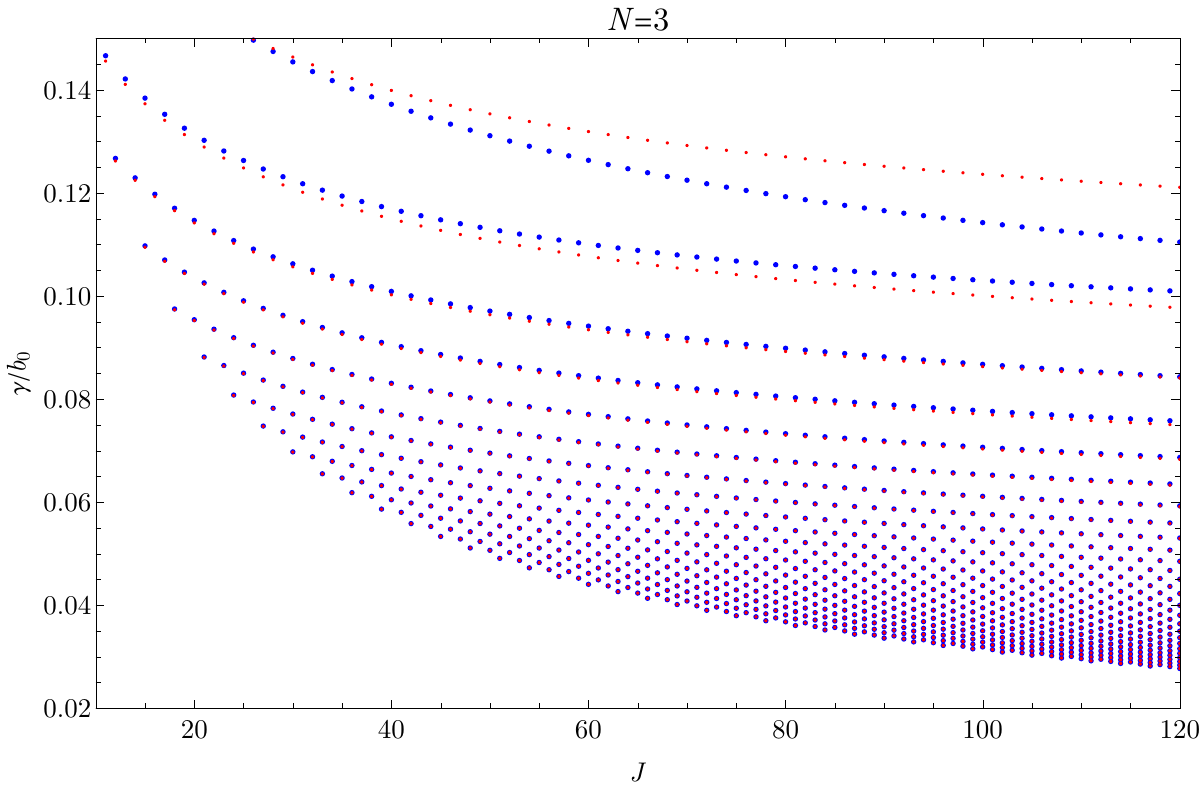}
	\caption{
		A comparison of the exact (blue dots) and the semiclassical (red dots) spectra. The semiclassical spectrum is computed taking into account the $c_1(E)$ correction. The exact spectrum is the same as in figure~\ref{fig:sampleN3}, slightly zoomed in onto the semiclassical region. In this figure, $\De_\f = 1.234,\, \De_\s = 0.6734$.
	}
	\label{fig:N3testRulesemiclassics}
\end{figure}

\begin{figure}[t]
	\centering
	\includegraphics[scale=0.7]{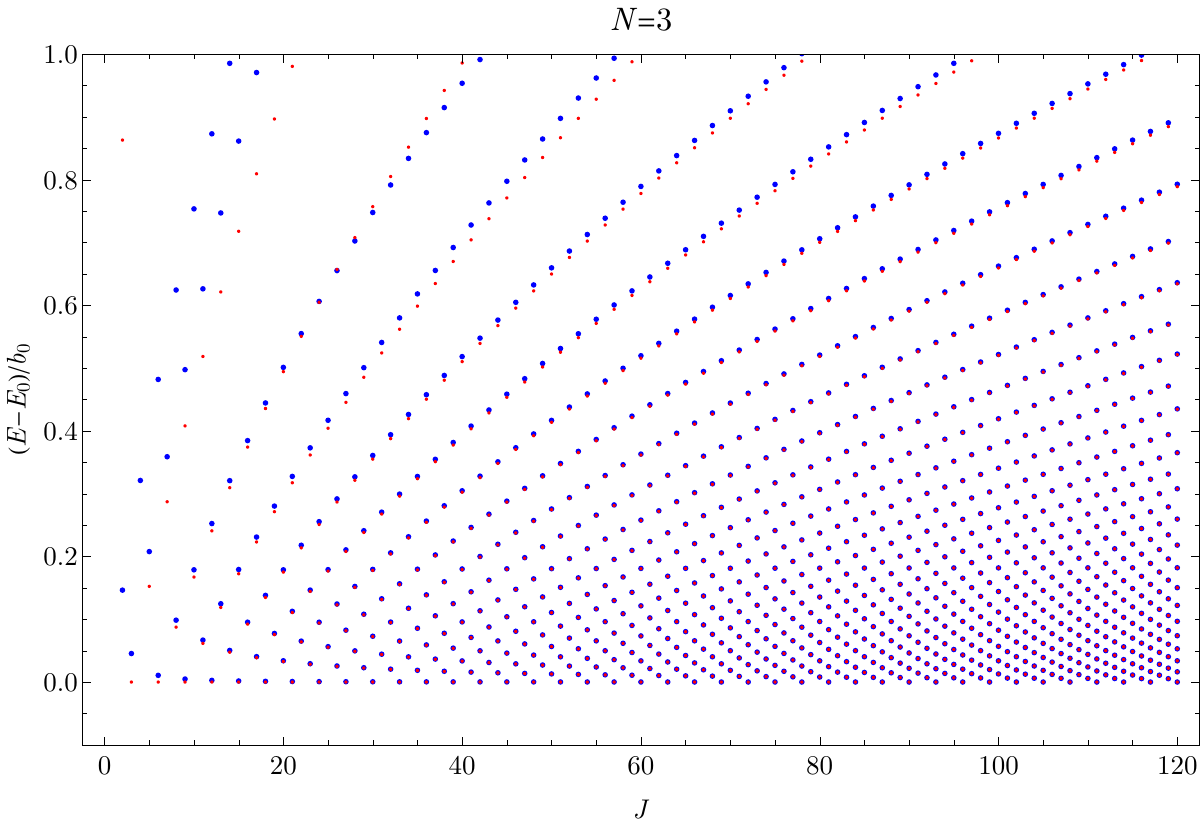}
	\caption{
		Same as figure~\ref{fig:N3testRulesemiclassics}, but shifted by the semiclassical $E_0$ (i.e.\ the energy of the $k=0$ state, which is possibly not $S_3$-invariant) and rescaled by $J^{\De_\s}$, in order to provide a better picture of the low-lying states.
	}
	\label{fig:N3testRulesemiclassicsScaled}
\end{figure}

We are now in a position to finally compare the semiclassical spectra derived in the previous subsections to exact numerical results. We consider two examples. The first is the case $\De_\f=1.234$ and $\De_\s=0.6734$ as in figures~\ref{fig:sampleN3} and~\ref{fig:GS_and_excited}. This comparison is shown in figures~\ref{fig:N3testRulesemiclassics} and~\ref{fig:N3testRulesemiclassicsScaled}. We can see good agreement for individual states, including the correct $J \mod 3$ dependence. Note that if one follows the Regge trajectories (approximately horizontal in this figure), the agreement becomes worse at large $J$. This is due to the large $J$ states along a Regge trajectory becoming hierarchical, with the separation between one pair of particles remaining finite.

An important caveat to keep in mind is that the ground-state energy turns out to be slightly lower than the minimum of $H^{(0)}_\text{symb}$, such that the equal-energy contour $\G^+_E$ cannot be defined. This is due to the negative correction from $H^{(1)}_\text{norm}$. In principle, this problem can be circumvented by rearranging the terms slightly and computing equal-energy contours for a perturbed Hamiltonian. We, however, circumvent it by linearly extrapolating $c_0(E)$ and $c_1(E)$ to energies at $O(J^{-1})$ below the minimum of $H^{(0)}_\text{symb}$.

The second case we consider is $\De_\f=\De_\s=2$, which is appropriate for the three-$\f$ states of $\f^3$ theory in $d=6-\e$ dimensions at one loop (here $\f$ is a real scalar). This case is not holographic, but it can be seen from the explicit description of the one-loop dilatation operator in~\cite[section~2]{Derkachov:1997uh} and \cite[section~4]{Derkachov:2010zza} that our analysis is still applicable. Specifically, the anomalous dimension is given by pair potentials that can be expanded in inverse powers of the two-particle Casimirs. Interestingly, the spectrum for odd $J$ can be determined analytically (see~\cite[section~4.1]{Derkachov:1997uh} or \cite[section~7.1]{Derkachov:2010zza}) and is given by
\be
	E_k = J^2\g_k = \frac{J^{2}}{2p(2p-1)}=\frac{9J^2}{(J-2k)(J+3-2k)},\quad p = \frac{J+3-2k}{6}.\label{eq:exactEk}
\ee
The quantization condition $J+k\equiv 0\mod 3$ on $k$ is the same as in the previous subsection, and $k$ is bounded by the condition $p\geq 1$.\footnote{For $p=1$ the eigenvalue is modified, but this is far from the semiclassical regime we are interested in here. See~\cite{Derkachov:2010zza} for details.} The relevant value of $b_0$ is $b_0= 1$. 

\begin{figure}[t]
	\centering
	\includegraphics[scale=0.7]{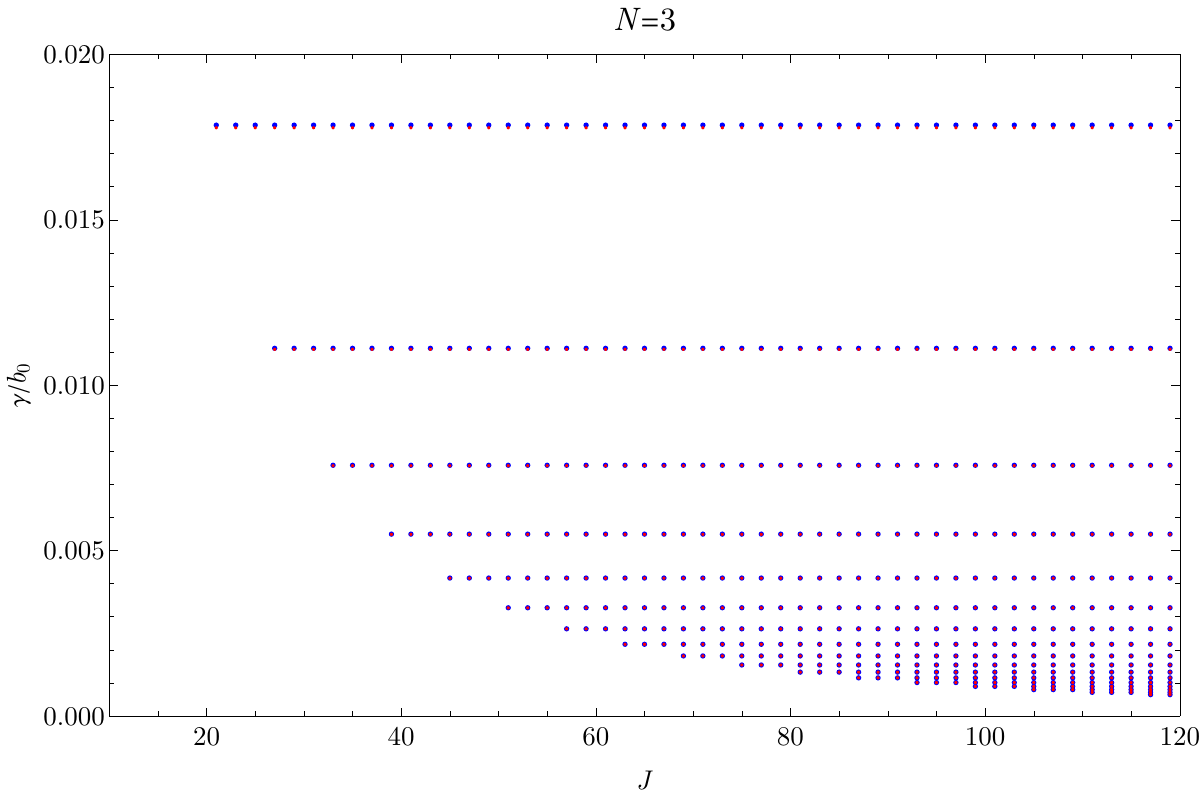}
	\caption{
		A comparison of the exact (blue dots) and the semiclassical (red dots) spectra, for odd $J$. The semiclassical spectrum is computed taking into account the $c_1(E)$ correction. In this figure, $\De_\f = 2,\, \De_\s = 2$.
	}
	\label{fig:N3exactRulesemiclassicsOdd}
\end{figure}

From the knowledge of the exact spectrum we can extract the exact expressions for the functions $c_0(E)$ and $c_1(E)$, which turn out to be
\be
	c_0(E) = J\frac{\pi(\sqrt{E}-3)}{\sqrt{E}},\quad c_1(E) = \frac{5\pi}{2}.\label{eq:c0c1exact}
\ee
We have been able to reproduce these results numerically by evaluating $c_0(E)$ and $c_1(E)$ according to their definitions from the Berezin-Toeplitz quantization. The comparison of the semiclassical and the exact spectra is given in figure~\ref{fig:N3exactRulesemiclassicsOdd} for odd $J$, showing a perfect agreement. 

\begin{figure}[t]
	\centering
	\includegraphics[scale=0.7]{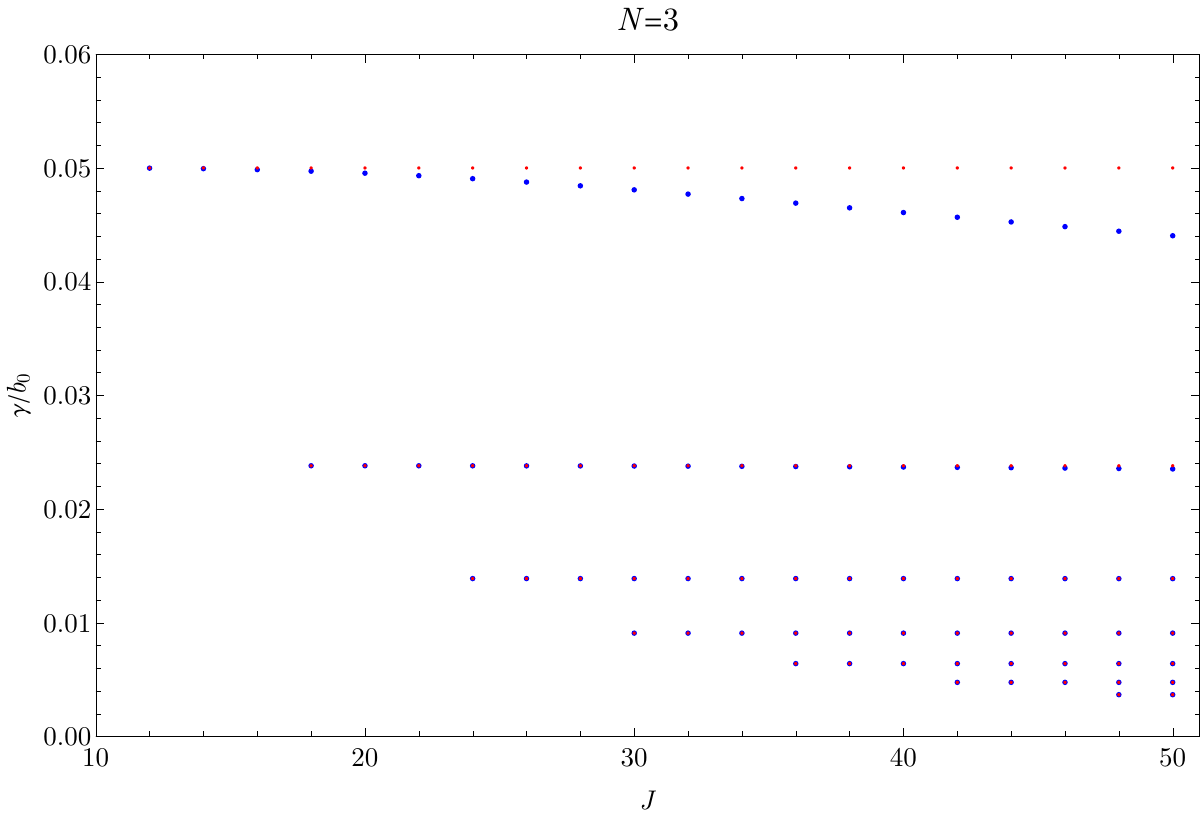}
	\caption{
		A comparison of the exact (blue dots) and the semiclassical (red dots) spectra, for even $J$. The semiclassical spectrum is computed taking into account all $J^{-n}$ corrections, i.e.\ using the resummed expression~\eqref{eq:resummedEk}. In this figure, $\De_\f = 2,\, \De_\s = 2$.
	}
	\label{fig:N3exactRulesemiclassicsEven}
\end{figure}

A somewhat surprising feature of this result is that while~\eqref{eq:c0c1exact} has been derived from the exact odd-spin spectrum, there is nothing in our semiclassical analysis that distinguishes odd and even values of $J$. Therefore, the same $c_0(E)$ and $c_1(E)$ can be used to compute the semiclassical approximation to the even-$J$ spectrum. Even more is true: we can use the exact odd-$J$ spectrum~\eqref{eq:exactEk} to derive the higher-order corrections $c_2(E),\,c_3(E),\,\cdots$ to the Bohr-Sommerfeld condition and use them to determine the even-$J$ spectrum. This of course just means that~\eqref{eq:exactEk} is valid also for even $J$, but now up to non-perturbative errors,
\be
	E_k = J^2\g_k=\frac{9J^2}{(J-2k)(J+3-2k)} + O(J^{-\oo}).\label{eq:resummedEk}
\ee
Here, the error is smaller than any power of $J^{-1}$ as long as we keep $k/J$ fixed and less than $1/6$. In figure~\ref{fig:N3exactRulesemiclassicsEven} we compare this prediction with the exact spectrum, finding perfect agreement. Note that the exact (approximately horizontal) Regge trajectories eventually deviate from~\eqref{eq:resummedEk} at large enough $J$, so that they are able to reproduce the expected double-twist behavior in the large-$J$ limit. The same phenomenon was observed in~\cite{Henriksson:2023cnh} in the case of four-$\f$ states in $\f^4$ theory in $d=4-\e$, see the discussion in their section 4.2 and their figure 11.

Finally, we remark that the function $c_0(E)$, modulo a straightforward normalization of $E$, only depends on the value of $\De_\s$, and in fact only on the twist of $\s$. Therefore, the analytic expression~\eqref{eq:c0c1exact} for $c_0(E)$ is valid in any system where the leading exchange has twist two. This observation may be relevant for four-dimensional CFTs.

\subsection{The lowest-lying states}
\label{sec:small_k}

In general, the full semiclassical analysis of the previous subsections cannot be performed analytically due to the non-trivial shape of the constant-energy contours $\G_E^+$. Nevertheless, it is possible to obtain analytic results in certain limits, in particular when the level number $k$ is much less than the spin $J$. In this case, the contour $\G_E^+$ is approximately circular since $H^{(0)}_\text{symb}$ can be approximated by a quadratic function. Effectively, our quantum system can be approximated by a harmonic oscillator in this limit. In this section, we will derive the small-$k$ energy levels directly from the Bohr-Sommerfeld condition (see~\cite{LeFlochElliptic, LeFlochHyperbolic} for a discussion of the Bohr-Sommerfeld condition near critical points of $H^{(0)}_\text{symb}$). 

Working in $w$-coordinate, we find for the leading symbol
\be\label{eq:quadraticH}
	H^{(0)}_\text{symb}=H^{(0)}_\text{symb}(0)+\ptl_w\ptl_{\bar w}H_\text{symb}^{(0)}(0)w\bar w+\cdots.
\ee
If we define $\de E=E-H^{(0)}_\text{symb}(0)$, then the contour $\G_E^+$ in $w$ coordinate is, to the leading order at small $\de E$, the circle defined by
\be
	|w|^2=\frac{\de E}{\ptl_w\ptl_{\bar w}H_\text{symb}^{(0)}(0)}.
\ee
The symplectic form is $\w=2i dw\wedge d\bar w/(1-w\bar w)^2\approx 2i dw\wedge d\bar w$, and thus we find 
\be
	c_0(E)=\frac{4\pi J\de E}{\ptl_w\ptl_{\bar w}H_\text{symb}^{(0)}(0)}+\cdots.
\ee
The leading Bohr-Sommerfeld rule $c_0(E)=2\pi k$ then implies $\de E\sim k/J$.

The subleading Bohr-Sommerfeld rule is now equivalent to
\be
	\frac{4\pi \de E}{\ptl_w\ptl_{\bar w}H_\text{symb}^{(0)}(0)}+\frac{c_1(E)}{J}=\frac{2\pi}{J}\p{k+\thalf}.
\ee
We are presently interested in deriving $O(k/J)$ and $O(1/J)$ corrections to $E$. Note that $c_1(E)$ is obtained by contour integrals of connection forms over $\G_E^+$. These can be turned into area integrals over $U_E^+$ of curvature forms. If the curvatures are regular, this would imply $c_1(E)=c_{1,1}\de E+\cdots$ for some constant $c_{1,1}$. However, as discussed in section~\ref{sec:subleadingBS}, there is at least one singular curvature contribution, and we instead expect $c_1(E)=c_{1,0}+c_{1,1}\de E+\cdots$ for some constant $c_{1,0}$. The first term contributes to the Bohr-Sommerfeld rule at $O(1/J)$, whereas the second contributes at $O(k/J^2)$ and can be dropped. We therefore find
\be
	\frac{4\pi \de E}{\ptl_w\ptl_{\bar w}H_\text{symb}^{(0)}(0)}+\frac{c_{1,0}}{J}=\frac{2\pi}{J}\p{k+\thalf},
\ee
which gives
\be
	\de E=\frac{\ptl_w\ptl_{\bar w}H_\text{symb}^{(0)}(0)}{2J}(k+\thalf)-\frac{\ptl_w\ptl_{\bar w}H_\text{symb}^{(0)}(0)c_{1,0}}{4\pi J}+\cdots.
\ee

To determine $c_{1,0}$, we can follow the discussion below equation~\eqref{eq:kappa}. Translating it to $w$ coordinate, we find that for very small contours $\G_E^+$,
\be
	c_{1,0}=\oint_{\G_E^+} \kappa=2\pi i\frac{2i H^{(1)}_\text{norm}(0)}{\ptl_w\ptl_{\bar w} H^{(0)}_\text{symb}(0)},
\ee
so that the final result is
\be
	\de E=\frac{\ptl_w\ptl_{\bar w}H_\text{symb}^{(0)}(0)}{2J}(k+\thalf)+\frac{1}{J}H^{(1)}_\text{norm}(0).
\ee
Restoring the error~\cite{LeFlochElliptic}, the result in terms of energy eigenvalues $E_k$ is
\be\label{eq:generalHO}
	E_k=H_\text{norm}(0)+\frac{\ptl_w\ptl_{\bar w}H_\text{symb}^{(0)}(0)}{2J}(k+\thalf)+O(J^{-2}),
\ee
where there error term is valid for fixed $k$. Plugging in explicit expressions, we find
\be
E_k=\frac{3^{1+\De_\s/2}b_0\G(\De_\f+\De_\s/2-1)^2}{\G(\De_\f-1)^2}\p{1+\frac{\De_\s(\De_\s+2)k}{2J}+\frac{\De_\s(\De_\s+4-6\De_\f)}{4J}+O(J^{-2})}.
\ee
We can observe the constant level spacing characteristic of the harmonic oscillator. Note that $k$ is quantized as before: $k+J\equiv 0\mod 3$. 

\paragraph{A general classical Hamiltonian} The general conclusion of~\eqref{eq:generalHO} is that the energy levels are given by the value of the normalized symbol $H_\text{norm}$ at the minimum of the classical Hamiltonian, plus integer-spaced excitations corresponding to the frequency of the quadratic approximation to the classical Hamiltonian. The expression~\eqref{eq:generalHO} relies on the form~\eqref{eq:quadraticH} of the quadratic approximation. Since the second term in~\eqref{eq:generalHO} arises from the computation of the symplectic volume, one can check that for a more general Hamiltonian (i.e. including $w^2$ and $\bar w^2$ terms), it is enough to seek for a (not necessarily holomorphic) coordinate change that puts the classical Hamiltonian into the form~\eqref{eq:quadraticH} while preserving the symplectic form at $w=0$.

Let us go through this logic in more detail. The general form of~\eqref{eq:quadraticH} is
\be\label{eq:generalQuadraticH}
	H^{(0)}_\text{symb}(w) = H^{(0)}_\text{symb}(0)+\Omega(w)+O(w^3), 
\ee
where $\Omega(w)$ is a positive-definite quadratic form on $\C$, $\Omega(w)=\Omega_{ww}w^2+\Omega_{\bar w\bar w}\bar w^2+2\Omega_{w\bar w}w\bar w$. The symplectic form at $w=0$ is $\w|_{w=0} = 2idw\wedge d\bar w$. We now search for a coordinate transformation $u=\a w+\b \bar w$ (commonly known as Bogoliubov transformation) such that
\be
	\Omega(w) &= \Omega_1 u\bar u,\\
	\w|_{w=0}&=idu\wedge d\bar u,
\ee
for some $\Omega_1>0$. It is easy to check that such a transformation always exists and 
\be
	\Omega_1=\sqrt{\Omega_{w\bar w}^2-\Omega_{w w}\Omega_{\bar w\bar w}}.
\ee
The spectrum of $H$ is then given by~\cite{LeFlochElliptic}
\be
	E_k=H_\text{norm}(0)+J^{-1}\Omega_1(k+\thalf)+O(J^{-2}).
\ee

In section~\ref{sec:general_N_HO} we will need the natural generalization of this result to $n$-dimensional phase space with $n=N-2$. Suppose that there exist (not necessarily holomorphic) coordinates $u_a$ near the minimum $P$ of $H^{(0)}_\text{symb}$, such that $P$ is at $u_a=0$ and near $P$
\be\label{eq:Hsymb_to_Omega_a}
	H^{(0)}_\text{symb}=H^{(0)}_\text{symb}(P)+\sum_{a=1}^n \Omega_a u_a\bar u_a+O(u^3),
\ee
while the symplectic form at $P$ is
\be
	\w\vert_P = idu_a\wedge d\bar u_a.
\ee
Then the spectrum is given by
\be
	E_k=H_\text{norm}(0)+J^{-1}\sum_{a=1}^n\Omega_a(k_a+\thalf)+O(J^{-3/2}),
\ee
where $k_a\geq 0$ are independent mode numbers and the error term is valid when $k_a$ are kept fixed. If the harmonic oscillator spectrum is non-degenerate, then the first correction is at $O(J^{-2})$ and $E_k$ admits an expansion in integer powers of $J$. See~\cite{DeleporteHO} for a proof and more detailed statements about the error terms. 

Such coordinates $u_a$ always exist. Indeed, suppose the original coordinates are $x_a$ and let $\tl x = (\Re x, \Im x)$ be the $2n$ real components. Similarly, set $\tl u = (\Re u,\Im u)$. Suppose that the symplectic form in $x$ coordinates already has the form
\be
\w|_P=i dx_a\wedge d\bar x_a=J^{ij}d\tl x_i \wedge d\tl x_j,
\ee
where $J=\begin{pmatrix}0 & I_n\\-I_n & 0\end{pmatrix}$. Furthermore, suppose that the quadratic form $\Omega(x)$ defined from
\be
	H^{(0)}_\text{symb}(x) = H^{(0)}_\text{symb}(0)+\Omega(x)+O(x^3) 
\ee
is given by $\Omega(x) = S^{ij}\tl x_i \tl x_j$ for a symmetric positive-definite matrix $S$.

We can write $\tl x=R\tl u$ for a real symplectic matrix $R$, so that $R^TJR=J$. We then require that $R^T S R=D\oplus D$ with diagonal $D$. The eigenvalues of $D$ are the frequencies $\Omega_a$. Existence of an $R$ satisfying the above conditions is guaranteed by the Williamson theorem~\cite{Williamson}. It is easy to check that $R^{-1}JSR=J(D\oplus D)$, and thus the eigenvalues of $JS$ are $\pm i$ times the eigenvalues of $D$. Therefore, the frequencies $\Omega_a$ can be determined as the positive eigenvalues of $iJS$.

\subsection{Breakdown of semiclassics}
\label{sec:breakdown}

As can be clearly seen from figure~\ref{fig:N3testRulesemiclassics}, not all states in the spectrum are described well by the Bohr-Sommerfeld rule~\eqref{eq:BSsubleading_connected}. In other words, the semiclassical expansion breaks down for sufficiently high level numbers $k$.

Crucially, this breakdown does not come from an intrinsic limitation of the semiclassical analysis of Berezin-Toeplitz quantization. Indeed, if one considers Berezin-Toeplitz quantization of a compact phase space with a smooth and bounded classical Hamiltonian, the entire spectrum can be understood semiclassically. Our setup differs from this more simple scenario, most importantly in that our leading-order Hamiltonian $H^{(0)}_\text{symb}$ is singular---see figure~\ref{fig:potential}.

The exact symbol $H_\text{symb}=J^{\De_\s}\cU_{N,J}$ is smooth and finite. Therefore, the singular behavior of $H^{(0)}_\text{symb}$ means that the large-$J$ expansion of $H_\text{symb}$ breaks down near the singular locus, i.e.\ near the boundary of the acute region. Indeed, one can verify that the effective expansion parameters are the products $R_k J$, where $R_k$ is defined in~\eqref{eq:Rfunction}. A crucial property of the functions $R_k$ is that they vanish at the boundary of the acute region.

As we increase the energy $E$ of an eigenstate, the level set $H_\text{symb}^{(0)}=E$ on which the eigenstate is supported is pushed toward the boundary of the acute region. Near a generic point of the boundary, the classical Hamiltonian diverges as $H_{\mathrm{symb}}^{(0)}\sim R_k^{-\De_\s/2}$.   Therefore, at large $E$ at least one function $R_k$ on the equal-energy slice must scale as $R_k\sim E^{-2/\De_\s}$. The effective expansion parameter becomes $E^{2/\De_\s}J^{-1}$, and we require
\be
	E\ll J^{\De_\s/2}
\ee
in order for the expansion of $H_\text{symb}$ to be valid. Relatedly, at $E\sim J^{\De_\s/2}$ the obtuse region becomes classically-accessible, which is another symptom of our approximations breaking down.

One can estimate that the phase volume outside $H_\text{symb}^{(0)}>E$ scales as $E^{-1/\De_\s}$. This means that in order for our approximations to be valid, we require
\be
	\dim \cH_{3,J}^\text{primary}-k\gg \sqrt{J}.
\ee
Since $\dim \cH_{3,J}^\text{primary}\sim J/6$, we find that the fraction of states to which the semiclassical analysis applies goes to one at large $J$.

\section{$N$-body problem at large spin}
\label{sec:Nbody}

In this section, we will show that the semiclassical description of the $N$-body problem is a generalization of the $N=3$ case: the classical Hamiltonian is a positive-definite function on a phase space given by a K\" ahler manifold. Like in the $N=3$ case, the phase space is disconnected and has an action of the permutation group $S_N$; the largest permutation subgroup that acts on a single connected component is $\Z_N$. In each component, the classical Hamiltonian has a unique minimum located at the fixed point of the $\Z_N$ action, and it diverges at the boundary of the phase space. 

The crucial difference between $N=3$ and $N>3$ is that the classical system now has more than one degree of freedom. As a result, there is no generalization of the Bohr-Sommerfeld conditions that allows for a systematic semiclassical expansion of the spectrum. Nonetheless, we can approximate the total number of states below a given energy, given by the symplectic volume enclosed in the equal-energy slice at leading order. Moreover, just like in the case of $N=3$, the low-lying excited states can be described by a system of harmonic oscillators and a systematic $k/J$ expansion can be constructed.

\subsection{Momentum space representation of minimal twist states}
As we mentioned in section~\ref{sec:linebundles_and_BT}, the construction of the phase space in section~\ref{sec:three-body} does not work for $N>3$. A simple way to see this is to note that the symplectic form is always given by
\be
	\w=i\bar\ptl\ptl \log r^{-2},
\ee
where $r$ as before is the radius of the smallest disk containing all $N$ points $\a_i$. Generically, only three points will be on the boundary of the disk, and therefore (after modding out by $\C^\x\ltimes \C$) the function $r$ only depends on one of the coordinates on $\CP^{N-2}$. This means that the symplectic form generically has rank $1$ and is therefore degenerate almost everywhere on $\CP^{N-2}$ if $N>3$. We expect the rank to grow as additional points approach the boundary of the disk, and therefore the symplectic volume form $\w^{\wedge (N-2)}/(N-2)!$ should be supported in the neighborhood of the concyclic configurations (i.e.\ the configurations where all $N$ points lie on one circle). Relatedly, it is only in this region that the effective potential scales as $J^{-\De_\s}$: if one particle is at a finite distance away from the smallest circle, the potential will scale as $J^{-\De_\s/2}\gg J^{-\De_\s}$, while if two or more particles are away from the smallest circle, the potential will remain $O(1)\gg J^{-\De_s/2}$.

While it might be possible to zoom in on this concyclic region in the limit $J\to \oo$ and obtain the classical phase space in ``position representation'', we found it easier to work in a ``momentum representation'', where the lowering operator $L_+=\ptl_\a$ becomes a multiplication operator. This representation is commonly used in the perturbative CFT literature, see e.g.~\cite{Derkachov:1995zr,Derkachov:1997uh,Braun:1999te,Derkachov:2010zza}. However, note that this representation is not achieved by simply Fourier-transforming the wavefunctions $\psi(\a)$.

Instead, the momentum space wavefunctions $\tilde \psi(z)$ can be defined through the wavefunctions $\psi(\a)$ by the identity
\be
\psi(\alpha)=\tilde{\psi}(\ptl_{\bar \a}) (1-\a \bar \a)^{-\De} \vert_{\bar \a=0}.
\ee
This defines a one-to-one map between polynomials $\psi(\a)$ and $\tilde\psi(z)$. The wavefunctions $\tilde\psi(z)$ have a simple interpretation: in a generalized free theory, the primary operator represented by a wavefunction $\tilde\psi(z_1,\cdots,z_N)$ can be written as~\cite[section~3]{Derkachov:2010zza}
\be
	\cO(x,u) = \tilde\psi(\ptl_{\a_1},\cdots,\ptl_{\a_N}):\!\f(x+\a_1 u)\cdots \f(x+\a_N u)\!:\Big\vert_{\a_1=\cdots=\a_N=0},
\ee
where $u$ is a null polarization vector. Furthermore, when restricted to $\sum_i z_i=0$, the wavefunction $\tilde\psi(z_1,\cdots, z_N)$ coincides with the Fourier transform of the light-ray operators wavefunctions as studied in~\cite{Henriksson:2023cnh,Homrich:2024nwc}. We use the name ``momentum space'' due to these two connections.

In the momentum space, the generators of $\mathfrak{so}(2,1)$ act as
\begin{equation}
	({L}_-\tilde{\psi})(z) = z \tilde{\psi}(z), \quad ({L}_0\tilde{\psi})(z) = (z\ptl_z+\De_\phi/2)\tilde{\psi}(z),\quad ({L}_+\tilde{\psi})(z) = (z\ptl_z+\De_\phi)\ptl_z\tilde{\psi}(z).
	\label{eq:Lgen_momspace}
\end{equation}
The lowest weight vector is again a constant function, as it satisfies
\begin{equation}
	{L}_+\tilde{\psi}  = 0,\quad {L}_0 \tilde{\psi} = \frac{\De_\phi}{2} \tilde{\psi} \Longrightarrow \tilde{\psi} = \mathrm{const},
\end{equation}
while the repeated action of $L_-$ generates arbitrary holomorphic functions of $z$. For this representation, it was shown in e.g.~\cite{Derkachov:1997pf,Derkachov:2010zza} that the inner product can be computed as
\begin{equation}\label{eq:SP_momspace_diff}
	\<\psi_1 |\psi_2\>=\<\tilde{\psi}_1 |\tilde{\psi}_2\> = \tilde{\psi}_2(\ptl_{\bar\a}) \bar{\tilde{\psi}_1(\a)}|_{\a=0}.
\end{equation}

To summarize, in the momentum space, the elements of ${\mathcal{H}}_{N,J}$ are represented by homogeneous degree-$J$ polynomials $\tilde\psi(z_1,\cdots,z_N)$, while the primary states in $\cH^\text{primary}_{N,J}$ satisfy additionally the constraint
\begin{equation}
	({L}_+\tilde{\psi})(z_1,\dots,z_N) =0.
	\label{eq:primary_Npart_momspace}
\end{equation}
Note that in the momentum space this constraint is difficult to solve explicitly for more than two particles since $L_+$ acts as a second-order differential operator. In the next sections, we avoid this technicality by exploiting the hermiticity condition $L_+^\dagger = L_-$ which acts as multiplication by $z$. 

\subsection{$N$-body problem in momentum space}
The common definition of the scalar product in~\eqref{eq:SP_momspace_diff} is ill-suited for the large-spin expansion of the $N$-body problem in ${\mathcal{H}}_{N,J}$. For this reason, we introduce an alternative integral representation of the scalar product in momentum space, which takes the form
\begin{equation}
	\<\tilde{\psi}_1 |\tilde{\psi}_2\> = \frac{2}{\pi\, \Gamma(\De)} \int_{\C} d^2z\, \mathcal{K}_{1-\De}(|z|^2) \, \bar{\tilde{\psi}_1(z)}\,\tilde{\psi}_2(z), \quad \mathcal{K}_\nu(x)  := \frac{1}{2}\int_0^\infty \frac{dt\, e^{-(t+t^{-1}x)}}{t^{1+\nu}}.
	\label{eq:SP_momspace_int}
\end{equation}
The function $\mathcal{K}_\nu$ in the measure is called the modified Bessel-Clifford function, and is related to the modified Bessel function $K_\nu$ of the second kind via $\mathcal{K}_\nu(x) =x^{-\nu/2} K_{\nu}(2\sqrt{x})$. The measure is normalized such that the lowest-weight vector $\tilde{\psi}=1$ has unit norm. This formulation of the scalar product can be seen as the local operator analogue of the scalar product in \cite[eq.~(3.86)]{Henriksson:2023cnh} for the Fourier transform of wavefunctions of light ray operators.

To complete the momentum space formulation of the $N$-body problem, we need a realization of the two-particle anomalous dimension as an explicit operator in momentum space. The simplest way to do this is to express it as a function of two-particle Casimir operators $L_{ij}$, defined in~\eqref{eq:casimir_def}. The existence of such an expression follows from the representation theory of $\mathfrak{so}(2,1)$ discussed in section~\ref{sec:diagonalization}. Concretely, if the two-particle anomalous dimension on $\cH_{2,J}$ is of the form
\begin{equation}
	\gamma_{2,J} = U_{\mathrm{Cas}}\left((\De_\phi+J)(\De_\phi+J-1)\right)
\end{equation}
for some function $U_\text{Cas}$, then the Hamiltonian on $\cH_{N,J}$ can be expressed as
\begin{equation}
	H_N(\a,\ptl_\a) = \sum_{1\leq i<j\leq N}U_{\mathrm{Cas}}\left(L_{ij}(\a,\ptl_\a)\right),
\end{equation}
where $L_{ij}(\a,\ptl_\a)$ is the quadratic Casimir for the action~\eqref{eq:Laction} of $\mathfrak{so}(2,1)$ on the particles $i,j$ in position space. We can then go to momentum space using the corresponding realization of the $\mathfrak{so}(2,1)$ generators in~\eqref{eq:Lgen_momspace}. The resulting momentum space Casimir is a second-order differential operator with the following normal-ordered form:
\begin{equation}
	{L}_{ij}(z_i,z_j,\ptl_i,\ptl_j) = -z_iz_j\ptl_{ij}^2+\De_\phi z_{ij}\ptl_{ij}+\De_\phi(\De_\phi-1),
	\label{eq:cas_momspace}
\end{equation}
where $z_{ij}:=z_i-z_j$ and $\ptl_{ij}:=\ptl_i-\ptl_j$. As a result, we can express the Hamiltonian in the momentum space representation as
\begin{equation}
	{H}_N(z,\ptl) = \sum_{1\leq i<j\leq N}U_{\mathrm{Cas}}\left({L}_{ij}(z,\ptl_z)\right).
	\label{eq:ham_momspace_pdo}
\end{equation}
Here, note that we are restricting ourselves to Hamiltonians that are sums of two-particle potentials. We leave the more general case for future work.

\subsection{Reduction to $\CP^{N-2}$}
In the previous section, we defined the momentum space $N$-body problem in terms of the scalar product~\eqref{eq:SP_momspace_int} and the anomalous dimension operator~\eqref{eq:ham_momspace_pdo}. After restriction to the subspace $\cH_{N,J}^{\mathrm{primary}}$, we will now show that they admit the same formulation as section~\ref{sec:linebundles_and_BT}, that is to say a scalar product and a Toeplitz operator on $\CP^{N-2}$.

First, let us explain how primary wavefunctions in momentum space correspond to sections of the holomorphic line bundle $\cO(J)$ on $\CP^{N-2}$, equipped with a scalar product of the form
\begin{equation}
\<\tilde{\psi}_1|\tilde{\psi}_2\> = \int_{\CP^{N-2}} d\tilde{\mu}_J\, \tilde{h}_J(\tpsi_1,\tpsi_2),
\label{eq:SP_CP_momspace}
\end{equation}
To see how this arises from the scalar product~\eqref{eq:SP_momspace_int} on $\C^N$, recall the hermiticity condition 
\begin{equation}
	\< \tilde{\psi}_1 | {L}_{+} \tilde{\psi}_2\>= \< L_-\tilde{\psi}_1 | \tilde{\psi}_2\>=\< (z_1+\dots+z_N) \tpsi_1|\tilde{\psi}_2\> = 0.
\end{equation}
From this relation, we expect that the integral over $\C^N$ can be reduced to the hypersurface $z_1+\dots+z_N=0$. Viewing $(z_1,\dots,z_N)$ as projective coordinates of $\CP^{N-1}$, the hypersurface then defines the domain of integration in the scalar product~\eqref{eq:SP_CP_momspace}. 

Let us prove this statement. We begin by defining
\be
F(\b,\bar \b)=\<e^{-i\b L_-}\tilde\psi_1  | e^{i\b L_-}\tilde\psi_2\>, \quad \b \in \C.
\ee
Using the $\mathfrak{so}(2,1)$ commutation relations and the primary constraint for $\tilde\psi_{1,2}$, we can find the following system of two first-order differential equations:
\be
\ptl_\b F = -\bar\b(2\bar h + \bar\b\ptl_{\bar \b}) F,\quad \ptl_{\bar \b} F = -\b(2\bar h + \b\ptl_{ \b}) , \quad \bar h = N \frac{\De_\phi}{2} +J.
\ee
Given the initial condition $F(0,0)=\<\tilde \psi_1 | \tilde\psi_2\>$, the system is solved by
\be
\<e^{-i\b L_-}\tilde\psi_1 | e^{i\b L_-}\tilde\psi_2\>=F(\b,\bar \b)=(1-\b\bar\b)^{-2\bar h}\<\tilde \psi_1|\tilde\psi_2\>.
\ee
Integrating both sides over $d^2\b$ then yields 
\begin{equation}\label{eq:localized_inner}
	\<\tilde{\psi}_1 |\tilde{\psi}_2\>=C_{\tilde{M}}(\De_\phi,J)\int_{\C^N}d^{2N}z \prod_{k=1}^N\mathcal{K}_{1-\De_\phi}(|z_k|^2) \,\delta^{(2)}\left(\sum_{k=1}^N z_k\right)\bar{\tilde{\psi}_1(z)}\tilde{\psi}_2(z),
\end{equation}
where $C_{\tilde{M}}(\De,J)=2^N \pi^{-(N+1)}\Gamma(\De)^{-N} (N\De+2J-1)$ is an overall multiplicative constant that will not enter any further calculations.

The final step towards~\eqref{eq:SP_CP_momspace} is integrating over the orbits of $\C^\x$ given by complex rescalings $z\rightarrow \l z$. This procedure is simpler than that of section~\ref{sec:dmu_h1_cU} because $\C^\x$ is unimodular, with a unique left- and right-invariant Haar measure given by $d^2\l/|\l|^2$. Using this measure and homogeneity of $\tpsi_{1,2}(z)$, we can recast the integral~\eqref{eq:localized_inner} into
\begin{align}
&\<\tilde{\psi}_1 |\tilde{\psi}_2\> =C_{\tilde{M}}(\De_\phi,J) \int_{\C^N} \frac{d^{2N} z}{\vol(\C^\x)} \de^{(2)}\left(\sum_{k=1}^N z_k\right) \tilde{M}_J(z) \bar{\tilde{\psi}_1( z)} \tilde{\psi}_2(z), \label{eq:SP_volCx_momspace} \\
& \tilde{M}_J(z) = \int_0^\infty \frac{d |\l |}{|\l|^{2(N+J)-1}}\prod_{k=1}^N\mathcal{K}_{1-\De_\phi}(|\l|^{-2} |z_k|^2).
\label{eq:MJ_momspace}
\end{align}

The function $\tilde{M}_J$, which is manifestly homogeneous of degree $2-2N-2J$ in $|z|$, ensures that the integrand in~\eqref{eq:SP_volCx_momspace} is invariant under $\C^\x$. One can obtain explicit expressions of the form~\eqref{eq:SP_CP_momspace} by gauge-fixing the $\C^\x$ action. Comparing~\eqref{eq:SP_volCx_momspace} with~\eqref{eq:SP_CP_momspace}, we find
\begin{equation}
d\tilde{\mu}_J \tilde{h}_J(\psi_1,\tpsi_2) = C_{\tilde{M}} \frac{d^{2N}z}{\vol\C^\x} \tilde{M}_J(z) \bar{\tpsi_1(z)}\tpsi_2(z).
\label{eq:MJ_to_muh_momspace}
\end{equation}

Having now reduced primary wavefunctions and their scalar product to $\CP^{N-2}$, the second and final task of this section is rewriting $H_N(z,\ptl)$ in~\eqref{eq:ham_momspace_pdo} as a Toeplitz operator. Intuitively, we can obtain such a formulation via integration by parts in matrix elements. Indeed, consider the matrix element $\<\tpsi_1 | H_N(z,\ptl) \tpsi_2\>$ written in terms of the integral representation~\eqref{eq:SP_volCx_momspace}. Since the Casimir operators are holomorphic and commute with $L_-=z_1+\dots+z_N$, we can express the latter as
\begin{equation}
\<\tpsi_1 | H_N(z,\ptl) \tpsi_2\> = \frac{C_{\tilde{M}}(\De_\phi,J)}{\vol(\C^\x)} \int_{\C^N} d^{2N} z\,  \tilde{M}_J(z) H_N(z,\ptl) \left\{ \de^{(2)}\left(\sum_{k=1}^N z_k\right) \bar{\tilde{\psi}_1(z)} \tilde{\psi}_2(z)\right\}.
\end{equation}
If $H_N(z,\ptl)$ were a differential operator, then we could integrate by parts to have its transpose ${}^t H_N(z,\ptl)$ act on $\tilde{M}_J(z)$, assuming the boundary terms vanish. For a normal ordered differential operator like $L_{ij}(z,\ptl)$ in~\eqref{eq:cas_momspace}, the transpose amounts to the composition of anti-normal ordering with $\ptl\rightarrow -\ptl$. The transpose of the operator~\eqref{eq:ham_momspace_pdo}, as a function of the Casimirs~\eqref{eq:cas_momspace}, is then given by
\begin{equation}
	{}^t H_N = \sum_{i<j} U_{\mathrm{Cas}}\left({}^tL_{ij}\right),\quad {}^tL_{ij}(z,\ptl_z) = -\ptl_{z_{ij}}^2z_iz_j-\De_\phi \ptl_{z_{ij}}z_{ij}+\De_\phi(\De_\phi-1),
	\label{eq:transpose_ham_momspace}
\end{equation} 
where $z_{ij}=z_i-z_j$ and $\ptl_{z_{ij}}=\ptl_{z_i}-\ptl_{z_j}$. Since $U_{\mathrm{Cas}}(L)$ is not a polynomial, the operator $H_N(z,\ptl_z)$ is not a differential operator. However, the large-$L$ expansion of $U_{\mathrm{Cas}}(L)$ with exponents in $-\De_\s/2+\N$ makes it a pseudodifferential operator in the sense of H\"ormander \cite{HormanderPDO}. For this class of operators, the integration by parts of $H(z,\ptl)$ does generalize to the action of~\eqref{eq:transpose_ham_momspace} on $\tilde{M}_J$, leading to the equality of matrix elements
\begin{equation}
\<\tpsi_1 | H_N(z,\ptl_z)\tpsi_2\>=\<\tpsi_1| \tilde{\cU}_{N,J}(z) \tpsi_2\>,\quad 	\tilde{\cU}_{N,J} (z) \tilde{M}_J(z) = {}^t\tilde{H}_N(z,\ptl_z) \tilde{M}_J(z),
	\label{eq:cU_momspace}
\end{equation}
for any $\tpsi_{1,2}\in \cH_{N,J}$. In conclusion, $H_N(z,\ptl_z)$ is a Toeplitz operator with the symbol $\tilde{\cU}_{N,J}$.

\subsection{Large-spin expansion and pseudodifferential operators}
\label{sec:generalNsymbols}
Having reduced the $N$-body Hamiltonian in momentum space to a Toeplitz operator for a function on $\CP^{N-2}$, we can now study the large-spin limit as a semiclassical expansion in the framework of Berezin-Toeplitz quantization, introduced in section~\ref{sec:BT_general}. To this end, we compute the large-spin expansion of the measure $d\tilde{\mu}_J$, the Hermitian form $\tilde{h}_J$, and the symbol $\tilde{\cU}_{N,J}$ of the Toeplitz operator.

Let's start with $\tilde{M}_J$, defined by the integral~\eqref{eq:MJ_momspace}, which determines the measure and Hermitian form via~\eqref{eq:MJ_to_muh_momspace}. At large spin, the factor of $|\l|^{-2J}$ ensures that the integral is dominated by the small $\l$ region of the integrand, which corresponds to the large-argument limit of the Bessel-Clifford functions:
\begin{equation}
	\mathcal{K}_{1-\De_\phi}(x^2) = \frac{\sqrt{\pi}}{2} x^{\De_\phi-\frac{3}{2}} e^{-2 x} \left(1+O(x^{-1})\right).
\end{equation}
Plugging  this expansion into~\eqref{eq:MJ_momspace} yields a Gamma function integral in $|\l|$, from which we obtain the leading large-spin expansion of $\tilde{M}_J$:
\begin{equation}
	\tilde{M}_J(z) =C'_{\tilde{M}}(\De_\phi,J) \left(\sum_{k=1}^N |z_k|\right)^{2(1-N-J)} \prod_{k=1}^N \left(\frac{|z_k|}{\sum_{\ell=1}^N |z_\ell|}\right)^{\De_\phi-\frac{3}{2}}\left(1+O(J^{-1})\right),
	\label{eq:MJ_momspace_LS}
\end{equation}
where $C'_{\tilde{M}}(\De,J) = \pi^{N/2} 2^{2-(\De+3/2)N-2J} \Gamma(2J+N(\De+1/2)-2) C_{\tilde{M}}(\De,J)$ is another multiplicative constant that will not matter for physical observables. Higher-order corrections can also be systematically computed from the large-argument expansion of the Bessel function, but we will not need them in this work. 

Based on~\eqref{eq:MJ_momspace_LS} we define
\begin{equation}
\tilde{h}_J(\tpsi_1,\tpsi_2) = e^{-J\cK} \bar{\tpsi_1} \tpsi_2,\quad \cK(z):= 2\log\sum_{k=1}^N |z_k|.
\label{eq:hermform_momspace}
\end{equation}
This is chosen so that $e^{J\cK}\tilde{M}_J$ does not have factors exponential in $J$. Note that $\tilde{h}_J=\tilde{h}_1^{\otimes J}$ and, as discussed in section~\ref{sec:three-body}, the classical symplectic form in Berezin-Toeplitz quantization is identified with the curvature of then Chern connection of $\tilde{h}_{1}$. In other words, $\cK$ can be viewed as the K\"ahler potential on the phase space,\footnote{Strictly speaking, $\cK$ is not a function on $\CP^{N-2}$ because it is not homogeneous in $z_k$. However, if we choose a local holomorphic embedding of $\vf:\CP^{N-2}\to \C^N$, the pullback $\vf^*\cK$ defines a function locally on $\CP^{N-2}$. It is easy to check that if $\chi$ is a different holomorphic embedding, then $\chi^*\cK=\vf^*\cK+f+\bar f$, where $f$ is a holomorphic function. Therefore, $\ptl\bar\ptl \cK$ is independent of the embedding.} and the symplectic form is given by
\be
	\w = i\ptl \bar\ptl \cK.
\ee
The large-spin expansion of $d\tilde{\mu}_J$ is then determined from~\eqref{eq:MJ_to_muh_momspace} and~\eqref{eq:MJ_momspace_LS}:
\begin{align}
 &d\tilde{\mu}_J = C'_{\tilde{M}}(\De_\phi,J) \frac{d^{2N} z}{\vol \C^\x} \left(\sum_{k=1}^N |z_k|\right)^{2(1-N)} \prod_{k=1}^N \left(\frac{|z_k|}{\sum_{\ell=1}^N |z_\ell|}\right)^{\De_\phi-\frac{3}{2}} \left(1+O(J^{-1})\right).  \label{eq:meas_momspace}
\end{align}

We would now like to compute the large-spin expansion of $\tilde{\cU}_{N,J}(z)$ in~\eqref{eq:cU_momspace}. Note that $\tilde{M}_J=\mathrm{const} \times e^{-J\cK}f$, where $f$ has an expansion in power of $1/J$. Concretely,
\be
	f(z)=\left(\sum_{k=1}^N |z_k|\right)^{2(1-N)} \prod_{k=1}^N \left(\frac{|z_k|}{\sum_{\ell=1}^N |z_\ell|}\right)^{\De_\phi-\frac{3}{2}} \left(1+O(J^{-1})\right).
\ee
We can then rewrite~\eqref{eq:cU_momspace} as
\begin{equation}
	\tilde{\cU}_{N,J}(z) f(z) = \sum_{i<j}U_{\mathrm{Cas}}\left({}^t L_{ij}(z,\ptl_z -J \ptl_z \cK)\right) f(z),
\end{equation}
where have effectively conjugated by $\exp(-J\cK)$. At large $J$, each two-particle Casimir will scale like $J^2$, such that we can replace $U_{\mathrm{Cas}}$ by its large-argument expansion. In general, the leading term is of the form 
\begin{equation}
	U_{\mathrm{Cas}}(L) = b_{\mathrm{Cas},0}\,L^{-\De_\s/2} \left(1+O(L^{-1})\right).
\end{equation}
For the anomalous dimension of the AdS toy model, we have specifically $b_{\mathrm{Cas},0}=b_0\, \Gamma(\De_\phi+\De_\s/2-1)^2 \Gamma(\De_\phi-1)^{-2}$, where $b_0$ is given by~\eqref{eq:b0defn}.

Our computation therefore reduces to the large-spin expansion of ${}^t L_{ij}(z,\ptl_z-J\ptl_z\cK)^{-\De_\s/2} f(z)$. This expansion is well-defined on any domain of $\CP^{N-2}$ where ${}^t L_{ij}(z,-J\ptl_z \cK)>0$. Indeed, since ${}^t L_{ij}$ is a second-order differential operator, its expansion takes the form
\begin{equation}
	{}^t L_{ij}(z,\ptl_z-J\ptl_z\cK) = J^2 a_0(z,\ptl_z\cK) + J \cD_1(z,\ptl_z\cK,\ptl_z^2\cK,\ptl_z)+\cD_2(z,\ptl_z),
\end{equation}
for some function $a_0$ and some first- and second-order differential operators $\cD_1$ and $\cD_2$, respectively. In this case, the action of ${}^t L_{ij}(z,\ptl_z-J\ptl_z\cK)^{-\De_\s/2} $ on $f$ to subleading order is
\begin{equation}
	{}^t L_{ij}(z,\ptl_z-J\ptl_z\cK)^{-\De_\s/2}  f = J^{-\De_\s}\left( a_0^{-\De_\s/2} -\tfrac{\De_\s}{2} J^{-1} a_0^{-\De_\s/2-1} \cD_1 +O(J^{-2})\right)f.
\end{equation}

Using this strategy, we obtain the large-spin expansion of $\tilde{\cU}_{N,J}$ to subleading order (the precision of~\eqref{eq:MJ_momspace_LS} is enough for this). Defining the symbol $H_{\mathrm{symb}}:= J^{\De_\s} \tilde{\cU}_{N,J}$, we find
\be
H_{\mathrm{symb}} &=H^{(0)}_\text{symb} +J^{-1}H_{\mathrm{symb}}^{(1)}+O(J^{-2}), \label{eq:Hsymb_expansion_momspace}\\
H^{(0)}_\text{symb}&= b_{\mathrm{Cas},0}\sum_{1\leq i<j\leq N} \left( \frac{2 |z_iz_j|-z_i\bar z_j-\bar z_i z_j}{(|z_1|+\dots+|z_N|)^2}\right)^{-\frac{\De_\s}{2}}.
\label{eq:Hsymb_momspace}
\ee
We omit the explicit expression for the subleading term $H_{\mathrm{symb}}^{(1)}$, which will only enter the calculation of the ground state energy. While we will not use them in this work, higher orders can be systematically computed using the calculus of pseudodifferential operators, summarized in appendix~\ref{app:pseudodiff}.

In conclusion, we have assembled all the ingredients to apply semiclassical methods in Berezin-Toeplitz quantization: the Hermitian form~\eqref{eq:hermform_momspace}, the measure~\eqref{eq:meas_momspace}, and the symbol~\eqref{eq:Hsymb_expansion_momspace} of the Hamiltonian. In particular, the classical limit is specified by the K\"ahler potential $\cK$ and the Hamiltonian $H_{\mathrm{symb}}^{(0)}$. 

\subsection{Geometry of the classical phase space}

In the previous sections we derived a phase space of complex dimension $N-2$ given by the hypersurface $z_1+\dots+z_N=0$ in $\CP^{N-1}$, with K\"ahler potential $\cK=2\log( |z_1|+\dots+|z_N|)$. We will denote it by $\cM$. A useful approach to this K\"ahler manifold is to view it as a quotient of the projective null cone
\be
	\cC = \{W\in \CP^{N-1}|W^2=0\},
\ee
where $W$ is the vector of homogeneous coordinates on $\CP^{N-1}$. We have
\be
	\cM = \cC/(\Z_2^N/\Z_2^\text{diag}),
\ee
where $(s_1,\cdots, s_N)\in \Z_2^N$ acts on $\cC$ by $W_k\mapsto s_k W_k$. The element $(-1,-1,\cdots,-1)\in \Z_2^N$ acts trivially on $\cC$ since $W\sim -W$, hence the quotient by the diagonal $\Z_2^\text{diag}$.

The manifold $\cC$ is a K\"ahler manifold equipped with the K\"ahler potential $\cK=2\log ||W||^2$ which is $4\x$ the standard  K\"ahler potential $\half \log ||W||^2$ induced from $\CP^{N-1}$. In fact, $\cC$ is isomorphic to the real Grassmannian of oriented two-planes in $\R^N$, which is a Hermitian symmetric space associated to $\SO(N)$:
\begin{equation}
	\cC\cong \mathrm{Gr}(2,N) \cong \frac{\SO(N)}{\SO(2)\times \SO(N-2)} \cong \left\{ W\in \C^N \,|\, W\sim \l W,\, W^2=0\right\}.
	\label{eq:grass_to_SO}
\end{equation}
To see this, set $e_1=\Re W$, $e_2=\Im W\in \R^N$ and note that $W^2=0$ is equivalent to $|e_1|=|e_2|$ and $e_1\.e_2=0$. By positive rescalings of $W$ we can set $|e_1|=|e_2|=1$, such that the pair $(e_1,e_2)$ forms and oriented orthonormal basis of a two-plane in $\R^N$. After modding out by the remaining phase rotations of $W$, we lose the information about the choice of basis and only the two-plane with its orientation remain. Note also that the K\"ahler potential $\cK$ is invariant under the action of $\SO(N)$.

To summarize, we now view the phase space $\cM$ as $\mathrm{Gr}(2,N)/(\Z^{N}_2/\Z_2^\text{diag})$.

We will say that $z\in \cM$ is regular if all its components $z_k$ are non-zero and have distinct phases; almost all elements of $\cM$ are regular. We say that $W\in \mathrm{Gr}(2,N)$ is regular if it projects to a regular element in $\cM$. Equivalently, if all $W_k$ are non-zero and none are a real multiple of another.  Each regular $z\in \cM$ defines a cyclic ordering of integers $\{1,\cdots,N\}$ which is obtained by reading $z_k/|z_k|$ counterclockwise on the unit circle. We denote by $\cM_{(a\cdots b)}$ the open set of regular elements with the cyclic ordering $(a\cdots b)$.

We claim that every $\cM_{(a\cdots b)}$ is isomorphic to the positive Grassmannian $\mathrm{Gr}_+(2,N)$. The positive Grassmannian $\mathrm{Gr}_+(2,N)$ is the subset of $\mathrm{Gr}(2,N)$ on which the Pl\"ucker coordinates $\Im (W_j \bar W_i)$ satisfy 
\be\label{eq:positiveGr}
\Im (W_j\bar W_i)>0,\quad \forall i,j\text{ such that }i<j.
\ee
Note that this condition is invariant under complex rescalings of $W$ and that $\mathrm{Gr}_+(2,N)$ contains only regular elements.

It is enough to prove our claim that $\cM_{(a\cdots b)}\simeq \mathrm{Gr}_+(2,N)$ for the standard cyclic ordering $(1\cdots N)$ since every $\cM_{(a\cdots b)}$ can be related to $\cM_{(1\cdots N)}$ by a $S_N$ permutation of $z_k$'s.

Let $\pi:\mathrm{Gr}(2,N)\to \cM$ be the canonical projection. We want to show that for every $z\in \cM_{(1\cdots N)}$, $\pi^{-1}(z)$ contains precisely one point from $\mathrm{Gr}_+(2,N)$, and that $\pi(\mathrm{Gr}_+(2,N))=\cM_{(1\cdots N)}$. First, given a $z\in \cM_{(1\cdots N)}$ we can assume that $z_1=1$ and define $W_k=\sqrt{z_k}$, where we place the branch cut of the square root just below the real axis. Note that $W\in \pi^{-1}(z)$. It is easy to check that $W_k=r_k \,e^{i\f_k}$ with $r_k>0$ and $\f_k\in [0,\pi)$ strictly increasing, which immediately implies~\eqref{eq:positiveGr} and thus $W\in \mathrm{Gr}_+(2,N)$. All other $W'\in \pi^{-1}(z)$ are obtained from $W$ by non-trivial elements $(s_1,\cdots,s_N)\in \Z_2^N/\Z_2^\text{diag}$. If the element $(s_1,\cdots,s_N)$ is non-trival, then there is a pair $s_i,s_j$ such that $s_is_j=-1$, which means $\Im(W'_j\bar W'_i)=-\Im(W_j\bar W_i)<0$ and so $W'\not\in \mathrm{Gr}_+(2,N)$.

Second, if $W\in \mathrm{Gr}_+(2,N)$ then (assuming, as we can, $W_1=1$) equation~\eqref{eq:positiveGr} implies that $W_k=r_k\, e^{i\f_k}$ with $r_k>0$ and $\f_k\in [0,\pi)$ strictly increasing. This implies that $z_k=W_k^2$ is a regular element with the cyclic ordering $(1\cdots N)$, i.e.\ $z\in \cM_{(1\cdots N)}$ as required. This finishes the proof of our claim.

As we will soon see, the classical Hamiltonian blows up near the boundaries of the sets $\cM_{(a\cdots b)}$. Therefore, only the set $\cM^\text{reg}$ of regular elements is classically accessible . Note that $\cM^{\text{reg}}$ is the disjoint union of the sets $\cM_{(a\cdots b)}$ for all cyclic orderings $(a\cdots b)$. The permutation group $S_N$ generally acts by permuting the connected components $\cM_{(a\cdots b)}$. However, some permutations can preserve a cycling ordering $(a\cdots b)$. In particular the group $\Z_N$ of cyclic permutations of $(1\cdots N)$ preserves $\cM_{(1\cdots N)}$ and has a unique fixed point $z\in\cM_{(1\cdots N)}$ where $z_k=e^{2i\pi k/N}$. Equivalently, the fixed point is at $W\in \mathrm{Gr}_+(2,N)$ with $W=e^{i\pi k/N}$.

Let us consider the example $N=3$ in more detail. In this case we obtain a construction of the $N=3$ classical phase space that is different from the one described in section~\ref{sec:LO_semiclass}. Let us first consider the manifold $\mathrm{Gr}(2,3)$. We can parameterize it as\footnote{The map between $W\in \mathrm{Gr}(2,3)$ and $w\in \CP^1$ is one-to-one. Indeed, from $W_1,W_2$ we can determine $w_1^2$ and $w_2^2$. Out of $2^2$ choices of signs for $w_1,w_2$, only two are consistent with $W_3$, and these two are related by $w\to -w\sim w$.}
\be
	W_1=w_1^2-w_2^2,\quad W_2=iw_1^2+iw_2^2,\quad W_3=2w_1 w_2,
\ee
which explicitly solves the constraint $W^2=0$ and establishes the isomorphism of $\mathrm{Gr}(2,3)$ and $\CP^1$ with homogeneous coordinates $w$. The K\"ahler potential becomes
\be
	\cK=4\log \|w\|^2+\text{const},
\ee
which is 8$\x$ the standard $\CP^{1}$ K\"ahler potential $\half\log\|w\|^2$ that defines the Fubini-Study metric. The symplectic volume of $\mathrm{Gr}(2,3)$ is therefore $\vol \mathrm{Gr}(2,3)=8\pi$, and the volume of $\cM$ is $\vol\cM=2^{-2}\vol \mathrm{Gr}(2,3)=2\pi$, in agreement with section~\ref{sec:LO_semiclass}.

A somewhat surprising point is that the phase space $\cM$ constructed here has positive curvature, whereas the phase space in section~\ref{sec:three-body} is negatively-curved.  However, we do not have any reason to expect that the Hermitian or complex structures of the two spaces should agree, and by Darboux's theorem there are no local symplectic invariants. Indeed, an explicit symplectomorphism is given by
\be
	z=\frac{y-\bar y^3}{\bar y(1-y\bar y)},
\ee
where $y=w_2/w_1$ and $z$ is the coordinate used in section~\ref{sec:three-body}, see figure~\ref{fig:acuteobtuse}. In terms of $y$, $\mathrm{Gr}_+(2,3)$ is the set $\{y\in \C|\Re y<0,\,\Im y<0,\,|y|<1\}$. The three natural boundary components of $\mathrm{Gr}_+(2,3)$ are mapped to the points $z\in \{0,1,\oo\}$, and the interior is mapped to the acute region with $\Im z>0$ in figure~\ref{fig:acuteobtuse}. This symplectomorphism can also be checked to correctly map the classical Hamiltonians~\eqref{eq:Hsymb_momspace} and~\eqref{eq:Hsymb0} in the two pictures one to another.

\subsection{Relation to a classical LLL phase space}

Before we proceed to the semiclassical expansion of the spectrum, it is helpful to make contact with the position space analysis of section~\ref{sec:three-body}. There is of course no direct transformation between the momentum space $z_k$ and position space $\a_k$. Nevertheless, it is reasonable to expect that the classical systems obtained in both cases should be related. In this subsection, we will show (rather schematically) how the phase space $\cM$ discussed above can be formally obtained from a classical position space picture of the kind discussed in~\ref{sec:intuition}.

We begin by expressing the points $\a_k$ on the hyperbolic disk as the stereographic projections of vectors on the future-directed hyperboloid in $\R^{2,1}$:
\begin{equation}
	X_k = \frac{2}{1-|\a_k|^2}\left(\frac{1+|\a_k|^2}{2},\Re\a_k,\Im\a_k\right), \quad (X_k^0)^2-(X_k^1)^2-(X_k^2)^2=1.
\end{equation} 
If we assume that $\a$ parameterize the phase space $\D^N$ with the symplectic form on each $\D$ given by the standard hyperbolic volume form (this is the classical phase space for LLL centres), then $\sum_{k=1}^N X_k$ is the momentum map for the $\SO(2,1)$ action. Symplectic reduction by $\SO(2,1)$ can be performed by restricting to $\sum_{k=1}^N X_k = (J,0,0)$ and modding out by the stabilizer of $(J,0,0)$.

As we discussed previously, we expect that the important limit at large $J$ is when all $\a_k$ are close to the unit circle. Let us therefore parameterize
\be
	\a_k = (1-J^{-1}p_k^{-1})e^{i\theta_k},
\ee
in terms of which we find 
\be
	X_k = Jp_k(1,\cos\theta_k,\sin\theta_k)+O(1).
\ee
The condition $\sum_{k=1}^N X_k = (J,0,0)$ becomes
\be
	\sum_{k=1}^N p_k=1,\quad \sum_{k=1}^N p_k e^{i\theta_k}=0,
\ee
and the symplectic quotient is obtained by modding out shifts in $\theta_k$. The symplectic form induced from $\D^N$ is
\be
	\w_\text{LLL}=J\sum_{k=1}^N dp_k\wedge d\theta_k+O(1).
\ee

 This effective large-spin phase space can be identified with $\cM$ via
\begin{equation}
	\theta_k = \arg(z_k),\quad p_k = \frac{|z_k|}{|z_1|+\dots+|z_N|}, \quad k=1,\dots,N.
	\label{eq:z_to_theta}
\end{equation}
In particular, this identifies $J^{-1}\w_\text{LLL}$ with the symplectic form on $\cM$.

\subsection{Leading density of states}

The classical Hamiltonian in the momentum space picture is given in~\eqref{eq:Hsymb_momspace}. Note that the factor
\be
	2|z_i z_j|-z_i \bar z_j-\bar z_i z_j\geq 0
\ee
vanishes precisely when $z_i$ and $z_j$ are collinear. Thus for $\De_\s>0$, the Hamiltonian $H^{(0)}_\text{symb}$ blows up as we approach the non-regular points on the boundaries of the regions $\cM_{(a\cdots b)}$. This means that, as mentioned previously, the classically accessible phase space is the set $\cM^\text{reg}$ of regular points, which is a disjoint union of the connected components $\cM_{(a\cdots b)}$ over all cyclic orderings $(a\cdots b)$. These regions correspond to the particles arranged on a large circle in $\D$ in various cyclic orderings. 

Of course, when the particles are identical as in our toy model, we need to quotient $\cM^\text{reg}$ by the permutation group $S_N$. This is equivalent to restricting to just one region $\cM_{(1\cdots N)}$ and modding out by the residual $\Z_N$ cyclic permutations that act on $\cM_{(1\cdots N)}$.

Since each region $\cM_{(a\cdots b)}$ is isomorphic to the positive Grassmannian $\mathrm{Gr}_+(2,N)$, it is useful to consider $H^{(0)}_\text{symb}$ as a function on $\mathrm{Gr}_+(2,N)$. It takes the form
\begin{equation}\label{eq:leading_symb_Gr}
	H_{\mathrm{symb}}^{(0)}(W)=  b_{\mathrm{Cas},0}\sum_{1\leq i<j\leq N} \left(2 \frac{\Im W_j\bar W_i}{||W||^2} \right)^{-\De_\s},
\end{equation}
where, by definition, $\Im W_j\bar W_i>0$ on $\mathrm{Gr}_+(2,N)$. The unique minimum of the Hamiltonian is the unique $\Z_N$ fixed point of the positive Grassmannian, at $W_k=e^{\pi i k/N}$. In terms of the variables~\eqref{eq:z_to_theta}, this corresponds to the classical configuration where the points on the circle form a regular polygon at angles $\theta_k=2\pi i k/N$, as anticipated in~\cite{Fardelli:2024heb}.  As the energy increases, the equal-energy slices move away from the minimum and toward the singular locus $\Im W_j\bar W_i = 0$ for some $j>i$, where the Hamiltonian diverges.

Now that we have a good understanding of the phase space and the classical Hamiltonian, we can study the semiclassical approximation of the integrated density of states $n_E$, that is to say the number of states with energy below $E$. As in section~\ref{sec:three-body}, let $U_E$ be the set in $\cM^\text{reg}$ where $H^{(0)}_\text{symb}<E$. At leading order in $\hbar = J^{-1}$, $n_E$ is the symplectic volume enclosed by $U_E$ in units of phase space volume $(2\pi \hbar)^{N-2}$:
\begin{equation}
n_E^{\text{all}}= \left(\frac{J}{2\pi} \right)^{N-2} \int_{U_E} \frac{\omega^{\wedge (N-2)}}{(N-2)!} + O(J^{N-3}).
\end{equation}
Here we are counting all states, i.e.\ not only $S_N$-invariant states. When we are dealing with identical particles, we have to restrict to $S_N$-invariant states. The analysis of this parallels the discussion in section~\ref{sec:permutations} for $N=3$ and is discussed in detail in the next subsection.

For now, we expect the leading density of $S_N$-invariant states to be obtained from the volume in the quotient $\cM^\text{reg}/S_N$. Equivalently, it is enough to compute the volume in $\mathrm{Gr}_+(2,N)/\Z_N$. In other words, if we denote by $U_N^+$ the set in $\mathrm{Gr}_+(2,N)$ where $H^{(0)}_\text{symb}<E$, then the number of $S_N$-invariant states below energy $E$ is
\be
n_E= \frac{1}{N}\left(\frac{J}{2\pi} \right)^{N-2} \int_{U_E^+} \frac{\omega^{\wedge (N-2)}}{(N-2)!} + O(J^{N-3}).
\label{eq:nE_class}
\ee
Here, recall that $\omega=i \ptl\bar\ptl\cK$ is the symplectic form corresponding to the K\"ahler potential $\cK=2\log ||W||^2$ on the Grassmannian $\mathrm{Gr}(2,N)$. 

In particular, the total number of semiclassical states is given by 
\be
n_\oo = \p{\frac{J}{2\pi}}^{N-2} \frac{\vol \cM}{N!}+O(J^{N-3})=\p{\frac{J}{2\pi}}^{N-2} \frac{\vol \mathrm{Gr}(2,N)}{2^{N-1}N!}+O(J^{N-3}).
\ee
The symplectic volume of $\mathrm{Gr}(2,N)$ can be computed as follows. We previously identified $\mathrm{Gr}(2,N)$ as the degree-$2$ variety in $\CP^{N-1}$ defined by $W^2=0$. This implies
\be
	\int_{\mathrm{Gr}(2,N)} \a^{\wedge(N-2)}=2,
\ee
where $\a$ is the multiple of $\w=2i\ptl\bar\ptl \log \|W\|^2$ that represents the generator of $H^2(\CP^{N-1},\Z)$, i.e.\ $\int_{\CP^1} \a = 1$.\footnote{Here we are relying on the fact that the integral cohomology ring of $\CP^{N-1}$ is generated by $\a$, and thus $\a^{\wedge (N-2)}$ gives $1$ when paired with $\CP^{N-2}$.} It is easy to check that $\a=\w/4\pi$, and thus
\be
	\vol \mathrm{Gr}(2,N) = \int_{\mathrm{Gr}(2,N)}\frac{\w^{\wedge(N-2)}}{(N-2)!}=\frac{2(4\pi)^{N-2}}{(N-2)!}.
\ee
We conclude
\be
	n_\oo = \p{\frac{J}{2\pi}}^{N-2} \frac{\vol \mathrm{Gr}(2,N)}{2^{N-1}N!}+O(J^{N-3})=\frac{J^{N-2}}{N!(N-2)!}+O(J^{N-3}).
\ee

This number agrees precisely with the expansion~\eqref{eq:dimHNJ} of $\dim \cH_{N,J}^\text{primary}$ to leading order at large $J$. Consequently, we expect that the fraction of states well approximated by semiclassics goes to $1$ as $J\to \oo$.

At generic energies, we do not know how to compute the integral in~\eqref{eq:nE_class} analytically. However, it is easy to compute it numerically using Monte-Carlo integration. For this, we generate random pairs of orthogonal unit vectors, each of which defines a plane in $\R^N$, and therefore a point on $\mathrm{Gr}(2,N)$. We then use the $S_N\ltimes \Z_2^N$ action to map these points into the positive Grassmannian. Computing the fraction of the resulting points that satisfy $H^{(0)}_\text{symb}<E$ and multiplying by $\vol \mathrm{Gr}_+(2,N)$ then yields an approximation to the integral in~\eqref{eq:nE_class}.

\begin{figure}[t]
	\centering
	\includegraphics[scale=.75]{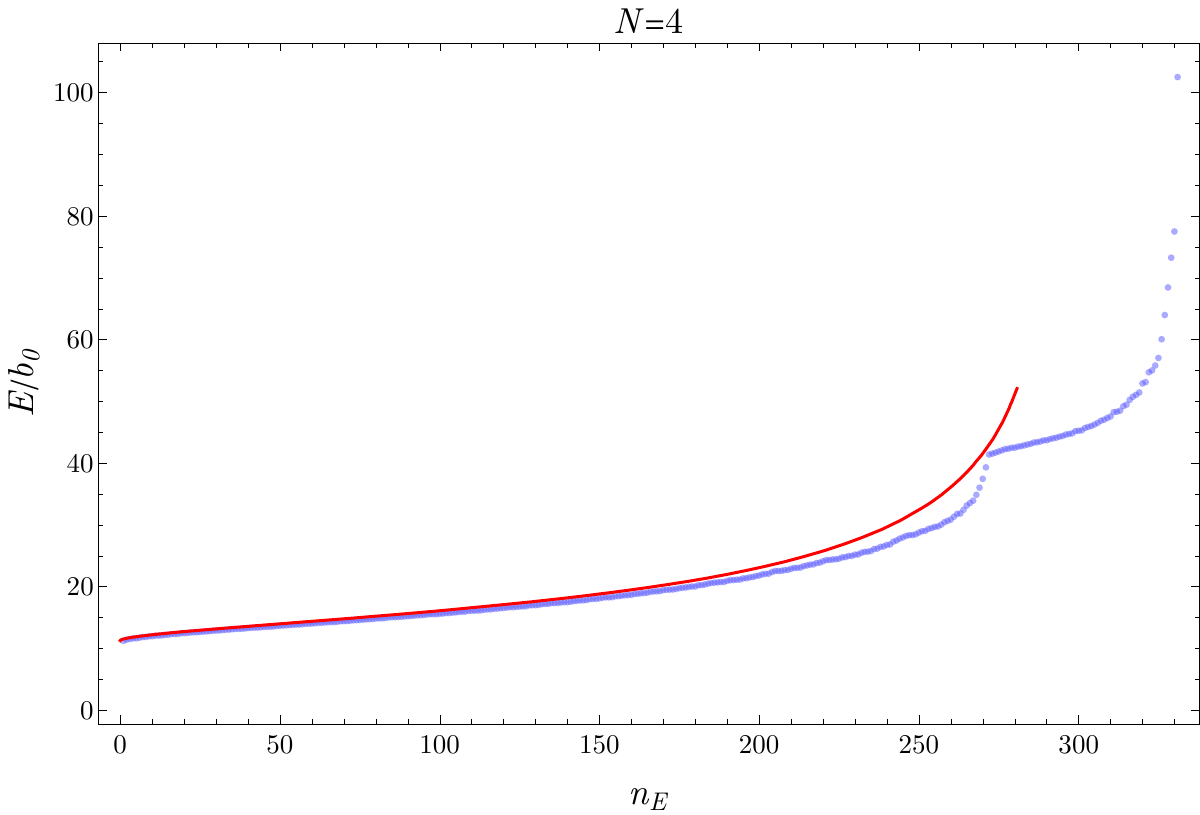}
	\caption{Energy as a function of mode number at $N=4$, $(\De_\phi,\De_\s)=(1.234,0.6734)$. Blue points are the exact eigenvalues of $J^{\De_\s} H_N$ at $J=120$ from numerical diagonalization, while the red curve is an interpolation of the Monte Carlo approximation of $n_E$ in~\eqref{eq:nE_class} with $10^5$ sample points.}
	\label{fig:montecarlo_vs_exact_4part}
\end{figure}

\begin{figure}[t]
 	\centering
 	\includegraphics[scale=.75]{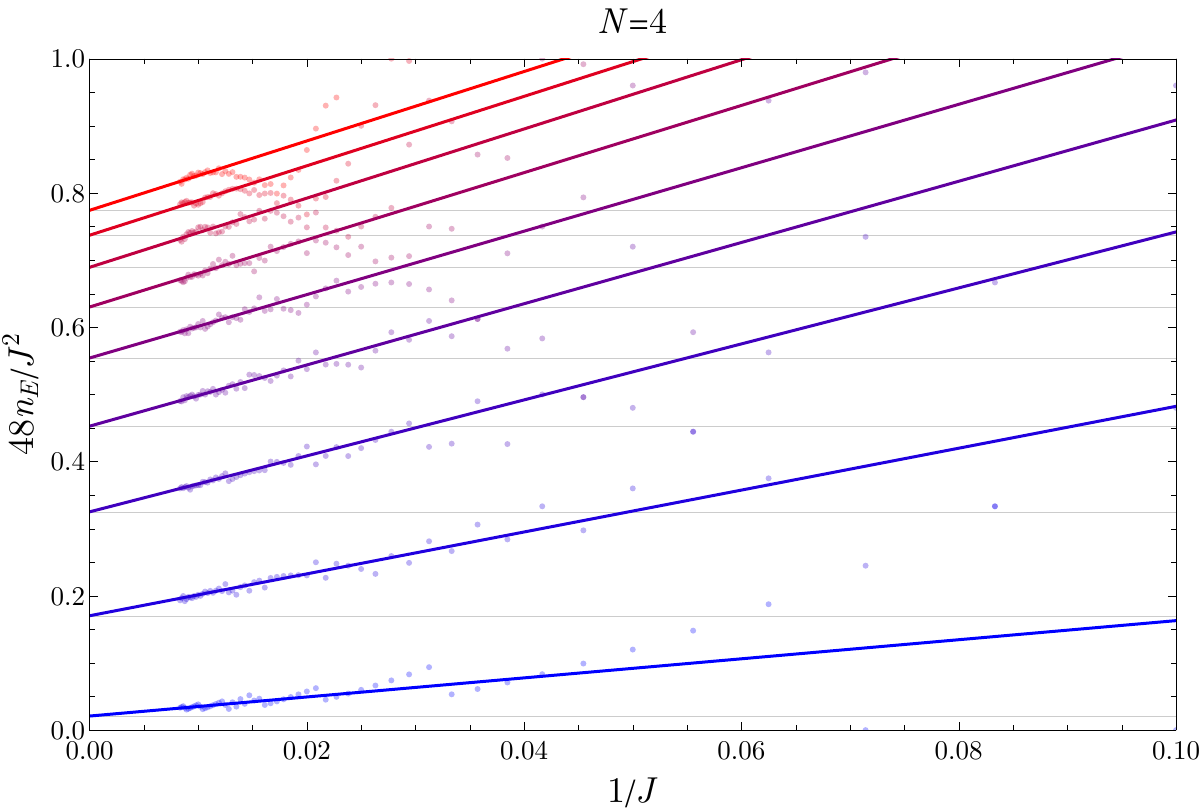}
 	\caption{Proportion of the number of semiclassical states below fixed energies $E$ as a function of inverse spin. The energies $E=12,14,\dots,26,28$, covering roughly eighty percent of the total symplectic volume, are indicated by a color gradient from blue (lower $E$) to red (higher $E$). Data points correspond to the exact counting function at even spins $J\leq 120$ with the parameters $(\De_\phi,\De_\s)$ of figure~\ref{fig:montecarlo_vs_exact_4part}. The solid lines represent linear fits that are conditioned to reproduce the leading term~\eqref{eq:nE_class} at $J=\oo$. Despite the non-smoothness in $n_E$, these fits show that the numerical data is consistent with~\eqref{eq:nE_class}.}
 	\label{fig:nEJ_fits}
 \end{figure}

 In figure~\ref{fig:montecarlo_vs_exact_4part} we compare the leading-order result for $n_E$ with the exact diagonalization data at $J=120$, finding good agreement at low energies. The distinct feature in the exact data starting at $n_E\approx 270$ seemingly corresponds to the three-body states of the form $[\f\f\f^2]_J$. In figure~\ref{fig:nEJ_fits} we make the same comparison, now as a function of $J$ for a few sample values of $E$. We see that the numerics support the leading-order result~\eqref{eq:nE_class}.
 
 It would be interesting to compute subleading $1/J$ terms in the expansion~\eqref{eq:nE_class} for $n_E$. However, one must keep in mind that already the subleading term in~\eqref{eq:nE_class} cannot be easily determined in general systems. Indeed, the exact function $n_E$ is not smooth. Instead, it is piecewise-constant with jumps at the energy eigenvalues $E_k$. In general, this ``discrete'' behavior can show up already at the subleading order, as is evident in the example of an isotropic harmonic oscillator with $N-2$ degrees of freedom. To see this, note that the energy level $E_k\sim \hbar k=k/J$ has $O(k^{N-3})$ degeneracy, which implies that $n_E$ is of the order $J^{N-2}$ while discontinuities in $n_E$ are of the order $J^{N-3}$.
 
The Gutzwiller trace formula shows that the discontinuous behavior of $n_E$ is controlled by the periodic orbits of the classical Hamiltonian---see for instance~\cite{gutzwiller2013chaos,Balian:1974ah}.\footnote{One way to interpret the (subleading) Bohr-Sommerfeld condition in systems with one degree of freedom is that periodic trajectories are easy to control, and their contribution can be resummed.} Therefore, one may hope to obtain sensible subleading corrections under the assumption that the Hamiltonian is sufficiently generic and does not have too many periodic orbits. For instance, it is possible to obtain a subleading term in the Weyl law for the Laplace-Beltrami operator under similar assumptions~\cite{MR575202}. We are not aware of a readily-available result for the semiclassical limit of Berezin-Toeplitz quantization, but one can likely be obtained using the methods of~\cite{Charles:1999qq,Douglas:2009fvp} (see also~\cite{MR2876259}).

\subsection{Semiclassical states under the action of $S_N$}
\label{sec:generalNperms}

In this section we examine in more detail the behavior of semiclassical states under $S_N$.

The set $U_E\subset \cM^\text{reg}$ has in general $(N-1)!$ connected components $U_{E,(a\cdots b)}=U_E\cap \cM_{(a\cdots b)}$, and the semiclassical states can localize to any of these connected components. This gives a total of $(N-1)!$ semiclassically-degenerate states for each energy level. As per usual, this degeneracy may be broken by instanton corrections. We have verified that, for example, the approximate six-fold degeneracy is indeed present in the exact $N=4$ spectrum of all (not necessarily $S_4$-invariant) states.

For a given energy eigenstate, the semiclassical state localized in $\cM_{(1\cdots N)}$ transforms in some representation $\r=e^{2\pi i m/N}$ of $\Z_N$. Similarly to section~\ref{sec:permutations}, the representation of $S_N$ that acts on the full set of $(N-1)!$ nearly-degenerate energy levels is then the induced representation $\mathrm{Ind}^{S_N}_{\Z_N}\r$. The multiplicity of the trivial representation of $S_N$ in $\mathrm{Ind}^{S_N}_{\Z_N}\r$ is one when $\r$ is trivial and zero otherwise. This follows immediately from Frobenius reciprocity.

More generally, the $S_N$ representation content of $\mathrm{Ind}^{S_N}_{\Z_N}\r$ can be determined using theorems 8.8 and 8.9 in~\cite{Reutenauer}: the irreducible representation of $S_N$ with Young diagram $\mu$ appears with multiplicity equal to the number of standard Young tableaux $T$ of shape $\mu$ with major index $\mathrm{maj}(T)$ satisfying $\mathrm{maj}(T)\equiv m\mod N$.

Recall that a standard Young tableau $T$ of shape $\mu$ is an assignment of integers from $1$ to $N$ to the cells of $\mu$ such that the integers in each row and each column are strictly increasing. An integer $i$ in $T$ is called a descent if $i+1$ appears strictly below $i$ in $T$.  The major index $\mathrm{maj}(T)$ of $T$ is the sum of all of its descents. It satisfies $\mathrm{maj}(T)\in\{0,\cdots,\binom{N}{2}\}$.

For instance, at $N=4$, it is easy to check that
\be
	\mathrm{Ind}^{S_4}_{\Z_4}e^{2\pi i m/4}=\begin{cases}
		\myng{(4)}_1\oplus \myng{(2,2)}_2\oplus\myng{(2,1,1)}_3&,\quad m=0\\
		\myng{(3,1)}_3\oplus\myng{(2,1,1)}_3&,\quad m=1\\
		\myng{(1,1,1,1)}_1\oplus\myng{(2,2)}_2\oplus\myng{(3,1)}_3 &,\quad m=2\\
		\myng{(3,1)}_3\oplus\myng{(2,1,1)}_3&,\quad m=3
	\end{cases}
\ee
where we labeled each irrep of $S_4$ by its dimension. Note that for each $m$ the dimensions add up to $6$. Just like in section~\ref{sec:permutations}, the states belonging to one irrep are protected from splitting due to instanton corrections. In our numerics for $N=4$, we indeed observed the splitting of the 6 nearly degenerate levels into $3+3$ or $1+2+3$, depending on the energy level.

\subsection{The lowest-lying states}
\label{sec:general_N_HO}
In this section, we derive the analogue of the result of section~\ref{sec:small_k} for low-lying states in the case of general $N$.

Recall that in the semiclassical limit $J\to\oo$, the low-lying states localize around the minimum of the classical Hamiltonian, near which the potential can be approximated by that of $N-2$ coupled harmonic oscillators. We first discuss the spectrum of all (not necessarily $S_N$-invariant) states, and later discuss their $S_N$ representation content based on the previous subsection.

It is enough to focus on the region $\cM_{(1\cdots N)}$ or, equivalently, the positive Grassmannian $\mathrm{Gr}_+(2,N)$. As discussed at the end of section~\ref{sec:small_k}, the energy eigenvalues for small mode numbers $k_1,\dots,k_{N-2}\ll J$ take the form
\begin{equation}
	E_k = H_\text{norm}\vert_P+ J^{-1}\sum_{a=1}^{N-2} \Omega_a \left(k_a+\frac{1}{2}\right) +O\left(J^{-3/2}\right),
	\label{eq:low_lying_E}
\end{equation}
where $P\in \mathrm{Gr}_+(2,N)$ is the position of the minimum of $H^{(0)}_\text{symb}$. The characteristic frequencies $\Omega_a$ are defined in section~\ref{sec:small_k}, around equation~\eqref{eq:Hsymb_to_Omega_a}. 

The first term in~\eqref{eq:low_lying_E} can be straightforwardly computed and is given by
\be
	H_\text{norm}\vert_P&=b_{\mathrm{Cas},0}\, \varepsilon_0(N,\De_\s)\p{1-\thalf\De_\s(N \De_\phi-1)J^{-1}}+O(J^{-2}), \\ 
	\varepsilon_0(N,\De_\s) &=\left(\frac{2}{N}\right)^{-\De_\s} \sum_{q=1}^{N-1}(N-q) (\sin \tfrac{\pi q}{N})^{-\De_\s}. 
\ee
In particular, $\varepsilon_0(3,\De_\s)=3^{1+\De_\s/2}$ and $\varepsilon_0(4,\De_\s) = 2^{1+3\De_\s/2}(2+2^{-\De_\s/2})$. The leading in $J$ term comes from $H^{(0)}_\text{symb}$ in~\eqref{eq:leading_symb_Gr} and agrees with the prediction in~\cite{Fardelli:2024heb}. Recall also that the coordinates of $P$ are given by
\be\label{eq:minlocation}
W_k=e^{\pi i k/N}.
\ee
The subleading term is obtained using~\eqref{eq:normsymb} and the subleading symbol $H^{(1)}_\text{symb}$, which is computed following section~\ref{sec:generalNsymbols}.

It remains to determine the frequencies $\Omega_1,\dots,\Omega_{N-2}$ by expanding the Hamiltonian to quadratic order around the minimum and separating the corresponding quadratic form into a sum of decoupled harmonic oscillators, as described in section~\ref{sec:small_k}. This can be simplified by using the action of $\Z_N$ on $\mathrm{Gr}_+(2,N)$. We define the coordinates $X_0,\cdots, X_{N-1}$ as
\begin{equation}
	X_n=\sum_{k=1}^N W_k^2 \, e^{-2\pi i k n/N},
\end{equation}
such that the cyclic action $W_k\to W_{k+1}$ now becomes $X_n\to e^{2\pi i n/N}X_n$. Note that the condition $W^2=0$ implies $X_0=0$, and the coordinates of the minimum~\eqref{eq:minlocation} become $X_1=N$ and all other $X_n=0$.

Based on the variables $X_n$, we can now define affine coordinates 
\be
	x_n = \frac{1}{\sqrt{2}}\frac{X_{n+1}}{X_1},\quad n=1,\cdots, N-2.
\ee
Notice that $x_n$ has charge $n$ under $\Z_N$ and the minimum is at $x_n=0$. In terms of the $x_n$, the symplectic form at $P$ is
\be
	\w|_P = i\sum_{n=1}^{N-2} dx_n\wedge d\bar x_n.
\ee

Recall that in section~\ref{sec:small_k} we defined the quadratic form $\Omega(x)$ as
\be
	H^{(0)}_\text{symb} = H^{(0)}_\text{symb}\vert_P+\Omega(x)+O(x^3).
\ee
Using $\Z_N$ symmetry, we find that the general form of $\Omega(x)$ is given by
\be
	\Omega(x) = \sum_{n=1}^N A_n x_n\bar x_n+\sum_{n=2}^{\lfloor N/2\rfloor} \p{B_n x_n x_{N-n}+\bar B_n x_n x_{N-n}}.
\ee
We therefore deduce that $\Omega(x)$ separates into three types of blocks.
\begin{enumerate}
	\item $A_1 x_1\bar x_1$, which gives the frequency $\Omega_1=A_1$.
	\item $A_n x_n \bar x_n+A_{N-n} x_{N-n} \bar x_{N-n}+B_n x_n x_{N-n}+\bar B_n \bar x_n \bar x_{N-n}$, for $1<n<N/2$. In this case we get two frequencies which we can assign as
	\be
		\Omega_n&=\sqrt{\left(\frac{A_n+A_{N-n}}{2}\right)^2-|B_n|^2}+\frac{A_n-A_{N-n}}{2},\\
		\Omega_{N-n}&= \sqrt{\left(\frac{A_n+A_{N-n}}{2}\right)^2-|B_n|^2}+\frac{A_{N-n}-A_{n}}{2}.
	\ee
	\item $A_{N/2}x_{N/2}\bar x_{N/2}+B_{N/2}x_{N/2}^2+\bar B_{N/2}\bar x_{N/2}^2$ when $N$ is even. In this case the frequency is 
	\be
		\Omega_{N/2}=\sqrt{A_{N/2}^2-4|B_{N/2}|^2}.
	\ee
\end{enumerate}
The frequencies for each type of block have been determined following the discussion in section~\ref{sec:small_k}. We note that the separated variables $u_a$, $a=1,\cdots, N-2$ can be constructed so that the $\Z_N$ charges are the same as those of $x_a$. This is possible because both the symplectic form and the classical Hamiltonian are $\Z_N$-invariant.

For example, at $N=3$ we find
\be
	\frac{\Omega_1}{b_{\mathrm{Cas},0}\, \varepsilon_0(N,\De_\s)}=\De_\s(\De_\s+2)/2,
\ee
in agreement with section~\ref{sec:small_k}. For $N=4$ we find
\be\label{eq:Omega1_N4}
	\frac{\Omega_1}{b_{\mathrm{Cas},0}\, \varepsilon_0(4,\De_\s)} &= \De_\s \left(\frac{2\De_\s}{2+2^{-\De_\s/2}}+1\right),\\
 	\frac{\Omega_2}{b_{\mathrm{Cas},0}\, \varepsilon_0(4,\De_\s)} &= \frac{2\De_\s}{2+2^{-\De_\s/2}} \sqrt{(1+\De_\s/2)(2+2^{-\De_\s/2}(1+\De_\s))}. \label{eq:Omega2_N4}
\ee
Though we do not have an analytic formula for all frequencies for general $N$, the above discussion can be straightforwardly applied to compute $\Omega_1,\dots,\Omega_{N-2}$ at any fixed value of $(N,\De_\s)$.

Let us now determine the $\Z_N$ charge of the states in terms of the mode numbers $k_1,\dots,k_{N-2}$. Similarly to section~\ref{sec:permutations}, one can check that the $\Z_N$ charge gets a contribution $J$ from the line bundle $L$. We expect the eigenfunctions of the harmonic oscillator to have the multiplicative structure
\be
	\psi \sim u_1^{k_1}\cdots u_{N-2}^{k_{N-2}}.
\ee
Of course, the coordinates $u_1$ are not holomorphic functions on $\mathrm{Gr}_+(2,N)$ and so this cannot be understood literally. Instead, this product should be viewed as an operator acting on the ground state wavefunction. Nevertheless, we can still read off the charge due to the excitations and conclude that the total $\Z_N$ charge is
\be
		J+\sum_{a=1}^{N-2} a\, k_a \mod N.
\ee

We now recall that each state on $\mathrm{Gr}_+(2,N)$ gives rise to $(N-1)!$ approximately degenerate states on $\cM$. The representation of $S_N$ on these states has been computed in section~\ref{sec:generalNperms} in terms of the $\Z_N$ charge. In particular, if we are interested in $S_N$-invariant states, we need to impose
\be\label{eq:ZN_selection_singlet}
		J+\sum_{a=1}^{N-2} a\, k_a=0 \mod N.
\ee

\begin{figure}[t]
	\centering
	\includegraphics[scale=0.65]{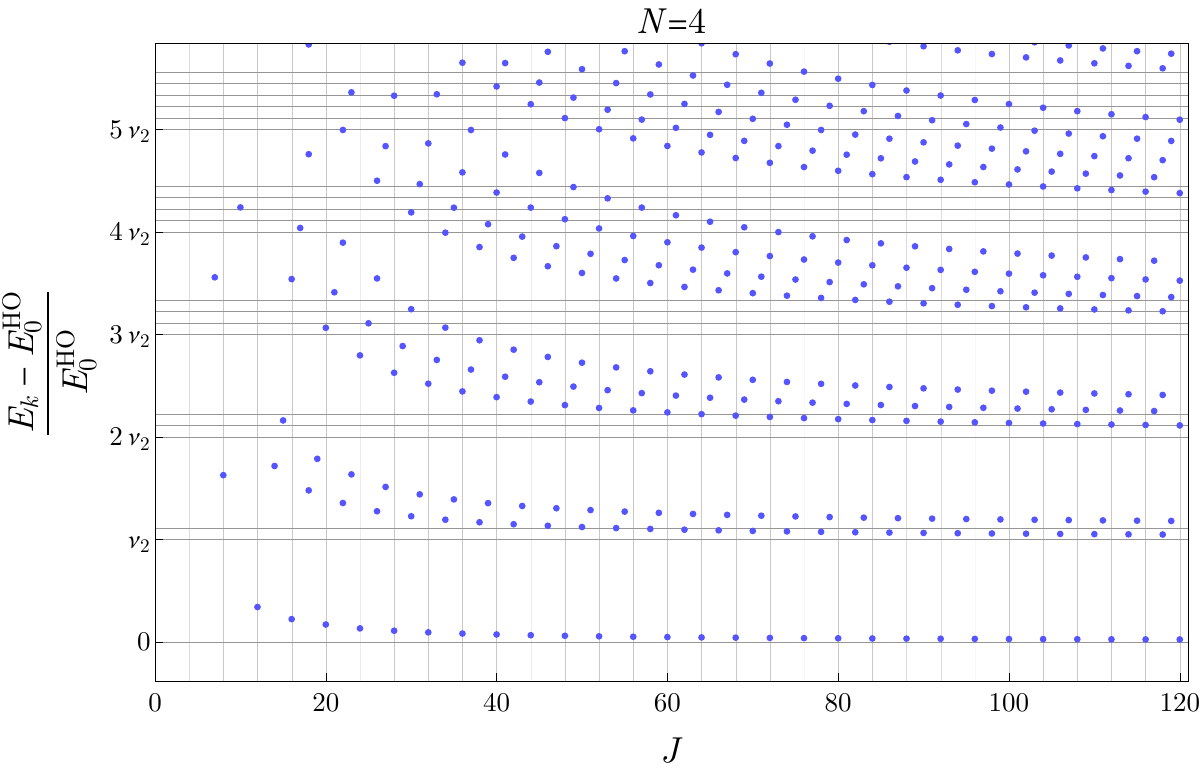}
	\caption{The ratio $(E_k-E_0^{\mathrm{HO}})/E_0^{\mathrm{HO}}$ as a function of spin for $(\De_\phi,\De_\s)=(1.234,1.386)$. The blue dots are the values from exact diagonalization, while the horizontal lines correspond to the energies $E_k^\text{HO}$. The expressions for $\nu_a:=\Omega_a/\left(b_{\mathrm{Cas},0}\, \varepsilon_0(4,\De_\s)\right)$ are given by~\eqref{eq:Omega1_N4},\eqref{eq:Omega2_N4}. The vertical grid lines are at $J\equiv 0\mod 4$}
	\label{fig:lowlyingN4J}
\end{figure}
\begin{figure}[t]
	\centering
	\includegraphics[scale=0.65]{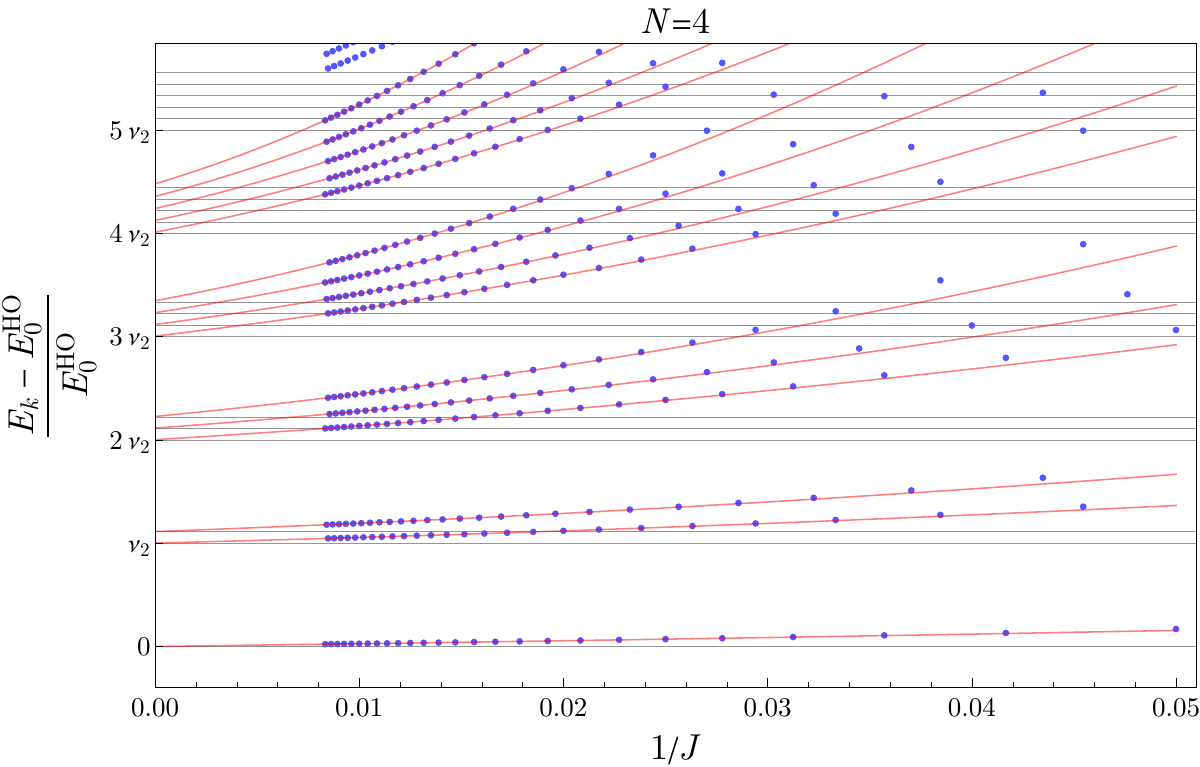}
	\caption{Same as figure~\ref{fig:lowlyingN4J}, but as function of $1/J$. The red curves represent fits of the numerical data points at $J\geq 80$ by a quadratic function $a+b/J+c/J^2$.}
	\label{fig:lowlyingN4invJ}
\end{figure}

Let $E^\text{HO}_k$ be the eigenstate energies in the harmonic oscillator approximation at~$O(J^{-1})$,
\be
E_k^\text{HO} \equiv H^{(0)}_\text{symb}\vert_P+J^{-1}H^{(1)}_\text{norm}\vert_P+ J^{-1}\sum_{a=1}^{N-2} \Omega_a \left(k_a+\frac{1}{2}\right),
\ee
with the selection rule~\eqref{eq:ZN_selection_singlet} understood. In figures~\ref{fig:lowlyingN4J} and~\ref{fig:lowlyingN4invJ} we show a comparison of the energies $E_k^\text{HO}$ for $N=4$ with the exact numerical spectrum at $(\De_\phi,\De_\s)=(1.234,1.386)$. In these figures, we plot $(E_k-E_0^\text{HO})/E_0^\text{HO}$. These ratios are $J$-independent in the harmonic oscillator approximation and appear as horizontal lines. They arrange into groups of $1,2,3,\cdots$ due to the breaking of the degeneracy present for $\Omega_1=\Omega_2$ (in our example, the difference between $\Omega_1$ and $\Omega_2$ is $\approx 10\%$). The fits in figure~\ref{fig:lowlyingN4invJ} indicate a clear convergence of the numerics towards $E_k^\text{HO}$ with a correction that scales like $1/J^{2}$, as expected\footnote{Recall that the $O(J^{-3/2})$ correction in~\eqref{eq:low_lying_E} can only exist if there is a degeneracy at leading order, as explained in~\cite[Sec.~5.1]{DeleporteHO}.}. Figure~\ref{fig:lowlyingN4J} can be examined to confirm the selection rule~\eqref{eq:ZN_selection_singlet}; for instance, the $k_1=k_2=0$ states only exist at $J\equiv 0\mod 4$.

\section{Discussion}
\label{sec:discussion}

In this paper we have analyzed the large-$J$ limit of the leading-twist $N$-particle states in $\mathrm{AdS}$. We found that the quantum-mechanical problem that describes them becomes semiclassical for the majority of states, with $\hbar=1/J$. We have developed methods for studying the semiclassical spectrum, relying on existing results from Berezin-Toeplitz quantization. We found that all our analytical results are in good agreement with the exact numerics.

There are many questions that we have not addressed in this work. It will be important to understand more general interactions between the individual particles. For instance, in a multi-twist state with both $\f^2$ and $\f$ constituents, an exchange of $\f$ can swap the positions of $\f^2$ and $\f$. If our methods can be applied to general CFTs, this would be important for the $[\s\e\e]_J$ triple-twist operators in 3d Ising CFT, where $\e\sim\s^2$. A related perturbative problem is $[\f\f\f\f]_J$ quadruple-twist states in Wilson-Fisher theory at one loop, where the only non-zero anomalous dimensions correspond to states of the form $[\f^2\f\f]_J$.

Another interesting problem is to understand if, and under which conditions, one can derive subleading corrections to the density of states~\eqref{eq:nE_class} for $N\geq 4$. Although we have not explored this question with any degree of rigour, this might be feasible to do, at least formally, using the path integral methods of~\cite{Charles:1999qq,Douglas:2009fvp}. 

Similarly, it would be interesting to derive systematic corrections to the harmonic oscillator approximation discussed in sections~\ref{sec:small_k} and~\ref{sec:general_N_HO}. One can see from figure~\ref{fig:lowlyingN4invJ} that the leading term does not work very well, even at $J=120$, but that the next correction would likely remedy this.

We have focused on the limit where $J$ is the largest parameter in the problem. There are several other natural limits to explore. Perhaps the most intriguing one is the double-scaling limit $N^2\sim J\gg 1$. Recall from section~\ref{sec:intuition} that the lowest-twist states can be interpreted as the lowest Landau level (LLL) states on the hyperbolic disk. The ratio $N^2/(2J)$ can be interpreted as the filling fraction of the LLL. The physics of interacting fermions in the LLL in flat space at finite filling fraction is well studied in the field of fractional quantum Hall effect (FQHE)~\cite{laughlin1983anomalous,Haldane:1983xm}.\footnote{In fact, early in this project we drew a lot of intuition from the Laughlin state which is often discussed in FQHE context.} It seems clear that in principle the FQHE setup can be obtained in an appropriate limit of our problem (swapping scalars for fermions and taking large-$\AdS$ radius limit). It is therefore interesting to ask whether some remnant of FQHE physics can be observed in the large-spin multi-twist spectra of various CFTs. Interpreting $N$ as a $U(1)$ charge $Q$, we note that the regime $J\sim Q^2$ is between the Regge phase and the giant vortex phase studied in~\cite{Cuomo:2022kio}. One issue with this limit is that the inter-particle distances do not grow and we do not expect two-body potentials to dominate, or the interactions to be universal. However, it is possible that some universal features of FQHE can still be observed. At the very least, one can try engineering a FQHE setup in holographic theories such as planar $\cN=4$ Super Yang-Mills.

Another interesting limit is to consider fixed $N$ with large $\De_\f\sim J$. In this case the semiclassical limit is given by the classical LLL particles on the hyperbolic disk, and somewhat more general dynamics are allowed. Independently of spin, large scaling dimensions correspond to large $\AdS$ radii, and the physics of the classical $N$-body problem in flat space can be recovered. Can anything interesting be said about this problem from this point of view? In this context, note that the relation between near-threshold and large-spin expansions of flat space amplitudes was studied in~\cite{Correia:2020xtr} from the Froissart-Gribov formula. 

Finally, perhaps the most important open question is to derive the quantum-mechanical problems of the kind considered in this work in general CFTs and without appealing to the $\AdS$ picture. One approach is to try to formally derive the interaction Hamiltonian for $N=3$ from crossing equations for six-point functions. Besides the technical complexity, one aspect of this approach that differs from $N=2$ is that large inter-particle separation (and hence universality of interactions) is guaranteed by the large $J$ limit at $N=2$, irrespective of the state. For $N=3$, only a subclass of states has large inter-particle separations, however large this subclass is. Therefore, such a derivation would have to rely on some assumption about the states, and we hope that our analysis will be useful in identifying their key features. 

An alternative approach would be to construct (perhaps using ideas from~\cite{Alday:2007mf}) an effective large-spin theory in which the appropriate Wilson coefficients can be matched to various limits of CFT correlation functions. Based on the intuition in section~\ref{sec:intuition}, we expect that the appropriate theory should be phrased in terms of fields that second-quantize the LLL. We leave these and other questions for future study.

\appendix

\section*{Acknowledgements}

The authors thank Alix Deleporte, Simon Ekhammar, Victor Gorbenko, Nikolay Gromov, Shota Komatsu, Gregory Korchemsky, Sameer Murthy, Slava Rychkov, Volker Schomerus, David Simmons-Duffin, and especially Jean Lagac\'e for discussions. 

The work of PK was funded by UK Research and Innovation (UKRI) under the UK government's Horizon Europe funding Guarantee [grant number EP/X042618/1] and the Science and Technology Facilities Council [grant number  ST/X000753/1]. 

J.A.M. was supported by the Royal Society under grant URF$\backslash$R1$\backslash$211417 and by the European Research Council (ERC) under the European Union’s Horizon 2020 research and innovation program – 60 – (grant agreement No. 865075) EXACTC.

\section{Conformal algebra and embedding space conventions}
\label{app:conventions}

In Euclidean signature, $\AdS_{d+1}$ can be identified with the locus of solutions to the equation $X^2=-1$, where $X\in \R^{d+1,1}$. We label different components of $X$ by $X^A$ with $A=-1,0_E, 1,\cdots, d$, while the metric is taken to be $\eta_{AB}=\mathrm{diag}\{-1,+1,\cdots,+1\}$. We use the index $0_E$ to stress that this is a Euclidean component. Using $\mathbf{e}_A$ to denote the standard basis vectors of $\R^{d+1,1}$, we define the generators $L_{AB}$ of $\mathfrak{so}(d+1,1)$ by
\be
L_{AB}\mathbf{e}_{C}=\eta_{BC}\mathbf{e}_A-\eta_{AC}\mathbf{e}_B.
\ee
For fixed $A,B$, the function $X\mapsto L_{AB}X$ defines a vector field in $\R^{d+1,1}$ that is tangent to $\AdS_{d+1}\subset \R^{d+1,1}$. By restriction, it defines a vector field on $\AdS_{d+1}$ that we denote by $\cL_{AB}$. The action of $L_{AB}$ on a local operator in $\AdS_{d+1}$ then takes the simple form
\be
[L_{AB},\cO(x)]=(\cL_{AB}\cO)(x).
\ee
The generators $L_{AB}$ can be identified with the standard conformal generators as ($\mu,\nu\in \{1,\cdots, d\}$)
\be
L_{0_E,-1}=D,\quad 
L_{0_E,\mu}=-\frac{P_\mu+K_\mu}{2},\quad
L_{-1,\mu}=\frac{P_\mu-K_\mu}{2},\quad
L_{\mu\nu}=M_{\mu\nu}.
\ee

The Wick rotation to Lorentzian signature is achieved by setting $X^{0}=-iX^{0_E}$, $X_{0}=iX^{0_E}$ and similarly for all other tensors. In particular, $L_{0\mu}=iL_{0_E,\mu}$. The Wick-rotated metric in the resulting $\R^{d,2}$ becomes $\mathrm{diag}\{-1,-1,+1,\cdots,+1\}$. In terms of the global coordinates, $\AdS_{d,1}$ is embedded~as
\be
X^{-1}&=\cos t\cosh\r,\\
X^{0}&=-\sin t\cosh \r,\\
X^{\mu}&=n^\mu\sinh \r,
\ee
where $\mu=1,\cdots, d$ and $n$ is a unit vector in $S^{d-1}$, $n\. n=1$. Note that in terms of the Euclidean time $t_E=it$ this becomes
\be
X^{-1}&=\cosh t_E\cosh\r,\\
X^{0_E}&=-\sinh t_E\cosh \r,\\
X^{\mu}&=n^\mu\sinh \r.
\ee
It is easy to check that in these coordinates the vector field $\cD$ corresponding to $D$ is $-i\ptl_t$. In other words, $[D,\cO(x)]=-i\ptl_t \cO(x)$ and thus $D$ plays the role of the Hamiltonian for global time translations on $\R\x S^{d-1}$.

The Euclidean boundary Poincar\'e patch can be identified using $X^\pm = X^{-1}\pm X^{0_E}$ and rescaling to $X^+=1$. Then $X^\mu=x^\mu$ with $x^\mu$ ($\mu=1,\cdots,d$) being the standard coordinates on $\R^d$. In this Poincar\'e patch, the generators $D, P_\mu, K_\mu, M_{\mu\nu}$ become the standard conformal transformations. Furthermore, it is easy to check that the unit sphere $|x|=1$ is embedded as $(X^+,X^-,X^\mu)=(1,1,x^\mu)$, which coincides with the boundary ($\r\to \oo$) limit of the $t=t_E=0$ slice of the $\AdS$ space.

\section{Derivation of the effective pair potential}
\label{app:pair_potential}
In this appendix, we derive the expansion~\eqref{eq:scalar_potential_result} of $U_2(s)$, starting from its integral expression $U_2(|\a_1|^2/(1-|\a_1|^2)) = F(\a_1,0)$, where $F(\a_1,0)$ is given by~\eqref{eq:Fsymmetry_fixed}. We will be working in $(t,\a,q)$ coordinates of AdS$_{d+1}$, where $t$ is the global time and $(\a,q)\in \D\times \R^{d-2}$ are related to global coordinates by
\begin{equation}
\a=e^{i(\varphi-t)}\cos\theta\tanh\r, \quad q^i = \sinh \rho\, n^i,\,\,\, i=3,\dots,d.
\end{equation}
In these coordinates, $F$ reduces to an integral over $q_1$:
\be\label{eq:F_in_qcoords}
F(\a_1,0)=\frac{\pi^2\l^2 C_{\De_\f,d}^2}{(\De_\phi-1)^2} \int_{\R^{d-2}}d^{d-2}q_1\,(1+q_1^2)^{1-\De_\f} I(0,\a_1,q_1).
\ee
By its definition~\eqref{eq:def_I}, the function $I(x_1)$ is the unique solution to the following scalar AdS wave equation with source:
\begin{equation}
	(\nabla^2-\De_\s(\De_\s-d)) I(t,\a,q)=\frac{\de^2(\a)}{(1+q^2)^{\De_\phi}}.
	\label{eq:wave_eqtn_source}
\end{equation}

The derivation proceeds in three steps: first, we solve the homogeneous wave equation away from $\a=0$ with a separation of variables ansatz. Second, we fix all undetermined constants in the separated solution by requiring that it reproduce the source term at $\a=0$. Having determined its exact form, we finally integrate $I(0,\a_1,q_1)$ over $q_1$ in~\eqref{eq:F_in_qcoords} to obtain the expansion in lightcone blocks $k_{\De_\s+2n}$, $n\in \N$.
\subsection{Separation of variables for AdS Klein-Gordon equation}
\label{app:klein_gordon_source}
Since the source in~\eqref{eq:wave_eqtn_source} is invariant under translations of $t$, phase rotations of $\a$ and $SO(d-2)$ rotations of $q^a$, we can restrict the dependence of $I$ to $I(t,\a,q)\equiv I_0(s,|q|)$, where $s=s_{12}(\a,0)=|\a|^2(1-|\a|^2)^{-1}$. In this case, the action of the AdS Laplacian reduces to
\begin{equation}
\nabla^2 I_0(s,|q|) = \mathcal{D}_{|q|} I(s,|q|) + 4 (1+|q|^2)^{-1} \mathcal{D}_s I(s,|q|),
\label{eq:laplacian_AdS}
\end{equation}
where the second-order, one-variable differential operators $\mathcal{D}_{|q|}$, $\mathcal{D}_s$ are given by
\begin{align}
& \mathcal{D}_s(s,\ptl_s) = s (1+s)\ptl_s^2+(1+2s)\ptl_s, \\
& \mathcal{D}_{|q|}(q,\ptl_q)= (1+q^2)\ptl_q^2+\left((d+1)q+(d-3)q^{-1}\right)\ptl_q.
\end{align}
Consequently, the wave equation $(1+q^2)(\nabla^2-\De_\s(\De_\s-d))I(s,|q|)= 0$ away from the source at $\a=0$ separates into a sum of two one-variable differential equations in $s$ and $|q|$, respectively. To motivate our ansatz for the full solution, we first look at the particular case of $d=2$.  

\paragraph{Solution in two dimensions.} In $d=2$, where there are no transverse directions $q\equiv 0$, the curves $\a=\mathrm{const}$ are geodesics in AdS$_3$. In this case, the function $I(x)$ coincides with the function $\varphi^{12}_{\De_\s}$ studied in~\cite[section~3.1]{Hijano:2015zsa}. After translating their coordinates to ours, the solution is then given by
\begin{equation}
I_0(s)=I_{\De_\s,2}(s):=\frac{\Gamma(\De_\s/2)^2}{4\pi \Gamma(\De_\s)} k_{\De_\s}(1/(s+1)), \quad k_{2h}(z):=z^h {}_2F_1(h,h;2h;z).
\label{eq:I0_2d}
\end{equation}
It is easy to check that $4\mathcal{D}_s k_{\De_\s}=\De_\s(\De_\s-2)k_{\De_\s}$ away from $s=0$, which solves the homogeneous wave equation. To check that the multiplicative prefactor correctly accounts for the source $\de^2(\a)$, we can expand the wave equation near $\a=0$, where $s=|\a|^2+O(|\a|^4)=0$. There, the reduced Laplacian $4\mathcal{D}_s$ tends to the radial part of the Laplacian on the complex $\a$-plane in polar coordinates:
\begin{equation}
4\mathcal{D}_s(\e s,\e^{-1}\ptl_s) = 4\e^{-1} \ptl_s s \ptl_s+O(1)=\e^{-1} |\a|^{-1} \ptl_{|\a|}|\a|\ptl_{|\a|}+O(1).
\label{eq:laplacian_flat}
\end{equation}
At the same time, it follows from $2\pi I_{\De_\s,2}(s)= \log(s)+O(s^0)$ that $I_0$ tends to the rotation-symmetric Green's function of the Laplacian~\eqref{eq:laplacian_flat} near $|\a|=0$. As a result, $I_{\De_\s,2}$ is the unique solution to the AdS$_3$ Klein-Gordon equation with geodesic source:
\begin{equation}
(\nabla^2-\De_\s(\De_\s-2)) I_{\De_\s,2}(|\a|^2(1-|\a|^2)^{-1})= \de^2(\a).
\end{equation}

\paragraph{Solution in higher dimension.} In dimensions $d>2$, we make the separation of variables ansatz
\begin{equation}\label{eq:sov_ansatz}
I_0(s,|q|)= \sum_{n=0}^\infty c_n\, Q_n(|q|) \,I_{\De_\s+2n,2}(s),
\end{equation}
where $Q_n \,k_{\De_\s+2n}$ at $s\neq 0$ solves the wave equation for any $n$. Since $4\mathcal{D}_s I_{\De_\s+2n,2}=(\De_\s+2n)(\De_\s+2n-2)I_{\De_\s+2n,2}$, this is true if and only if $Q_n$ solves the following first-order differential equation:
\begin{equation}
\left(\mathcal{D}_q(|q|,\ptl_{|q|}) + \frac{(\De_\s+2n)(\De_\s+2n-2)}{1+|q|^2}-\De_\s(\De_\s-d) \right)Q_n(|q|)=0.
\label{eq:ode_q}
\end{equation}
After the change of variables and parameter redefinition
\begin{equation}
x:=\frac{|q|^2-1}{|q|^2+1},\quad Q_n(|q|)=(1-x)^{\De_\s/2} P_n(x),\quad (a,b):=\left(\De_\s-\frac{d}{2},\frac{d}{2}-2\right),
\label{eq:odeq_to_jacobi}
\end{equation}
it is straightforward to check that \eqref{eq:ode_q} reduces to the Jacobi differential equation with parameters $(n,a,b)$:
\begin{equation}
(1-x^2)P_n''(x)+(b-a-(a+b+2)x)P_n'(x)-n(n+a+b+1)P_n(x)=0.
\end{equation}
There are two independent solutions to this differential equation: the Jacobi polynomials $P_n^{(a,b)}$, and another solution with asymptotic behavior $P_n(x)\sim (1-x)^{(d-2\De_\s)/2}$ as $x\rightarrow 1$, which translates to $Q_n(|q|)\sim |q|^{\De_\s-d}$ as $|q|\rightarrow\infty$. When inserting this latter solution into~\eqref{eq:F_in_qcoords}, the integral would diverge from the large $|q|$ region when $\De_\s$ is sufficiently large. On the other hand, the Jacobi polynomials correspond to solutions that decay like $Q_n(|q|)\sim |q|^{-\De_\s}$ as $|q|\rightarrow \infty$, and therefore provide the correct solutions to the separation of variables ansatz~\eqref{eq:sov_ansatz}. As a result, the function $I$ can be expressed as 
\begin{equation}\label{eq:I_to_cn}
I(t,\a,q)= \left(\frac{1-x}{2}\right)^{\De_\s/2}\sum_{n=0}^\infty c_n P_n^{(a,b)}(x) k_{\De_\s+2n}(1/(s+1)),
\end{equation}
where $s=|\a|^2(1-|\a|^2)^{-1}$, $P_n^{(a,b)}$ are the Jacobi polynomials, and $(x,a,b)$ are given by~\eqref{eq:odeq_to_jacobi}.

\subsection{Solving for the source}
We would now like to fix the coefficients $c_n$ in~\eqref{eq:I_to_cn} via the input of the source in~\eqref{eq:wave_eqtn_source}. The first step is very similar to the $d=2$ case: expanding $\nabla^2-\De_\s(\De_\s-d)$ near $\a=0$ using~\eqref{eq:laplacian_AdS} and~\eqref{eq:laplacian_flat}, we again obtain the Laplacian of the complex $\a$-plane at leading order, albeit multiplied by a factor of $(1+q^2)^{-1}=(1-x)/2$. After inserting the solution~\eqref{eq:I_to_cn} and using the fact that $I_{\De_\s+2n,2}\sim \log|\a|^2/(2\pi)$ near $\a=0$, the wave equation~\eqref{eq:wave_eqtn_source} in the separation of variables ansatz then reduces to
\begin{equation}
\left(\frac{1-x}{2}\right)^{\De_\s/2+1}\sum_{n=0}^\infty c_n \, P_n^{(a,b)}(x) \, \de^2(\a) = \left(\frac{1-x}{2}\right)^{\De_\phi} \de^2(\a).
\end{equation}
After factoring out $\de^2(\a)$ and dividing by the power of $(1-x)/2$ on the left-hand side, we can understand this equation as defining the vector with coefficients $c_n$ in the infinite-dimensional vector space spanned by the Jacobi polynomials $P_n^{(a,b)}$. This vector space is equipped with a scalar product for which the Jacobi polynomials form an orthogonal basis:
\begin{equation}
\<f,g\>:= \int_{-1}^1 \frac{dx}{2} \left(\frac{1-x}{2}\right)^{a} \left(\frac{1+x}{2}\right)^{b} f(x)g(x). 
\label{eq:sp_jacobi}
\end{equation}
We can therefore determine the coefficients $c_n$ by projecting with respect to this scalar product, leading to
\begin{equation}
c_n \<P_n^{(a,b)},P_n^{(a,b)}\> =\left\<P_n^{(a,b)},\left((1-x)/2\right)^{\De_\phi-\De_\s/2-1}\right\>.
\label{eq:cn_sp}
\end{equation}
These scalar products that determine $c_n$ are given explicitly by equations 1 and 7 of \cite[\S 7.391]{gradshteyn2014table} in Gradshteyn \& Ryzhik. 

\subsection{Expansion into lightcone blocks}
\label{app:U2exp_into_ks}
Now that we have an explicit solution~\eqref{eq:I_to_cn} for $I(0,\a_1,q_1)$, with coefficients $c_n$ given by~\eqref{eq:cn_sp}, we can determine $F(\a_1,0) = U_2(|\a_1|^2(1-|\a_1|^2)^{-1})$ by plugging the solution into~\eqref{eq:F_in_qcoords}. Assuming the integral over $q\in \R^{d-2}$ can be swapped with the sum over $n$, we can then identify the coefficients $b_n$ of the expansion~\eqref{eq:scalar_potential_leading} as
\begin{equation}
b_n = c_n \frac{\pi^2\l^2 C_{\De_\f,d}^2}{(\De_\phi-1)^2} \int_{\R^{d-2}} d^{d-2} q\, (1+q^2)^{1-\De_\phi-\De_\s/2} P_n^{(a,b)}(x(|q|)).  
\end{equation}
To compute this integral, we take spherical coordinates for $\R^{d-2}$, integrate over the $S^{d-3}$, and change of variables from $|q|$ to $x$. Using the formula
\begin{equation}
\int_{\R^{d-2}} \frac{d^{d-2} q}{(1+q^2)^{\De_\s}} F(x(|q|)) = \frac{\pi^{d/2-1}}{\G(d/2-1)} \int_{-1}^1 \frac{dx}{2} \left(\frac{1-x}{2}\right)^{\De_\s-d/2} \left(\frac{1+x}{2}\right)^{d/2-2} F(x),
\end{equation}
as well as the expression~\eqref{eq:cn_sp} for the coefficients $c_n$, we find
\begin{equation}
	b_n = \frac{\pi^2\l^2 C_{\De_\f,d}^2}{(\De_\phi-1)^2} \frac{\pi^{d/2-1}}{\G(d/2-1)} \frac{\left\<P_n^{(a,b)},\left((1-x)/2\right)^{\De_\phi-\De_\s/2-1}\right\>^2}{\<P_n^{(a,b)},P_n^{(a,b)}\>},
\end{equation}
where $\<f,g\>$ is defined in~\eqref{eq:sp_jacobi}. 

\section{Matching the two-body binding energy with the Lorentzian inversion formula}
\label{app:LIF_to_LLL}

This appendix details the explicit proof that the two-body binding energy $\gamma_{2,J}$ in~\eqref{eq:def_2bodybinding} is equal to the anomalous dimension $\gamma_{LIF}$ in~\eqref{eq:gamma_lif}. The latter gives the contribution of a $\sigma$ exchange in the $t,u$ channels of $\<\phi\phi\bar{\phi\phi}\>$ to the anomalous dimension of the double-twist operator $[\phi\phi]_{0,J}$, as predicted from the Lorentzian inversion formula.

First, in section~\ref{app:proof_geodesic_to_LCB}, we prove the formula~\eqref{eq:geodesic_to_LCblocks} that recasts the two-body binding energy into the form~\eqref{eq:two-body-binding}. The latter can then be rewritten as a ratio of two integrals:
\begin{equation}
\g_{2,J} =\frac{\int_0^\infty ds\,s^{-\De_\phi}k_{2\De_\phi+2J}\left(\frac{s}{s+1}\right) U_2(s)}{\int_0^\infty ds\,s^{-\De_\phi} k_{2\De_\phi+2J}\left(\frac{s}{s+1}\right) }=\frac{\int_0^1 \frac{dz}{z^2} \left(\frac{z}{1-z}\right)^{2-\De_\phi}k_{2\De_\phi+2J}(z) U_2\left(\frac{z}{1-z}\right)}{\int_0^1 \frac{dz}{z^2} \left(\frac{z}{1-z}\right)^{2-\De_\phi}k_{2\De_\phi+2J}(z)},
\end{equation}
where we used ${}_2 F_1(h,h;2h,-s)=(1+s)^{-h}{}_2 F_1(h,h;2h;s/(s+1))$ in the first equality and changed variables to $s=z/(1-z)\leftrightarrow z=s/(s+1)$ in the last equality. In this form, we see some first similarities with $\g_{LIF}$, except that $s^{\De_\phi} k_{\De_\s}^{(d)}(1-z)$ and its multiplicative prefactor are replaced with $s^{2-\De_\phi} U_2(s)$ in the numerator, while $s^{\De_\phi}$ is replaced with $s^{2-\De_\phi}$ in the denominator. The multiplicative prefactor is proportional to the square of the OPE coefficient $C_{\phi\phi\s}$, which is given by~\eqref{eq:Cphiphisigma} to leading order in $\l$, as we review in section~\ref{app:Cphiphisigma}. Since $U_2(s)$ is defined by an expansion into lightcone blocks $k_{\De_\s+2n}(1-z)$ in~\eqref{eq:scalar_potential_result}, we determine the same expansion for $k_{\De_\s}^{(d)}(1-z)$ in section~\ref{app:proof_2dblockdec}. The result is
\begin{equation}
	k_{\De_\s}^{(d)} = \sum_{n=0}^\infty \mathcal{B}_n \,k_{\De_\s+2n},\quad \mathcal{B}_n= \frac{(d/2-1)_n}{n! (\De_\s-d/2+1)_n} \frac{(\De_\s/2)_n^2}{(\De_\s+n-1)_n},
	\label{eq:dd_to_2d}
\end{equation}
where $(a)_k:=\G(a+k)/\G(a)$. After swapping the integrals over $z$ with the sum over lightcone blocks, the equality $\g_{2,J}=\g_{LIF}$ then reduces to 
\begin{equation}
\sum_{n=0}^\infty b_n \frac{\Omega_{\De_\phi+J,\De_\s/2,2-\De_\phi}}{\Omega_{\De_\phi+J,0,2-\De_\phi}}= -2\frac{C_{\phi\phi\s}^2 \G(\De_\s)}{\G(\De_\s/2)^2} \frac{\sin^2 \pi(\De_\phi-\De_\s/2)}{\sin^2\pi\De_\phi}\sum_{n=0}^\infty \mathcal{B}_n \frac{\Omega_{\De_\phi+J,\De_\s/2,\De_\phi}}{\Omega_{\De_\phi+J,0,\De_\phi}},
\label{eq:lll_to_lif_allorders}
\end{equation}
where 
\begin{equation}
	\Omega_{h,h',p} = \Omega_{h',h,2-p} = \int_0^1 \frac{dz}{z^2} \left(\frac{z}{1-z}\right)^p k_{2h}(z) k_{2h'}(1-z).
	\label{eq:Omega_h1h2p}
\end{equation}
In fact, the equality~\eqref{eq:lll_to_lif_allorders} holds order-by-order in each summand $n=0,1,2,\dots$ if the hypergeometric integral~\eqref{eq:Omega_h1h2p} satisfies the identity
\begin{equation}
	\frac{\Omega_{h_1,h_2,p}}{\Gamma(h_1+p-1)^2\Gamma(h_2+1-p)^2} = \frac{\Omega_{h_1,h_2,2-p}}{\Gamma(h_1+1-p)^2\Gamma(h_2+p-1)^2}.
\end{equation}
We prove this identity in section~\ref{app:hyp_id}.
\subsection{Proof of \eqref{eq:geodesic_to_LCblocks}}
\label{app:proof_geodesic_to_LCB}
To prove this formula, it will be useful to rewrite the integral in terms of the future-directed unit hyperboloid in Minkowski space:
\begin{equation}
	\H^+_ := \{ X \in \R^{2,1}\,|\, X^2=-1, \, X^t>0\}, \quad X^2:= -(X^t)^2+(X^x)^2+(X^y)^2.
\end{equation}
The latter is isomorphic to the hyperbolic disk via stereographic projection:
\begin{equation}
X(\a_k):=X_k = \frac{2}{1-|\a_k|^2}\left(\frac{1+|\a_k|^2}{2},\Re\a_k,\Im\a_k\right).
\end{equation}
The Lorentz-invariant measure on the hyperboloid must agree with the $\mathfrak{so}(2,1)$-invariant measure $d^2\a/(1-|\a|^2)^2$ on the disk up to a multiplicative constant, which is easily checked to be
\begin{equation}
\int_{\R^{2,1}} \widetilde{dX}_k:=\frac{1}{2} \int_{\R^{2,1}} d^3 X\, \de(X^2+1)\, \theta(X^0) = \int_{\D} \frac{d^2\a}{(1-|\a|^2)^2}.
\end{equation}
Moreover, scalar products on $\H^+$ are related to two-point invariants on the disk as follows:
\begin{align}
&C(X_1,X_2):=\frac{1-X_1\cdot X_2}{2}=\frac{(1-\a_1\bar\a_2)(1-\bar\a_1\a_2)}{(1-\a_1\bar\a_1)(1-\a_2\bar\a_2)}:=\cosh^2 \bd_{12}/2, \\
&S(X_1,X_2):=-\frac{1+X_1\cdot X_2}{2}=\frac{(\a_1-\a_2)(\bar\a_1-\bar\a_2)}{(1-\a_1\bar\a_1)(1-\a_2\bar\a_2)}:=\sinh^2\bd_{12}/2,
\end{align}
where $\bd_{12}$ is the hyperbolic distance on the disk. After introducing the point $\a_0:=0$ on the disk and the vector $X(0):=X_0=(1,0,0)$ on the hyperboloid, we can recast the left-hand-side of~\eqref{eq:geodesic_to_LCblocks} into the following form:
\begin{align}
\int_{\D^2} \prod_{k=1}^2 d^2\a_k (1-\a_k\bar\a_k)^{\De_k-2} \de(s-s_{12}) &= \int_{(\H^+)^2} \prod_{k=2}^2 \widetilde{dX}_k \, C(X_k,X_0)^{-\De_k} \,\de(s-S(X_1,X_2)) \nonumber \\
& =: G(s;X_0).
\end{align}
From Lorentz invariance of the integral, it is easy to check that $G(s;X_0)$ is invariant under any Lorentz transformation $X_0\mapsto \Lambda X_0$, $\Lambda \in SO(1,2)$. Since $\H^+$ is homogeneous for $SO(1,2)$, Lorentz invariance implies that $Y\mapsto G(s;Y)$ is a constant function. After integrating against $\widetilde{dY}$, we obtain\footnote{Here we define $\vol\H^+:= \int \widetilde{dX}$.}
\begin{align}
G(s;X_0) &= \frac{1}{\vol(\H^+)} \int_{(\H^+)^2} \prod_{k=1}^2 \widetilde{dX}_k \,\de(s-S(X_1,X_2)) \int_{\H^+} \widetilde{dY} \prod_{k=1}^2 C(X_k,Y)^{-\De_k} \\
&= \frac{1}{\vol(\H^+)} \int_{(\H^+)^2} \prod_{k=1}^2 \widetilde{dX}_k\, \de(s-s(X_1,X_2)) \,g(s(X_1,X_2)).
\end{align}
The integral $g(s(X_1,X_2))$ is manifestly Lorentz-invariant, and must therefore be a function of its unique non-trivial two-point invariant $X_1\cdot X_2= -2S-1=1-2C$. We will compute its explicit form shortly. In the meantime, as the integral over $(X_1,X_2)$ localizes to $s=s(X_1,X_2)$,  we can factor out $g(s)$ and compute
\begin{equation}
\frac{G(s;X_0)}{g(s)} =  \frac{1}{\vol(\H^+)} \int_{(\H^+)^2} \prod_{k=1}^2 \widetilde{dX}_k\, \de(s-s(X_1,X_2)).
\end{equation}
There always exists a Lorentz transformation $\L$ such that $X_2=\L X_0$. After the change of variables $X_1\rightarrow \L X_1$, we can then factor out the integral over $X_2$ into $\vol(\H^+)$ to obtain
\begin{equation}
\frac{G(s;X_0)}{g(s)} = \int_{\H^+} \widetilde{dX}_1 \, \de(s-s(X_1,X_0)) = \int_{\D} \frac{d^2\a_1}{(1-|\a_1|^2)^2} \,\de\left(s-\frac{|\a_1|^2}{1-|\a_1|^2}\right) = \pi.
\end{equation}

We now finish the proof by computing $g(s)$.  After introducing the stereographic projection $Y:= X(\b)$, we can re-express $g(s(X_1,X_2))$ as an integral over the disk. Using Lorentz invariance to set $X_2=X(0)=X_0$, we obtain 
\begin{equation}
g\left(\frac{|\a_1|^2}{1-|\a_1|^2}\right) = (1-\a_1\bar\a_1)^{\De_1} \int_{\D} d^2\b\,(1-\b\bar\b)^{\De_1+\De_2-2} (1-\a_1\bar \b)^{-\De_1}(1-\bar\a_1\b)^{-\De_1}.
\end{equation}
We now expand the factor $(1-\a_1\bar\b)^{-\De_1}$ and its complex conjugate into the binomial series
\begin{equation}
(1-x)^{-\De}=\sum_{k=0}^\infty \frac{(\De)_k}{k!} x^k,\quad |x|<1, \quad (\De)_k:=\frac{\Gamma(\De+k)}{\Gamma(\De)},
\end{equation}
and apply the orthogonality relation
\begin{equation}
\frac{\De-1}{\pi} \int_{\D} d^2\b \,(1-|\b|^2)^{\De-2}\, \bar\b^m \b^n = \de_{mn} \frac{k!}{(\De)_k},
\end{equation}
with $\De=\De_1+\De_2$, onto each monomial. We obtain in this way
\begin{equation}
g\left(\frac{|\a_1|^2}{1-|\a_1|^2}\right) =\frac{\pi}{\De_1+\De_2-1} (1-|\a_1|^2)^{\De_1}\, {}_2F_1(\De_1,\De_1;\De_1+\De_2;|\a_1|^2).
\end{equation}
Note that we would have obtained the exact same function with $(\a_1,\De_1)\leftrightarrow (\a_2,\De_2)$ if we had set $X_1=X_0$ instead of $X_2=X_0$. This permutation symmetry becomes manifest after applying the Pfaff transformation of the Gauss hypergeometric function:
\begin{equation}
{}_2F_1\left(\De_1,\De_1;\De_1+\De_2;\frac{z}{z-1}\right) = (1-z)^{\De_1} {}_2F_1(\De_1,\De_2;\De_1+\De_2;z).
\end{equation}
Using $G(s;X_0) = \pi g(s)$ and $|\a_1|^2=s/(s+1)$, we then obtain~\eqref{eq:geodesic_to_LCblocks}.

\subsection{Bulk coupling and boundary OPE coefficient}
\label{app:Cphiphisigma}
At leading order in perturbation theory,  the relation between the cubic bulk coupling constant $\l \Phi_1 \Phi_2 \Phi_3$ and the OPE coefficient $C_{\phi_1\phi_2\phi_3}$ follows from the tree-level approximation to the CFT three-point function:
\begin{equation}
	C_{\phi_1\phi_2\phi_3} \prod_{1\leq i<j\leq 3} (-2P_i\cdot P_j)^{-\de_{ij}} = \l \int d^{d+1} x \sqrt{-g} \prod_{i=1}^3 K_{\De_i}(X,P_i) + O(\l^2),
\end{equation}
where $\de_{12}:= \frac{\De_1+\De_2-\De_3}{2}$, and $\de_{23}$,  $\de_{13}$ are cyclic permutations thereof. The AdS integral on the right-hand side, corresponding to a scalar three-point contact diagram, was computed explicitly in e.g.~\cite[eq.~(3.7)]{Paulos:2011ie}.  In this paper, the relevant cubic coupling is $\phi_1=\phi_2=\phi$, $\phi_3=\s$, such that
\begin{equation}
	C_{\phi\phi\s}=  \l \frac{\pi^{d/2} }{2} \G\left(\frac{2\De_\phi+\De_\s-d}{2}\right) \frac{C_{\De_\phi,d}}{\G(\De_\phi)^2} \frac{C_{\De_\s,d}^{1/2}}{\G(\De_\s)} \G\left(\frac{\De_\s}{2}\right)^2\G\left(\De_\phi-\frac{\De_\s}{2}\right).
\end{equation}

\subsection{Expansion of $k_{\De_\s}^{(d)}$ into lightcone blocks}
\label{app:proof_2dblockdec}
We would like to prove the relation~\eqref{eq:dd_to_2d}, which can be written as
\begin{align}
&{}_2 F_1(h,h;2h-\varepsilon;z) = \sum_{n=0}^\infty \mathcal{B}_n^{h,\varepsilon} z^n {}_2 F_1(h+n,h+n;2h+2n;z),\label{eq:id_logblock}\\ 
&\mathcal{B}_n^{h,\varepsilon}= \frac{1}{n!} \frac{(\varepsilon)_n}{(2h-\varepsilon)_n} \frac{(h)_n^2}{(2h+n-1)_n},
\label{eq:Bn_logblock}
\end{align}
for $h=\De_\s/2$ and $\varepsilon=(d-2)/2$. In a power series expansion around $z=0$, it is easy to check that the equality holds at the first orders. To show the equality at all orders, we expand the right-hand side into a double sum:
\begin{equation}
\sum_{n=0}^\infty \mathcal{B}_n^{h,\varepsilon} z^n {}_2 F_1(h+n,h+n;2h+2n;z)= \sum_{n,k=0}^\infty \mathcal{B}_n^{h,\varepsilon} \frac{(h+n)_k^2}{k! (2h+2n)_k} z^{n+k}.
\end{equation}
We then change the summation variables to $\ell:=n+k\in \N$ and $n=0,1,\dots,\ell$. After some manipulation of Pochhammer symbols, the sum over $n$ at fixed $\ell$ turns into a ${}_4 F_3$ and the previous expression simplifies to
\begin{equation*}
\sum_{n,k=0}^\infty \mathcal{B}_n^{h,\varepsilon} \frac{(h+n)_k^2}{k! (2h+2n)_k} z^{n+k} = \sum_{\ell=0}^\infty \frac{(h)_\ell^2 z^\ell}{\ell! \,(2h)_{\ell}} {}_4 F_3\left(-\ell,\varepsilon,2h-1,h+1/2;2h-\varepsilon,2h+\ell,h+1/2;-1\right)
\end{equation*}
For $\ell\in \N$, this last ${}_4 F_3$ reduces to
\begin{equation}
{}_4 F_3\left(-\ell,\varepsilon,2h-1,h+1/2;2h-\varepsilon,2h+\ell,h+1/2;-1\right) = \frac{(2h)_\ell}{(2h-\varepsilon)_\ell},
\end{equation}
thereby establishing the identity~\eqref{eq:id_logblock}.

\subsection{Hypergeometric identity for pairing of lightcone blocks}
\label{app:hyp_id}
In this appendix, we will prove the identity
\begin{equation}
	\frac{\Omega_{h_1,h_2,p}}{\Gamma(h_1+p-1)^2\Gamma(h_2+1-p)^2} = \frac{\Omega_{h_1,h_2,2-p}}{\Gamma(h_1+1-p)^2\Gamma(h_2+p-1)^2},
	\label{eq:app:Omega_identity}
\end{equation}
where $\Omega_{h,h',p}$ was defined in eq.~\eqref{eq:Omega_h1h2p}.  This relation can be seen as a generalization of the $a\leftrightarrow b$ permutation symmetry for the Euler representation of the Gauss hypergeometric function:
\begin{equation}
	\int_0^1 \frac{dz}{z(1-z)} \frac{z^b(1-z)^{c-b}}{\mathrm{B}(b,c-b)}(1-x z)^{-a}= \int_0^1 \frac{dz}{z(1-z)} \frac{z^a(1-z)^{c-a}}{\mathrm{B}(a,c-a)} (1-x z)^{-b}.
\end{equation}
Here, we begin with a similar Euler-type integral:
\begin{align}
	& \frac{\Omega_{h_1,h_2,p}}{\mathrm{B}(h_1+p-1,h_2+1-p)} = \\
	& \int_0^1 \frac{dz}{z(1-z)} \frac{z^{h_1+p-1}(1-z)^{h_2+1-p}}{\mathrm{B}(h_1+p-1,h_2+1-p)} {}_2 F_1\begin{bmatrix}
		h_1,h_1\\2h_1
	\end{bmatrix}(z)\,\,{}_2F_1\begin{bmatrix}
		h_2,h_2\\2h_2
	\end{bmatrix}(1-z).\nonumber
\end{align}
Expanding each of the two Gauss hypergeometric functions into a power series, we obtain the following double-sum:
\begin{equation}
	\frac{\Omega_{h_1,h_2,p}}{\mathrm{B}(h_1+p-1,h_2+1-p)} = \prod_{i=1}^2 \sum_{n_i=0}^\infty \frac{(h_i)_{n_i}^2}{n_i! (2h_i)_{n_i}} \frac{(h_1+p-1)_{n_1} (h_2+1-p)_{n_2}}{(h_1+h_2)_{n_1+n_2}}.
\end{equation}
This same double-sum admits an alternative but equivalent integral representation:
\begin{align}
	& \frac{\Omega_{h_1,h_2,p}}{\mathrm{B}(h_1+p-1,h_2+1-p)} =\label{eq:app:intrep_alt} \\
	& \int_0^1 \frac{dz}{z(1-z)} \frac{z^{h_1}(1-z)^{h_2}}{\mathrm{B}(h_1,h_2)} {}_2 F_1\begin{bmatrix}
		h_1+p-1,h_1\\2h_1
	\end{bmatrix}(z)\,\,{}_2F_1\begin{bmatrix}
		h_2+1-p,h_2\\2h_2
	\end{bmatrix}(1-z).
	\nonumber
\end{align}
After applying te Euler transformation of the Gauss hypergeometric function,
\begin{equation}
	{}_2 F_1\begin{bmatrix}
		a,b\\c\end{bmatrix}(z) = (1-z)^{c-a-b}{}_2 F_1\begin{bmatrix}
		c-a,c-b\\c\end{bmatrix}(z),
\end{equation}
we can rewrite the equality~\eqref{eq:app:intrep_alt} as
\begin{align}
	& \frac{\Omega_{h_1,h_2,p}\,\mathrm{B}(h_1,h_2)}{\mathrm{B}(h_1+p-1,h_2+1-p)^2} = \\
	& \int_0^1 \frac{dz}{z(1-z)} \frac{z^{h_1+p-1}(1-z)^{h_2+1-p}}{\mathrm{B}(h_1+p-1,h_2+1-p)} {}_2 F_1\begin{bmatrix}
		h_1+1-p,h_1\\2h_1
	\end{bmatrix}(z)\,\,{}_2F_1\begin{bmatrix}
		h_2+p-1,h_2\\2h_2
	\end{bmatrix}(1-z).\nonumber
\end{align}
We obtain yet another double-sum after expanding each Gauss hypergeometric series, only now the coefficients are manifestly symmetric under $h_1\leftrightarrow h_2$, or equivalently under $p\leftrightarrow 2-p$:
\begin{equation}
	\frac{\Omega_{h_1,h_2,p}\,\mathrm{B}(h_1,h_2)}{\mathrm{B}(h_1+p-1,h_2+1-p)^2}  = \prod_{i=1}^2 \sum_{n_i=0}^\infty  \frac{(h_i)_{n_i} (h_i+p-1)_{n_i}(h_i+1-p)_{n_i}}{n_i! (2h_i)_{n_i}}.
\end{equation}
The identity~\eqref{eq:app:Omega_identity} then follows from the invariance of this expression under $p\rightarrow 2-p$.

\section{Large-spin expansion of the integrals over orbits}
\label{app:orbit_integrals}
In this appendix, we perform the explicit computation of $M_J(\a)$ in~\eqref{eq:Mexpression} and $\cU_{N,J}$ in~\eqref{eq:VeffJexpresion} to leading orders at large spin $J$ in the case of $N=3$. After recalling below some general properties of these integrals at large $J$, we detail the derivation of~\eqref{eq:Macute}, \eqref{eq:acute_Veff} for the acute region and~\eqref{eq:Mobtuse}, \eqref{eq:obtuse_r} for the obtuse region in the subsequent subsections.  

First, note that if $U_N(e^{i\varphi}\a)=U_N(\a)$ for $\varphi\in[0,2\pi)$, then the integrals are invariant under phase shifts of $\l$, such that we can integrate out $\arg(\l)$ into a factor of $2\pi$ and reduce the domain of integration to $(|\l|,\b)\in\R_{>0} \times \C$. Next, due to the factor of $|\l|^{-2J}$, it is easy to see that the integral at large $J$ is dominated by $(|\l|,\b)=(r(\a),b(\a))$, where $r$ and $b$ are the radius and center of the smallest enclosing circle of $(\a_1,\a_2,\a_3)$ in the complex plane. Below, we will show that an expansion of the integrands as
\begin{equation}
(|\l|,\b)=(r(\a),b(\a))+\e (\de|\l|,\de\b) + O(\e^2),
\label{eq:lambdabeta_to_omegaa}
\end{equation}
translates into the large-spin expansion of the integrals for $\e\sim 1/J$. This reproduces the formulas of section~\ref{sec:dmu_h1_cU}. 

The smallest circle depends on whether the three points form an acute or obtuse triangle, which separates $\CP^1$ into two regions. We will analyze the acute and obtuse regions separately, and show that they produce different large-spin expansions. 

\subsection{Acute region: leading order}
If $(\a_1,\a_2,\a_3)$ form an acute triangle, then all points lie on the smallest circle. The $\C^\times\ltimes\C$ transformation $\a_k\rightarrow \xi_k$, where
\begin{equation}
\xi_k=\frac{\a_k-b(\a)}{r(\a)} = 1/\bar\xi_k,
\end{equation}  
maps each point to the unit circle. One retrieves the functions $M_J,\cU_{3,J}$ in a generic acute configuraton from their values on the unit circle by the covariance relations
\begin{equation}
M_J(\a)=r(\a)^{2-2N-2J}M_J\left(\xi\right),\quad \cU_{3,J}(\a)=\cU_{3,J}\left(\xi\right).
\end{equation}

Now, if $(\xi_1,\xi_2,\xi_3)$ lie on the smallest circle, then $(r(\xi),b(\xi))=(1,0)$ by definition and the quantities $r_i' := 1-\l^{-2} |\xi_i-\b|^2$ are small in the neighborhood of $(|\l|,\b)=(1,0)$.  This yields a linear map between $(\de|\l|,\de\b,\de\bar \b)$ in~\eqref{eq:lambdabeta_to_omegaa} and $(r_1',r_2',r_3')$:
\begin{equation*}
	r_i'=\e(2\de|\l|+\xi_i^{-1}\de\b +\xi_i\,\de\bar \b)+O(\e^2),\quad i=1,2,3.
\end{equation*}
The linear map is straightforward to invert, and we obtain in particular
\begin{equation}
	\label{eq:omega_acute}
\de|\l| = \frac{\e}{2} \sum_{k=1}^3 r_k' R_k(\xi) + O(\e^2),
\end{equation}
where $R_k$ is given by~\eqref{eq:Rfunction}. The Jacobian for the measure turns into
\begin{equation}
\label{eq:d3r_acute}
d|\l| d^2\b = \e^3 d\de|\l| d^2\de\b +O(\e^4) = \frac{1}{4}  \left|\frac{\xi_1\xi_2\xi_3}{\xi_{12}\xi_{23}\xi_{31}}\right|d^3r'(1+O(\e)).
\end{equation}
Together, formulas~\eqref{eq:omega_acute} and~\eqref{eq:d3r_acute} recast $M_J,\cU_{3,J}$ into integrals over $r_i'$. Starting with $M_J$ in~\eqref{eq:Mexpression}, we can apply the change of variables to obtain
\begin{align*}
M_J(\xi)= \frac{(\De_\phi-1)^3}{12\pi^2}\left|\frac{\xi_1\xi_2\xi_3}{\xi_{12}\xi_{23}\xi_{31}}\right| \int d^3r' \,\prod_{i=1}^3  r'^{\De_\phi-2}_i\left(1+\frac{1}{2}\sum_{i=1}^3 r_i'R_i(\xi)+O(\e^2)\right)^{-(2J+2N+1)}.
\end{align*}
If the expansion parameter of the integrand goes like $r_i'\sim\e\sim J^{-1}$, then the integral reduces to a product of Gamma functions after a rescaling of $r_i'$ by $J$, and we obtain the formula~\eqref{eq:Macute} for $M_J(\xi)$ in the acute region. 

Moving on to $\cU_{3,J}$ defined by~\eqref{eq:VeffJexpresion}, note that the expansion of the integrand at $r_i'=O(\e)$ induces a large-distance expansion of the potential $U_N$. We assume that its leading large-distance behavior is a symmetrized sum of pair potentials $U_2(s_{ij})$ with leading asymptotics $U_2(s\gg 1) = b_0 \,s^{-\De_\s/2}(1+O(s))$. In this case, the $U_N$ factor in the integrand of~\eqref{eq:VeffJexpresion} can be expanded as 
\begin{equation}
U_3\left(\frac{\xi-\b}{\l}\right)=U_3\left(\frac{\xi-\e\,\de\b}{1+\e\,\de|\l|+O(\e^2)}\right) = b_0\sum_{1\leq i<j\leq 3}\frac{(r_i'r_j')^{\De_\s/2}}{|\xi_{ij}|^{\De_\s}} + O(\e^{\De_\s+2}),
\label{eq:U3_largeS}
\end{equation}
such that the overall integral takes the form
\begin{align*}
	M_J(\xi)\, \cU_{3,J}(\xi)=b_0 \frac{(\De_\phi-1)^3}{12\pi^2}\left|\frac{\xi_1\xi_2\xi_3}{\xi_{12}\xi_{23}\xi_{31}}\right| \int d^3r' \,\prod_{i=1}^3  r'^{\De_\phi-2}_i e^{-Jr_i' R_i(\xi)} \sum_{1\leq i<j\leq 3}\frac{(r_i'r_j')^{\De_\s/2}}{|\xi_{ij}|^{\De_\s}}  \left(1+O(\e)\right).
\end{align*}
The factors $(r_i'r_j')^{\De_\s/2}$ shift two out of the three Gamma function integrals, leading to the expression~\eqref{eq:acute_Veff} at leading order in the large-spin limit. 

\subsection{Acute region: first subleading order}
From the leading-order analysis of the previous section, we expect that the $1/J$ expansion of the functions $M_J(\xi)$, $\cU_{3,J}(\xi)$ can be efficiently computed via the change of variables $(|\l|,\b)\rightarrow (r_i')_{i=1,2,3}$ and the expansion of the integrand as $r_i'=O(1/J)$. At fixed order, this expansion culminates in a linear combination of Gamma function integrals of the form
\begin{equation}
\cI\left[\prod_{i=1}^3 y_i^{\nu_i}\right]:=\int_{\R_+^3}d^3y \prod_{i=1}^3  y_i^{\De_\phi-2+\nu_i} e^{-y_i R_i} =  \prod_{i=1}^3\Gamma(\De_\f+\nu_i-1) R_i^{-(\De_\phi-1+\nu_i)},
\label{eq:cI_Gamma}
\end{equation}
where $y_i:=r_i'J=O(1)$ and we introduced for future use the linear functional
\begin{equation}
\mathcal{I}[f]:=\int_{\R_+^3}d^3 y\prod_{i=1}^3 y_i^{\De_\phi-2+\nu_i}e^{-y_iR_i} f(y).
\end{equation}

For the next-to-leading order in $1/J$, we need to invert the relation $r_i'(|\l|,\b)=1-|\l|^{-2}|\xi_i-\b|^2$ to second order around $(|\l|,\b)=(1,0)$. This yields an expansion of the scale factor and Jacobian of the following form:
\begin{align*}
	& |\l|=1+\frac{1}{2}\sum_{i=1}^3 R_i r_i' + \sum_{1\leq i, j\leq 3} \L^{(2)}_{ij}r_i'r_j'+O(1/J^3),\\
	& d |\l| d^2\b =  \frac{1}{4}  \left|\frac{\xi_1\xi_2\xi_3}{\xi_{12}\xi_{23}\xi_{31}}\right|d^3r' \left(1+ \sum_{i=1}^3\cJ^{(1)}_i r_i' + O(1/J^2)\right).
\end{align*}
In terms of this data, the integrand of $M_J$ will have the expansion
\begin{align}
&\frac{d|\l|d^2\b}{|\l|^{2J+2N+1}} \prod_{i=1}^3 r_i'^{\De_\phi-2} = \prod_{i=1}^3 dr_i' r_i'^{\De_\phi-2}\,e^{-J R_i(\xi)r_i'} \left(1+J^{-1}\cM^{(1)}(Jr',\xi)+O(J^{-2})\right), \\
&\cM^{(1)}(y,\xi) =\sum_{i=1}^3 \left(\cJ^{(1)}_i(\xi)-4 R_i(\xi)\right)y_i + \sum_{1\leq i, j\leq 3} \left(\frac{1}{4} R_i(\xi)R_j(\xi)-2\L^{(2)}_{ij}(\xi)\right)y_iy_j. \nonumber
\end{align}
As a result, its large-spin expansion to subleading order is
\be
M_J(\xi) = \frac{J^{6-3\De_\f}(\De_\phi-1)^3}{12\pi^2}\left|\frac{\xi_1\xi_2\xi_3}{\xi_{12}\xi_{13}\xi_{23}}\right| \left(\cI[1]+J^{-1}\cI[\cM^{(1)}]+O(J^{-2})\right),
\label{eq:MJ_nlo}
\ee
where the action of the linear functional $\cI$ on a power series in $y_i$ follows from~\eqref{eq:cI_Gamma}.

The above derivation generalizes readily to the subleading correction of $\cU_{3,J}(\xi)$, defined from the integral~\eqref{eq:acute_Veff}. To retrieve this result, note that the corrections to the leading asymptotics of $U_N((\xi-\b)/\l)$ at $r_i'\sim 1/J$ are of relative order $1/J^2$, and therefore subleading to the $1/J$ corrections in~\eqref{eq:VeffJexpresion}. After inserting the leading-order form~\eqref{eq:U3_largeS} of $U_3$ at $r_i'=O(1/J)$, we obtain
\begin{align}
M_J(\xi)\,\cU_{3,J}(\xi)=  &\frac{(\De_\phi-1)^3}{12\pi^2 J^{3\De_\f+\De_\s-6}}\left|\frac{\xi_1\xi_2\xi_3}{\xi_{12}\xi_{13}\xi_{23}}\right|\times \nonumber \\
& \sum_{1\leq i<j\leq 3}|\xi_{ij}|^{-\De_\s} \left(\cI[(y_iy_j)^{\De_\s/2}]+J^{-1}\cI[(y_iy_j)^{\De_\s/2}\cM^{(1)}]+O(J^{-2})\right),
\label{eq:MJU3J_nlo}
\end{align}
where the functionals can again be evaluated from~\eqref{eq:cI_Gamma}. Dividing~\eqref{eq:MJU3J_nlo} by~\eqref{eq:MJ_nlo} and expanding to the subleading order $J^{-\De_\s-1}$, we finally obtain the full formula~\eqref{eq:acute_Veff}.

\subsection{Obtuse region: leading order}
If $(\a_1,\a_2,\a_3)$ form an obtuse triangle, then only two out of three points lie on the smallest circle; without loss of generality, we assume the latter two are $\a_1,\a_2$. In this case, the radius and center of the smallest circle are $(r(\a),b(\a))=(|\a_1-\a_2|,\a_1+\a_2)/2$. The $\C^\times\ltimes\C$ transformation $(\a_1,\a_2,\a_3)\rightarrow (1,-1,z)$ such that $(r,b)\rightarrow (1,0)$ is then given by
\begin{equation}
\a_3\rightarrow z  = \frac{2\a_3-\a_1-\a_2}{\a_1-\a_2}.
\end{equation}
In this gauge, the smallest circle is the unit circle which must enclose the third point, i.e. $|z|< 1$. We can again reconstruct the functions at a generic obtuse triangle configuration from $(1,-1,z)$ via the covariance relations
\begin{align}
M_J(\a)= \left|\frac{\a_1-\a_2}{2}\right|^{2-2N-2J} M_J(1,-1,z), \quad \cU_{3,J}(\a)=\cU_{3,J}(1,-1,z).
\end{align}

Consider now the integral $M_J(1,-1,z)$ in~\eqref{eq:Mexpression}. The integrand is a product of powers $r_i'^{\De_\phi-2}$, where
\begin{equation}
r_1'=1-\left|\frac{1-\b}{\l}\right|^2,\quad r_2'=1-\left|\frac{1+\b}{\l}\right|^2, \quad r_3'=1-\left|\frac{z-\b}{\l}\right|^2.
\end{equation}
In these expressions, it is only for $i=1,2$ that $r_i' \rightarrow 0$ as $(|\l|,\b)\rightarrow (1,0)$, while $r_3'\rightarrow 1-|z|^2$ is finite. At leading order in the expansion~\eqref{eq:omega_acute}, the integrand therefore factorizes as 
\begin{equation*}
M_J(1,-1,z)=\frac{(\De_\phi-1)^3}{3\pi^2} \int\frac{ d^2\l d^2\b}{|\l|^{2(J+4)}} \left(1-\left|\frac{1-\b}{\l}\right|^2\right)_+^{\De_\phi-2}\left(1-\left|\frac{1+\b}{\l}\right|^2\right)_+^{\De_\phi-2} (1-|z|^2)^{\De_\phi-2}(1+O(\e)).
\end{equation*}
To compute the integral in this approximation, we make the change of variables
\begin{equation}
(\l^{-1},\l^{-1}\b):=\frac{1}{2}(\a_2'-\a_1' ,\a_1'+\a_2'),\quad \frac{ d^2\l d^2\b}{|\l|^{2(J+N+1)}} = \frac{d^2\a_1'd^2\a_2'}{4^{J+1}} |\a_1'-\a_2'|^{2(J+1)},
\end{equation}
after which $r_i'=1-|\a_i'|^2$ for $i=1,2$. We can then rewrite the leading-order integral as
\begin{align*}
M_J(1,-1,z) &= \frac{(\De_\phi-1)^3}{12\pi^24^J} \int_{\D^2} \prod_{i=1}^2 d^2\a_i' (1-|\a_i'|^2)^{\De_\f-2} |\a_1'-\a_2'|^{2(J+1)} (1-|z|^2)^{\De_\phi-2} +\dots \\
&= \frac{\De_\phi-1}{12\,4^J} \<\psi_{J+1}|\psi_{J+1}\> (1-|z|^2)^{\De_\f-2}+\dots,
\end{align*}
where $\psi_J(\a_1',\a_2')=(\a_1'-\a_2')^J$ is the unique two-particle, minimal-twist, primary wavefunction at spin $J$ (up to a multiplicative constant). Its norm-squared $\<\psi_J|\psi_J\>$ can be computed using the methods of section~\ref{sec:binding_energies}, where it is mapped to the integral $\<s^J\>_{\De_\phi+J,\De_\phi+J}$ defined by~\eqref{eq:favg_formula}. In these consecutive changes of variables $(\l,\b)\rightarrow (\a_1',\a_2')\rightarrow s$, we can keep track of the original expansion of the integrand by noting that the two-point invariant $s$ diverges as
\begin{equation}
s=\frac{|\a_1'-\a_2'|^2}{(1-|\a_1'|^2)(1-|\a_2'|^2)} = \frac{4}{(|\l|^2-|1-\b|^2)(|\l|^2-|1+\b|^2)} = O(1/\e^2),
\end{equation} 
when $(|\l|,\b)=(1,0)+O(\e)$. We then retrieve the large-$J$ expansion of the norm-squared by setting $\e\sim 1/\sqrt{s}\sim 1/J$. Assuming this scaling applies for the full expansion of the integrand, we find 
\begin{equation}
M_J(1,-1,z)=\frac{(\De_\phi-1)^3}{12\pi^2} 4^{-J} \<s^{J+1}\>_{\De_\phi+J+1,\De_\phi+J+1} (1-|z|^2)^{\De_\f-2}\left(1+O(1/J)\right). 
\end{equation}
The function $\<s^J\>_{\De_1,\De_2}$, defined from the integral~\eqref{eq:favg_formula}, can be computed explicitly from the Mellin-Barnes representation of the hypergeometric function:
\begin{equation}\label{eq:sJavg}
\<s^J\>_{\De_1,\De_2}= \pi^2 \frac{\Gamma(J+1)\Gamma(\De_1-J-1)\Gamma(\De_2-J-1)\Gamma(\De_1+\De_2-1)}{\Gamma(\De_1)\Gamma(\De_2)\Gamma(\De_1+\De_2+J-1)}
\end{equation}
Its leading large-spin expression then follows from the Stirling formula for the Gamma functions. 

We now use the same procedure to determine $\cU_{3,J}(1,-1,z)$ in the obtuse region. Specifically, in the integral~\eqref{eq:VeffJexpresion}, the function $U_3((\a-\b)/\l)$ is expanded around $r_1',r_2'=0$ and $r_3'=1-|z|^2$ in the limit $(|\l|,\b)\rightarrow (1,0)$. In terms of hyperbolic distances, this limit translates to $s_{13},s_{23}=O(\e^{-1})$ and $s_{12}=O(\e^{-2})$. If we again assume a decomposition into a sum of pair potentials $U_2(s_{ij})$ with leading asymptotics $U_2(s)=b_0s^{-\De_\s/2}(1+O(1/s))$, then the function $U_N$ takes the form
\begin{align}
U_3\left(\frac{1-\b}{\l},\frac{-1-\b}{\l},\frac{z-\b}{\l}\right) =& b_0\left[\left(r_1' \frac{1-|z|^2}{|1-z|^2}\right)^{\De_\s/2} +\left(r_2' \frac{1-|z|^2}{|1+z|^2}\right)^{\De_\s/2} \right]\nonumber\\&+O(\e^{\De_\s},\e^{\De_\s/2+1}).
\end{align}
Plugging this back into the integral, we find
\begin{align}
\cU_{3,J}(1,-1,z)=& b_0\frac{\<s^{J+1}\>_{\De_\phi+\De_\s/2+J+1,\De_\phi+J+1}}{\<s^{J+1}\>_{\De_\phi+J+1,\De_\phi+J+1}} \left[\left( \frac{1-|z|^2}{|1-z|^2}\right)^{\De_\s/2} +\left( \frac{1-|z|^2}{|1+z|^2}\right)^{\De_\s/2} \right] \nonumber \\&+ O(J^{-\De_\s},J^{-\De_\s/2-2})
\end{align}
Applying the large-spin expansion of the functions $\<s^J\>$ in~\eqref{eq:sJavg}, we are left with 
\begin{equation}
\frac{\<s^{J+1}\>_{\De_\phi+\De_\s/2+J+1,\De_\phi+J+1}}{\<s^{J+1}\>_{\De_\phi+J+1,\De_\phi+J+1}} = \frac{\Gamma(\De_\f+\De_\s/2-1)}{\Gamma(\De_\f-1)} 2^{-\De_\s/2} J^{-\De_\s/2}(1+O(1/J)),
\end{equation}
which yields the final expression~\eqref{eq:obtuse_Veff} for the leading asmyptotics of the effective potential in the obtuse region. 

\section{Calculus of pseudodifferential operators}
\label{app:pseudodiff}
This appendix reviews aspects of the calculus of pseudodifferential operators relevant for applications to the semiclassical analysis of sections~\ref{sec:three-body} and~\ref{sec:Nbody}. 

A general definition of pseudodifferential operators is given in section~\ref{app:def_pdo}, which is then specialized to non-integer powers of differential operators in section~\ref{app:power_do}. Next, we use pseudodifferential calculus to compute the large-spin expansion of the function $\tilde{\cU}_{N,J}(z)=J^{-\De_\s} H_\mathrm{symb}(z)$ defined by~\eqref{eq:cU_momspace} in section~\ref{app:toeplitz_momspace}. Finally, section~\ref{app:BSderivation} outlines a formal derivation of Bohr-Sommerfeld conditions in the case of one degree of freedom.

\subsection{Definition and asymptotic expansion}
\label{app:def_pdo}
We consider spaces of functions $\psi$ on $z=(z_1,\dots,z_N)\in \C^N$ that admit a WKB expansion of the form 
\begin{equation}
	\psi(z) = e^{i J S_0(z)} f(z),
\end{equation}
Then our definition of a pseudodifferential operator $\cP(z,\ptl_z)$, following closely that of H\"ormander~\cite{HormanderPDO}, is a linear and holomorphic operator whose action $e^{-i J S_0} \cP e^{iJ S_0} f$ admits an asymptotic expansion around $J=\infty$. This expansion takes the form
\begin{equation}
e^{-i J S_0} \cP e^{iJ S_0} f=\sum_{k=0}^\infty \cP_k[f,S_0] \,J^{-\tau_k},
\label{eq:P_on_S0f}
\end{equation}
where $(\tau_k)_{k}$ is a monotonically increasing sequence with $\tau_k\rightarrow\infty$ as $k\rightarrow\infty$. Linearity of $\cP$ implies that $\cP_k$ is homogeneous of degree $-\tau_k$ in $S_0$ and linear in $f$. For any $S_0,f$, the action~\eqref{eq:P_on_S0f} can be reconstructed from the case
\begin{equation} 
f\equiv 1, \quad g(z)=\<z,p_z\>:=z_1p_{z_1}+\dots+z_Np_{z_N},\quad  (\cP)_k(z,p_z):=\cP_k[1,\<z,p_z\>],
\end{equation}
where we call $(\cP)_k$ the $k$'th symbol function, and $(\cP)_0$ is often called the principal symbol. The explicit formula relating $\cP_k$ and $(\cP)_k$ is
\begin{equation}
\sum_{k=0}^\infty  J^{-\tau_k} \cP_k[f(z),S_0(z)] = \sum_{\ell=0}^\infty \sum_{a\in \N^N} \frac{1}{a!} \ptl_{p_z}^a(\cP)_\ell(z,J\ptl_z S_0(z)) (-i\ptl_w)^{a} e^{iJh_z(w)}f(w)  \Bigl\vert_{w=z},
\label{eq:symbol_reconstruction}
\end{equation}
where we introduced the multi-index notation
\begin{equation*}
a!:=a_1!\dots a_N!,\quad \ptl_{p_z}^a:= \ptl_{p_{z_1}}^{a_1}\dots \ptl_{p_{z_N}}^{a_N},
\end{equation*} 
and the function
\begin{equation}
 h_z(w):= S_0(w)-S_0(z)-\sum_{k=1}^N(w_k-z_k)\ptl_{z_k} S_0(z).
 \label{eq:def_hz}
\end{equation}
To decompose the right-hand side of~\eqref{eq:symbol_reconstruction} into an asymptotic expansion at $J=\infty$ that matches the left-hand side, note that $(\cP_k)^{(a)}$ is homogeneous of degree $-\tau_k-|a|$ in $J$, where $|a|=a_1+\dots+a_N$. This function multiplies $(-i\ptl_w)^a e^{iJ h_z(w)} f(w)\vert_{w=z}$. Using the Leibniz rule and the fact that $h_z(z)=0$, the latter turns into a polynomial of degree $|a|$ in $J$. In fact, since $\ptl_{w_k} h_z(w)=0$ at $w=z$ by virtue of~\eqref{eq:def_hz}, this polynomial is actually of degree at most $|a|/2$ in $J$. Consequently, the restriction to $\cP_k$ at order $J^{-\tau_k}$ on the left-hand side implies a truncation of the sum to $\ell < k$ and $|a|\leq 2(\tau_k-\tau_\ell)$ on the right-hand side. For this paper, we only use the first two orders in the case where $\tau_k=\De_\s+k$, yielding
\begin{equation}
\cP_0[f,S_0] = (\cP)_0(z,\ptl_z S_0) \,f,
\label{eq:P0f}
\end{equation}
and
\begin{align}
\frac{\cP_1[f,S_0]}{f} =(\cP)_1(z,\ptl_z S_0)-\frac{i}{2}\sum_{|a|=2} \ptl_{p_z}^a(\cP)_0(z,\ptl_z S_0) \ptl_z^a S_0 -i \<\ptl_p (\cP)_0(z,\ptl_z S_0),\ptl_z\>.
\label{eq:P1f}
\end{align}

\subsection{Powers of a differential operator}
\label{app:power_do}
The operators appearing in~\eqref{eq:cU_momspace} are of the form $({}^tL_{ij})^u$, where ${}^tL_{ij}$ is a second-order differential operator and $u$ is a non-integer exponent. More generally, the class of pseudodifferential operators corresponding to complex powers $\cP=\cD^u$ of a differential operator $\cD$ was studied by Seeley in~\cite{seeley1967complex}. The latter makes sense as a pseudodifferential operator if $(\cD)_0(z,\ptl S_0(z))\neq 0$ on the domain of $f,S_0$. In this case, an explicit procedure to extract the symbol functions of $\cD^u$ from the those of $\mathcal{D}$ was given in \cite{seeley1967complex}, based on the residues of the resolvent:
\begin{equation}
	(\mathcal{D}^u)_k(z,p_z)=-\oint \frac{d\l}{2\pi i} \l^u \left(\frac{1}{\cD-\l}\right)_k(z,p_z).
	\label{eq:symbols_powers_resolvent}
\end{equation}
 The symbols of the resolvent can be computed from the formula in \cite[Thm.~4.3]{HormanderPDO} for the symbol functions $(\cP\cQ)_j$ of a product $\mathcal{P}\mathcal{Q}$ of pseudodifferential operators:
\begin{equation}
	(\mathcal{P}\mathcal{Q})_j = \sum_{k+\ell+|a|=j} \frac{1}{a!} (\mathcal{Q})_k^{(a)} (-i\ptl_z)^a (\mathcal{P})_\ell.
	\label{eq:PQ_k}
\end{equation}
If $\cP=(\cD-\l)^{-1}$ and $\cQ=\cD-\l$,  then $(\cP\cQ)_j=\de_{j0}$ and we can solve~\eqref{eq:PQ_k} order-by-order for the symbol functions of $\cP$. Plugging these back into~\eqref{eq:symbols_powers_resolvent}, the contour integral over $\l$ is performed by picking up the poles and applying the residue theorem. 

We will need explicit formulas for the leading and subleading symbol functions of $\cD^u$, i.e. $k=0,1$ in~\eqref{eq:symbols_powers_resolvent}. These, in turn, depend on the leading and subleading symbol functions of the resolvent:
\begin{equation}
\left(\frac{1}{\cD-\l}\right)_0 = -\frac{1}{\l-(\mathcal{D})_0}, \quad \left(\frac{1}{\cD-\l}\right)_1 = -\frac{(\mathcal{D})_1}{(\l-(\mathcal{D})_0)^2} + i \frac{\<\ptl_{p_z}(\mathcal{D})_0,\ptl_z(\mathcal{D})_0\>}{(\l-(\mathcal{D})_0)^3}.
\end{equation}
We thus obtain a leading symbol
\begin{equation}
(\cD^u)_0=(\cD)_0^u,
\label{eq:Du_0}
\end{equation}
and a subleading symbol
\begin{equation}
(\mathcal{D}^u)_1  = u (\mathcal{D})_0^{u-1} (\mathcal{D})_1 -\frac{i}{2}u(u-1)(\mathcal{D})_0^{u-2} \< \ptl_{p_z} (\mathcal{D})_0, \ptl_z (\mathcal{D})_0 \>.
\label{eq:Du_1}
\end{equation}
\subsection{Application to Toeplitz operator in momentum space}
\label{app:toeplitz_momspace}
We can now reformulate~\eqref{eq:cU_momspace} in the language of pseudodifferential operators and apply it to the expansion~\eqref{eq:Hsymb_expansion_momspace} of the Toeplitz operator's symbol. First, the function $\tilde{M}_J$ admits a large-spin expansion of the form
\begin{equation}
\tilde{M}_J(z) = \text{const} \times e^{i J S_0(z)} f(z), \quad S_0 =i\cK,
\label{eq:MJ_to_S0f}
\end{equation}
where $\cK$ is given by~\eqref{eq:hermform_momspace} and $f$ to leading order at large $J$ is given by~\eqref{eq:meas_momspace}. Next, the operator ${}^t H_N(z,\ptl_z)$ is the symmetrization of $U_{\mathrm{Cas}}\left({}^t L_{ij}\right)$, where ${}^tL_{ij}$ in~\eqref{eq:transpose_ham_momspace} is the transpose of the quadratic Casimir operator and the function $U_{\mathrm{Cas}}$ admits a series expansion of the form
\begin{equation}
U_{\mathrm{Cas}}(L)=\sum_{n=0}^\infty b_{\mathrm{Cas},n} L^{-\De_\s/2-n}.
\label{eq:Ucas_Lexpansion}
\end{equation}
We can therefore determine the action of ${}^t H_N$ on $\tilde{M}_J$ from the action of $\cD^u$ on $e^{iJ S_0} f$, where $\cD={}^t L_{ij}$ and $u=-\De_\s/2-n$. 

The quadratic Casimir operators $L_{ij}(z,\ptl_z)$ in momentum space are given by~\eqref{eq:cas_momspace}. As second-order differential operators, their exponents are $\tau_k(L)=2-k$ and their asymptotic expansion truncates for $k>2$. The three non-zero symbol functions are
\begin{align}
& (L_{ij})_0(z,p_z) = z_iz_j(p_{z_i}-p_{z_j})^2, \\
& (L_{ij})_1(z,p_z) = i\De_\phi(z_i-z_j)(p_{z_i}-p_{z_j}), \\
&  (L_{ij})_2(z,p_z)=\De_\phi(\De_\phi-1).
\end{align}
Similarly, the symbol functions of their transpose~\eqref{eq:transpose_ham_momspace} are
\begin{align}
	& \left({}^tL_{ij}\right)_0(z,p_z) = z_iz_j(p_{z_i}-p_{z_j})^2=(L_{ij})_0(z,p_z), \\
	&\left({}^tL_{ij}\right)_1(z,p_z) = i(2-\De_\phi)(z_i-z_j)(p_{z_i}-p_{z_j}), \\
	& \left({}^tL_{ij}\right)_2(z,p_z)=(2-\De_\phi)(1-\De_\phi).
\end{align}
From the three functions above, we can fully reconstruct the action of ${}^t H_N$ on~\eqref{eq:MJ_to_S0f} using the methods of the previous two sections. For calculations in this paper, we will only use the leading $k=0$ and subleading $k=1$ symbol functions of ${}^t H_N$, with corresponding exponents $\tau_k(H)=\De_\s+k$. The leading symbol is given by
\begin{equation}
\left({}^t H_N\right)_0(z,p_z) = b_{\mathrm{Cas},0} \sum_{1\leq i<j\leq N} \left(L_{ij}\right)_0(z,p_z)^{-\De_\s/2}.
\end{equation}
Plugging this into~\eqref{eq:P0f} for $\cP={}^t H_N$, we then obtain the leading term $H_{\mathrm{symb}}^{(0)}(z)$ in the expansion~\eqref{eq:Hsymb_expansion_momspace}. Since the differential operators ${}^tL_{ij}$ have leading exponent $\tau_0(L)=2$, the subleading symbol of ${}^t H_N$ is still captured by the leading power in the large-argument expansion~\eqref{eq:Ucas_Lexpansion} of $U_\mathrm{Cas}(L)$:
\begin{equation}
	\left({}^t H_N\right)_1(z,p_z) = b_{\mathrm{Cas},0} \sum_{1\leq i<j\leq N} \left({}^t L_{ij}^{-\De_\s/2}\right)_1(z,p_z).
\end{equation}
The symbol functions on the right-hand side are obtained by application of formula~\eqref{eq:Du_1} with $\cD={}^t L_{ij}$ and $u=-\De_\s/2$. Plugging this into~\eqref{eq:P1f} for $\cP={}^t H_N$, we obtain the subleading symbol $H_{\mathrm{symb}}^{(1)}(z)$ in the expansion~\eqref{eq:Hsymb_expansion_momspace}.

The symbols of the Toeplitz operator at higher orders $k>1$ will also depend on the coefficients $b_{\mathrm{Cas},n>0}$ in~\eqref{eq:Ucas_Lexpansion} and the higher-order expansion of $f=e^{J\cK} \tilde{M}_J(z)$, where $\tilde{M}_J$ is given by~\eqref{eq:MJ_momspace}. Given this data, the pseudodifferential calculus of the two previous sections allows for an algorithmic reconstruction $H_{\mathrm{symb}}^{(k)}$ at arbitrary order $k$. 

\subsection{A formal derivation of Bohr-Sommerfeld conditions}
\label{app:BSderivation}

In this section we sketch a formal derivation of the leading and the subleading Bohr-Sommerfeld conditions~\eqref{eq:BSsubleading_connected} for a special class of Hamiltonians. Specifically, we consider an eigenvalue problem
\be\label{eq:PDOeigenstateEq}
	H \psi = E\psi
\ee
for a holomorphic pseudodifferential operator $\cP$ acting on holomorphic functions $\psi(z)$ of one variable. We assume that the functions $\psi$ belong to a Hilbert space with the inner product
\be
	\<\psi_1|\psi_2\> = \int d^2 z \mu e^{-J\cK} \bar\psi_1\psi_2,
\ee
for some (non-holomorphic) functions $\cK$ and $\mu$ on $\C$. For simplicity, we assume that the exponents $\tau_k$ in the large-$J$ expansion~\eqref{eq:P_on_S0f} in the case of $H$ are given by $\tau_k=k$.

We will also assume that $H$ is equivalent to a Toeplitz operator via the logic described in section~\ref{sec:generalNsymbols}. Specifically, we will assume
\be
	\<\psi_1|H|\psi_2\>=\int d^2 z \mu e^{-J\cK} \bar\psi_1\psi_2 H_\text{symb}
\ee
for some symbol $H_\text{symb}=H_\text{symb}(z,\bar z)$. We have
\be
	H_\text{symb}=\p{\mu^{-1}e^{J\cK}}{}^tH\p{\mu e^{-J\cK}}.
\ee

The symbol $H_\text{symb}$ has an expansion of the form $H_\text{symb}=\sum_{k=0}^\oo J^{-k}H_\text{symb}^{(k)}$, where the symbols $H_\text{symb}^{(k)}$ can be determined from the symbol functions $(H)_k$ using the discussion in the preceding subsections. In particular, the leading symbol is given by (see~\eqref{eq:P0f})
\be\label{eq:leadingH0fromPDO}
	H_\text{symb}^{(0)}(z,\bar z)=({}^tH)_0(z, i\ptl_z\cK(z,\bar z))=(H)_0(z, -i\ptl_z\cK(z,\bar z)).
\ee
More generally, the relationship between the symbol functions of $H$ and its transpose ${}^tH$ is given in~\cite{HormanderPDO}.

Using the WKB ansatz 
\be
	\psi = f e^{i J S_0},
\ee
the leading-order part of~\eqref{eq:PDOeigenstateEq} becomes simply
\be\label{eq:leadingPDOeigenstateEq}
	(H)_0(z,\ptl_z S_0(z))=E.
\ee
This equation can be solved for $\ptl_z S_0(z)$, which can then be integrated to give $S_0(z)$. Single-valuedness of $\psi$ then requires
\be
	J\oint_\G \ptl_z S_0(z)dz= 2\pi k
\ee
along closed contours $\G$. Setting $\G=\G_E$ the constant-energy contour on which $H_\text{symb}^{(0)}=E$ and taking~\eqref{eq:leadingH0fromPDO} into account, we find
\be
	J\oint_{\G_E} \ptl_z S_0(z)dz=-iJ\oint_{\G_E}\ptl_z \cK(z,\bar z)dz=c_0(E).
\ee

This immediately leads to the leading-order Bohr-Sommerfeld condition
\be
	c_0(E)+\cdots = 2\pi k.
\ee
Furthermore, $c_0(E)$ gets the interpretation of the (leading-order) phase picked up by $\psi$ along the contour $\G_E$.

The subleading Bohr-Sommerfeld condition can be derived by analysing the subleading term in the equation~\eqref{eq:PDOeigenstateEq} using the methods described in this appendix.\footnote{We use the expansion scheme in which we define $S_0$ as the solution to~\eqref{eq:leadingPDOeigenstateEq} for the exact value of $E$. The latter then defines $f$, for which we seek a $1/J$ expansion.} Similarly to the above, the subleading term allows one to determine $\ptl_z \log f$ up to $O(1/J)$ corrections. The single-valuedness condition for $\psi$ is
\be
	J\oint_\G \ptl_z S_0 dz-i\oint_\G\ptl_z \log f=2\pi k.
\ee
We have already shown that for $\G=\G_E$ the first term reduces to $c_0(E)$. Our claim is that the second term reduces to $c_1(E)-\pi+O(1/J)$. This can be shown using derivatives of~\eqref{eq:leadingH0fromPDO} and the relation $\ptl_z H_\text{symb}^{(0)} dz+\ptl_z H_\text{symb}^{(0)} d\bar z=0$ on $\G_E$. The detailed calculation is straightforward but tedious, so we omit it here.

\newpage
\bibliographystyle{JHEP}
\bibliography{refs}

\end{document}